\documentclass{article}
\usepackage{arxiv}

\usepackage[utf8]{inputenc} 
\usepackage[T1]{fontenc}    
\usepackage{hyperref,url,booktabs,amsfonts,nicefrac,graphicx,doi,arxiv,JAM,enumitem,amsmath,amssymb,amsthm, varioref, caption, subfig,listings}

\title{\normalfont\spacedallcaps{Post-processing of coronary and myocardial spatial data}} 

\graphicspath{./figs/}

\renewcommand{\d}[1]{\partial #1} 
\renewcommand{\r}[1]{\mathcal{#1}} 
\newcommand{\R}[1]{\mathbb{R}^{#1}} 

\title{Post-processing of coronary and myocardial spatial data}

\author{Jay A.~Mackenzie \thanks{jay.mackenzie@glasgow.ac.uk} \\
    School of Mathematics \& Statistics\\	
    University of Glasgow\\    
	Glasgow, G12 8QQ\\
	\texttt{jay.mackenzie@glasgow.ac.uk} \\
 \And
	Megan J.~Miller\\
	Department of Applied Mathematics\\	
        Virginia Military Institute\\    
	Lexington, Virginia 24450\\
 	\texttt{chambersmj@vmi.edu}
  \And
	Mette S.~Olufsen\\
	Department of Mathematics\\	
        North Carolina State University\\    
	Raleigh, North Carolina 27607\\
 	\texttt{msolufse@ncsu.edu} \\
	\And
	Nicholas A.~Hill\\
	School of Mathematics \& Statistics\\	
        University of Glasgow\\    
	Glasgow, G12 8QQ\\
 	\texttt{nicholas.hill@glasgow.ac.uk} \\
}

\renewcommand{\headeright}{A Preprint}
\renewcommand{\undertitle}{A Preprint}
\renewcommand{\shorttitle}{Coronary Arterial Trees}

\hypersetup{
pdftitle={Post-processing of coronary and myocardial spatial data},
pdfsubject={q-bio.NC, q-bio.QM},
pdfauthor={Jay A.~Mackenzie, Megan J.~Miller, Mette S.~Olufsen, Nicholas A.~Hill},
pdfkeywords={Segmentation, Left Ventricle, Coronary, Vasculature, Myocardium},
}

\begin{document}
\maketitle

\begin{abstract}
Numerical simulations of real-world phenomena require a computational scheme and a computational domain. In the context of haemodynamics, the computational domain is the blood vessel network through which blood flows. Such networks contain millions of vessels that are joined in series and in parallel. It is computationally unfeasible to explicitly simulate blood flow throughout the network. From a single porcine left coronary arterial tree, we develop a data pipeline to obtain computational domains for haemodynamic simulations in the myocardium from a graph representing a partial coronary arterial tree. In addition, we develop a method to ascertain which subregions of the left-ventricular wall are more likely to be perfused via a given artery, using a comparison with the American Heart Association division of the left ventricle for validation.
\end{abstract}


\section{Introduction}
The mammalian circulatory system is commonly divided into pulmonary and systemic circulatory systems \cite{haynes1963physical}. The pulmonary circulation carries deoxygenated blood from the heart to the lungs where it is oxygenated, and returns it to the heart whence it is distributed throughout the systemic vasculature. The systemic arteries carry oxygenated blood around the body and the oxygen-depleted blood is returned to the right ventricle by the systemic arteries. The right and left coronary arteries, which supply the myocardium, are drained via the cardiac venous system \cite{HurstHeart} and are considered part of the systemic circulation. The size and configuration of the heart, lungs, and blood vessels varies between species and individuals and is influenced by diseases such as pulmonary hypertension and diabetes
\cite{hunter2011pulmonary,leitschuh1987vascular,akbari1999diabetes}. Furthermore, the structure, function, and morphometry of vascular networks vary throughout the body and within the organs. 

Heart disease is the leading cause of death worldwide \cite{lindstrom2022global}. The burden of heart disease, together with the complexity of the cardiovascular system, motivates the need for tools to gain a deeper understanding of the vasculature in health and disease. For example, haemodynamic modelling is an avenue of investigation that can yield rich results \cite{lee2016silico, muller2014venous,mynard2016influence,olufsen2012rarefaction,qureshi2014numerical} and insight into the function of a complex system. All computational haemodynamic models require equations to solve and a computational domain in which to solve them. The computational domains used in haemodynamic studies represent the vasculature in which blood is flowing. 

The analysis of vascular morphometry has a long history \cite{zamir1986branching}. Vascular morphometries are obtained in  one of two ways: by infusing a casting medium into the vasculature of interest and removing the tissue once the cast has set, or by infusing a contrast medium into the vasculature and imaging the organ. Image processing and computer vision techniques are used to separate the vasculature from the surrounding environment. After digitisation, the captured vasculature is represented by a set of voxels from which a skeletonisation can be obtained that traces the centre-lines of the vessels represented by the voxels. A skeletonisation is a set of vertices that are joined by edges; vessel radius is known at each vertex.

Vascular bed image collection and vessel segmentation are well documented; e.g., Nordsletten \textit{et al.} \cite{nordsletten2006kidney} and  Chambers \textit{et al.} \cite{chambers2020pulmonary} present segmented micro-CT images of the murine renal and pulmonary vasculature, respectively; Schuster \textit{et al.} \cite{schuster2010isolated} captured MR images of \textit{ex vivo} porcine coronary anatomy that were reconstructed by Goyal \textit{et al.} \cite{goyal2012vasculature}. Termeer \textit{et al.} \cite{termeer2010patient} extract the left epicardial surface and the three main coronary arteries from MRI images; they partition the left ventricle according to these three arteries using the Euclidean distance as a metric. Similarly, Ham \textit{et al.} \cite{ham2020myocardial} seek to subdivide the myocardium based on the largest vessel vasculature of the vasculature using the Euclidean distance between the myocardium and the vasculature as the metric for the division.

The morphology of the computational domain is a major determining factor in simulation prediction \cite{mackenzie2021thesis,chambers2020pulmonary,nordsletten2006kidney,hyde2014parameterisation,muller2014venous,vanbavel1992coronary,mynard2014scalability,mynard2016influence,qureshi2014numerical,olufsen2012rarefaction}. Therefore, when validating a computational model or moving toward patient-specific simulations, the precision of the computational domain compared to the physical vasculature is vital.

The work presented here is part of a larger effort to create detailed computational models of myocardial infarction and vascular rarefaction. A myocardial infarction is the blockage or partial occlusion of a coronary artery or vein that causes local damage. Here, we present methods for building computational domains for use in 1D haemodynamic models from graphs that represent a given vasculature. The filtration methods developed later can be tuned to produce a {pruned tree} that satisfies a given set of criteria on the number of branches, the number of generations, and the radius of the vessel. From the {reconstructed} vascular trees and the left ventricular surface mesh, we determine the region of the ventricle that is most likely to be perfused by a given artery. We find which vessels in the pruned tree give rise to the vessels excluded from it. By constructing surfaces around these, we can subdivide the ventricle into regions most likely to be perfused by a given vessel. These hulls are constrained to be non-intersecting. Several studies are seeking to subdivide the myocardium based on the geometry of blood vessels \cite{termeer2008visualization,termeer2010patient,ham2020myocardial}, however, to the best of our knowledge, none of them base their division on the location of small blood vessels.

Here, we generate several trees, and hence several subdivisions. Investigations into the impact of the vascular computational domain morphology on simulated haemodynamics require that trees be easily generated and conform to given constraints (such as on generation count and vessel size). A preliminary investigation of this type can be found in Mackenzie \cite{mackenzie2021thesis}. The work presented here is the culmination of the need to generate vascular trees and regions of the left ventricle perfused by each terminal vessel in an efficient and reliable manner.

\section{Materials \& Methods}

The raw data discussed here were shared with the authors by collaborators. Data are from a single 40\;kg Large White/Landrace-cross pig. Data were collected as described by Schuster \textit{et al.} \cite{schuster2010isolated} and were originally processed according to Goyal \textit{ et al.} \cite{goyal2012vasculature}. This dataset is also used by, e.g., Lee \textit{et al.} \cite{lee2016silico}. Further descriptions of data collection and processing can be found in Hyde \textit{et al.} \cite{hyde2014parameterisation} and Sinclair \textit{et al.} \cite{sinclair2015microsphere}. More detailed discussions of materials and methods can be found in \cite{horssen2009extraction, VanDenWijngaard2011porcine}. We assume that the data collection and processing methods are given.

The data given consists of a partial left coronary arterial tree and a left ventricular surface. The arterial tree is represented by a set of 17945 nodes $T$ each of which has a known location in $\mathbb{R}^3$. These nodes lie along the centre-lines of the coronary arteries. A positive radius value is given at each node. Let the location of the $i^{\rm th}$ node, $i \in [1, n = 17945]$ be $T(i)$ with radius $R(i)$. The nodes are joined by edges that are straight line segments between two adjacent nodes. Let the edge collection be $E$. There are 17944 unique edges. The vascular data form a graph, $G$.

The ventricular surface is represented by a set of nodes $V$. The nodes are joined by tetrahedral mesh elements to form a hollow ventricular surface. The sets $V$ and $T$ do not intersect in that neither contains a node in the other, but the spatial regions that they occupy do overlap -- see Fig.~\ref{fig:dorsal}.

\subsection{Error Correction and Graph Simplification}\label{sec:correction}

As the edges in $E$ are unique, it is straightforward to categorise a node given the number of times it appears in $E$. The nodes that appear exactly once in $E$ are \textit{terminal} nodes. Nodes that appear exactly twice are called \textit{body} nodes and lie between two junction nodes or between a junction and a terminal node. Nodes that appear more than twice are called \textit{junctions}. The \textit{order} of the junction refers to the number of daughters: a bifurcation is a one-to-two branching, in which case the junction node will appear three times in $E$.

\begin{figure}\centering
 	\includegraphics[height = 7 cm]{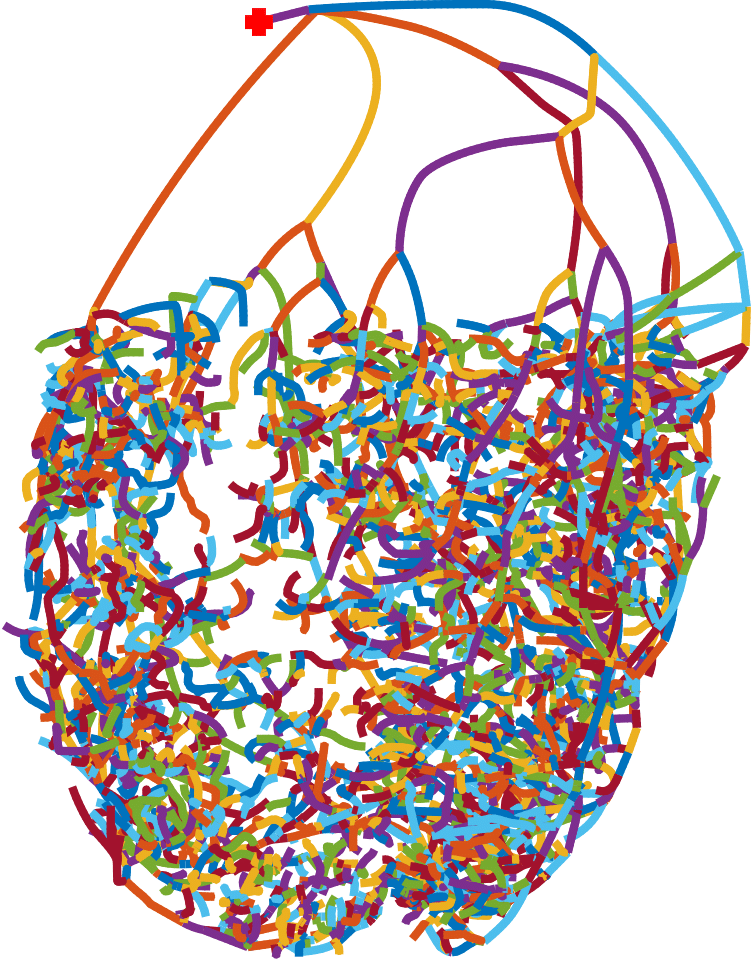}
\caption{A lateral view of the segments of the vascular tree. Line colour demarcates the segments. The root node is marked with a red plus.}\label{fig:segs}
\end{figure}

There is a unique terminal node, the \textit{inlet}, that is assumed to be at the root of the tree, representing the point at which the left coronary arteries arise from the aorta. This is chosen by inspection and is highlighted in Fig.~\ref{fig:segs}. 

Connected components are defined as those that can be reached by following edges from the initial node. A search for connected nodes is performed using the Dijkstra algorithm \cite{dijkstra1959note}. All nodes that cannot be reached from the initial node are defined to be disconnected and discarded -- there are 41 such nodes.  After disconnected nodes are discarded, their edges can also be discarded. There are now 17903 edges in $E$ and 17904 nodes in $T$.

To reduce the complexity of the graph, adjacent body nodes between two junctions or between a junction and a terminus are collected into \textit{segments}. Segments can be thought of as a set of edges $E'$ with the set of nodes $V'$ now only consisting of the junction and terminal nodes of $V$. Together, $E'$ and $V'$ also form a simply connected graph $G'$, but are significantly smaller than the original graph, so it is more convenient to work with --- there are 3910 segments but 17944 edges. Segments can be generated by choosing an inlet node and running Dijkstra's algorithm \cite{dijkstra1959note} to find paths from the root to all terminal nodes in the tree. This gives a path from the root to a given terminal node. These paths can be split at junction nodes to give paths between a junction and a junction or a terminus giving segments. As the collection of paths spans the tree, so does the collection of segments. Physiologically, coronary arterial flow is mostly one-directional \cite{mackenzie2021thesis}, so we endow the segments with an orientation to make a directed graph $G'$.

\begin{figure}[ht]\centering
    \subfloat[]{\includegraphics[scale = 0.45]{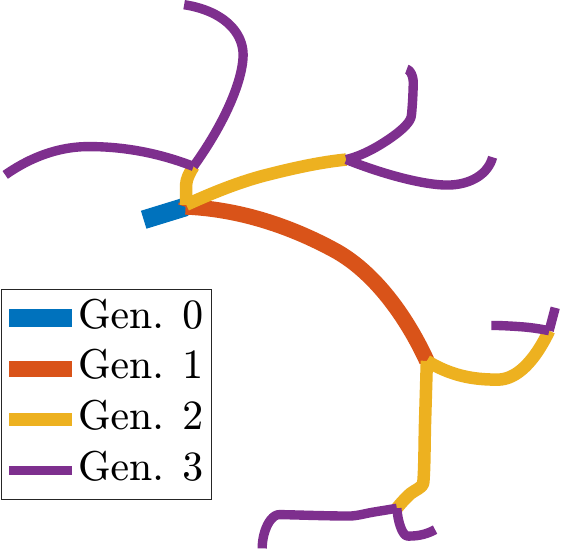}\label{fig:gens:bif}}\hspace{2cm}
    \subfloat[]{\includegraphics[scale = 0.45]{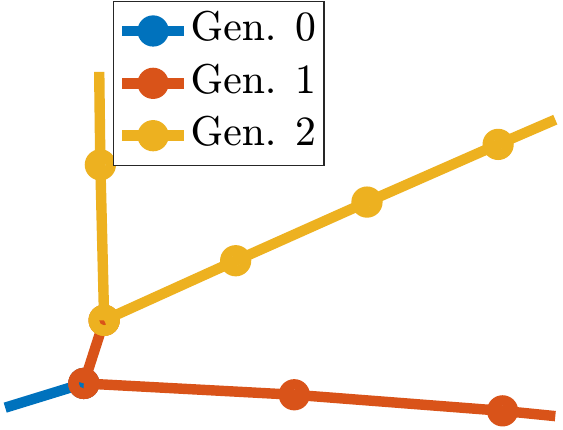}\label{fig:gens:bif:zoom}}\\
    \subfloat[]{\includegraphics[scale = 0.45]{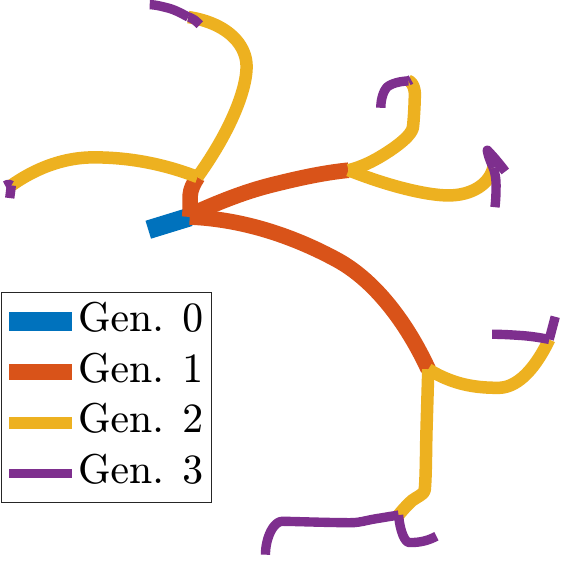}\label{fig:gens:trif}}\hspace{2cm}
    \subfloat[]{\includegraphics[scale = 0.45]{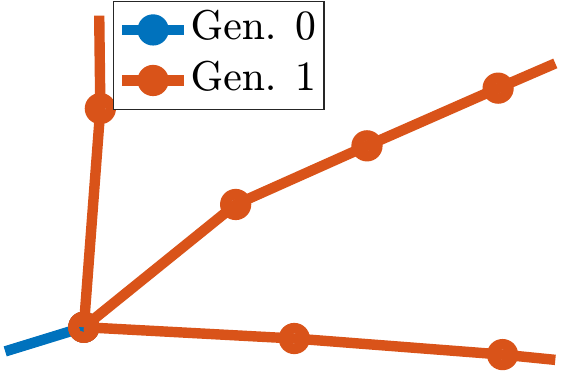}\label{fig:gens:trif:zoom}}\\
\caption{Coronal projection of the segments are sorted into generations. Line weight and colour signify generation affiliation. (a) All segments are sorted into generations without modification. (b) 30-fold magnification of the initial junction in the tree shown in (a); spatial nodes are highlighted with circles and the short segment is clear to see. (c) Pseudotrifurcations are removed and the resulting tree is shown with the same view as (a). (d) as in panel (b), but in the tree without pseudo-trifurcations.}
\end{figure}

By construction, a segment contains at least two nodes. However, any segment that contains only two nodes is very short. There are 761 segments that contain only two nodes. These are assumed to be artifacts of image segmentation and ought to be corrected. If a terminal branch is spuriously short, it is removed. If a spuriously short segment lies within the body of the tree, there will be two bifurcations in quick succession. We call this a \textit{pseudotrifurcation} and seek to replace these with trifurcations. Pseudotrifurcations are corrected by considering the daughters of the spuriously short segment and replacing the proximal node of each with the proximal node of the short segment. The spuriously short segment is discarded. The first two junctions of the tree form a pseudotrifurcation as can be seen in Fig.~\ref{fig:gens:bif:zoom}. The correction described here is applied to this pair of junctions to obtain the trifurcation seen in Fig.~\ref{fig:gens:trif:zoom}. To apply these rules, we must sort the segments into their respective generations.

\subsubsection{Generation Assignation}
The collection of segments can be hierarchically ordered according to the generation to which they belong. In addition to being helpful when removing spuriously short segments, generation affiliation can be used as a filter for segments. In either case, it is necessary to sort the nodes into generations. The $0^{\rm th}$ generation contains the segment with the inlet node. Given the $i$ sorted generations, the $(i+1)^{\rm st}$ generation contains segments that begin with the final nodes of the $i^{\rm th}$ generation. It is possible to sort any subset of the segments into generations. The 3910 segments in this data set can be sorted into 42 generations.

A coronal projection (onto the $(x, y)$-plane) of the first four generations is shown in Fig.~\ref{fig:gens:bif}. From the figure, it appears as if the vessels of generations 1 and 2 arise directly from the $0^{\rm th}$ generation. This occurs due to the presence of a spuriously short segment in $1^{\rm st}$ generation that is too short to be seen on such a scale. This short segment can be clearly seen in Fig.~\ref{fig:gens:bif:zoom} which is a 30 times magnification of Fig.~\ref{fig:gens:bif} centred at the first junction. The spuriously short segment contains only 2 nodes, while its sibling contains 51 nodes, and its daughters contain 9 and 39 nodes. In addition, the radii at both nodes of the segment are greater than its length. Given this, it is reasonable to presume that the short segment is an artifact of the initial skeletonisation. As discussed above, the initial node of spuriously short segments is appended to its daughter segments, so that the daughter segments are effectively moved up one generation, and the spuriously short segment can be removed without loss of information. {{The impact of this can be seen by comparing Fig.~\ref{fig:gens:bif:zoom} and Fig.~\ref{fig:gens:trif:zoom}.
It is by this procedure that we transform a pseudotrifurcation into a trifurcation. This is easily generalisable to higher junction orders.}}

Upon removing the 761 short segments, there are 3149 segments in 32 generations.

It is not clear if segments containing 3 nodes should be considered as spuriously short as none of the 381 such segments has Euclidean arc-length less than its mean radius. As such, no segments of length 3 or longer are considered spuriously short.

\subsection{Obtaining Principal Pathways} \label{sec:filter}
Here, we discuss the application of sorting and filtration methods to the 3910 segments to obtain a \textit{principal pathway}. Principal pathways are the collection of arteries that transport blood to the myocardium, where it is distributed by the small arterial network \cite{chambers2022morphological}. We call the process of finding the principal pathway as \textit{pruning}. Filters are used to select segments that are eligible for inclusion in a final principal pathway. A principal pathway is also a subtree of the original, unpruned tree, as it can only contain segments and nodes that were present in the original tree. All segments in the principal pathway must be accessible from the root, i.e.~the graph must be simply connected.

\subsubsection{Setting a Maximum Number of Generations}
\begin{figure}[ht]\centering
 	\includegraphics[height = 4 cm]{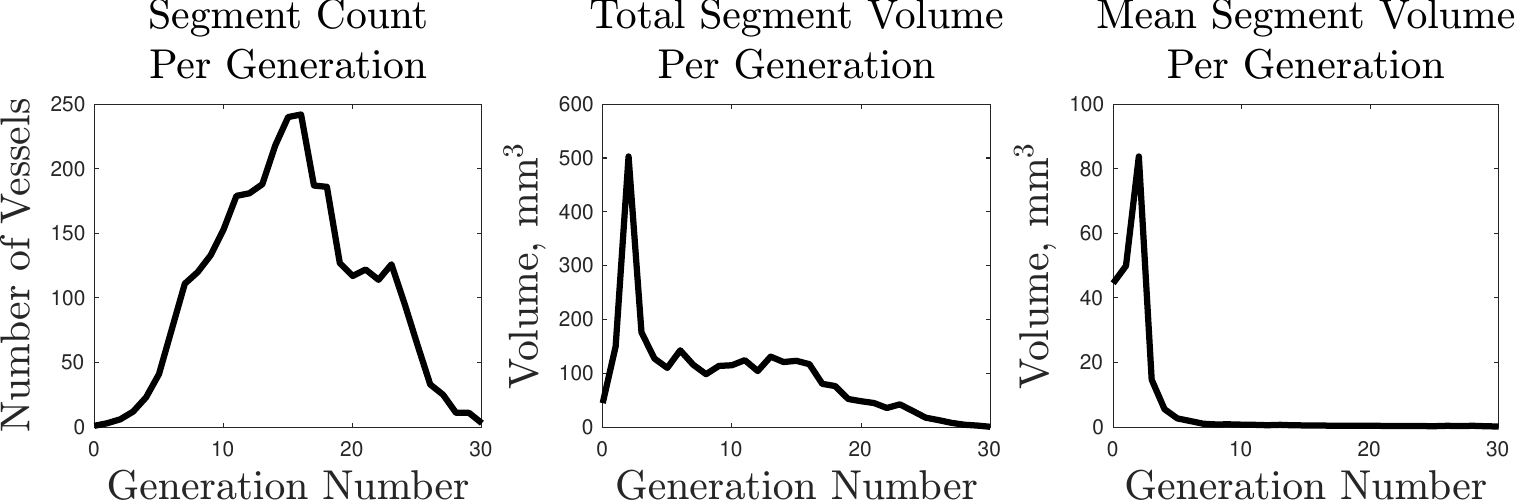}
    \caption{The panels show (a) the number of vessels, (b) the total volume of those segments, and (c) the mean volume of a segment in each of the 32 generations. The mean volume of a segment in each generation is the quotient of the total volume of a single generation by the number of segments in that generation.
    }\label{fig:info}
\end{figure}

One possible application of the trees generated using the current analysis is in haemodynamic flow models. In these, there is a trade off between model fidelity and computational cost. Flow simulations in trees with more branches are more computationally costly than those with fewer branches, not just because of the branch count itself, but because the junctions between vessels can be expensive to match \cite{mackenzie2021thesis}. In two-sided models, coupling arterial to venous networks, such as those discussed by Mackenzie \cite{mackenzie2021thesis} and Qureshi \textit{et al.} \cite{qureshi2014numerical}, a single additional generation of vessels can as much as quadruple computational costs. Further, as shown by Mackenzie \cite{mackenzie2021thesis}, the flow in all vessels is correlated to the total volume of the network. In the tree, segment volume typically decreases as a function of of increasing distance from the root. In this tree, the number of vessels per generation is increasing until the 16$^{\rm th}$ generation and contains a total of 1927 segments. Given this, it is reasonable to ask what the maximum number of generations that ought be included in the tree is. One possible metric that can be used to help answer this question is the information density, $I$. Let $n$ be the number of segments per generation and $v$ be the total volume of the segments in a generation. Generations are indexed by a non-negative integer $i$. Information density is defined to be $I(i)\coloneqq v(i)/n(i)$ in the $i^{\rm th}$ generation. Figure \ref{fig:info} shows graphs of $n(i)$, $v(i)$, and $I(i)$ as discrete functions of $i$. Information density has a peak when $i=2$ and decreasingly decreasing from then on. The additional information gained by adding another generation can be quantified by finding $\Delta I(i+1) = I(i+1)-I(i)$ for $i \geq 0$. This peaks at $i=0$, between the 0$^{\rm th}$  and 1$^{\rm st}$  generations. \textit{We choose} the threshold on $\Delta I$ for excluding a generation be two orders of magnitude lower than its peak, i.e. all generations after the first for which
\[
\vert \Delta I(i) \vert < \frac{{\rm max}(\Delta I (i))}{100}
\]
should not be included. This happens first when $i=5$. If one wishes to use a generation based filter to reduce the number of segments eligible for inclusion in a tree, a reasonable choice for an upper bound based on this metric is 5. The information density is used as a proxy for the balance between computational domain fidelity and computational time as a voluminous generation of few vessels should certainly be included in the computational domain, but a generation containing numerous small vessels will have a significant impact on computational time without a significant impact on simulation results.  The first 8 generations of the tree make up 51\% of the total volume of the tree. Hence, including more than 8 generations may increase the fidelity of a fluids simulation (subject to modelling assumptions), but beyond 8 generations computational costs will increase more quickly that computational domain fidelity increases.
 
\subsubsection{Radius-Based Metrics}
Radius is another natural metric by which to exclude a segment from the principal pathway. 
Every segment contains at least two nodes, and all nodes have an associated radius value. In radius-based pruning, a radius threshold is given. The radii in this coronary tree are not monotonically deceasing with distance from the root node, so applying a radius-based pruning condition is not as straightforward as finding the segments that contain a node that exceeds a threshold. There are several possible ways to apply a radius-based pruning condition. Here, we discuss a mean radius condition, a proportional condition, and a single-node condition.

\begin{figure}[ht]\centering
 	\subfloat[]{\includegraphics[height = 5 cm]{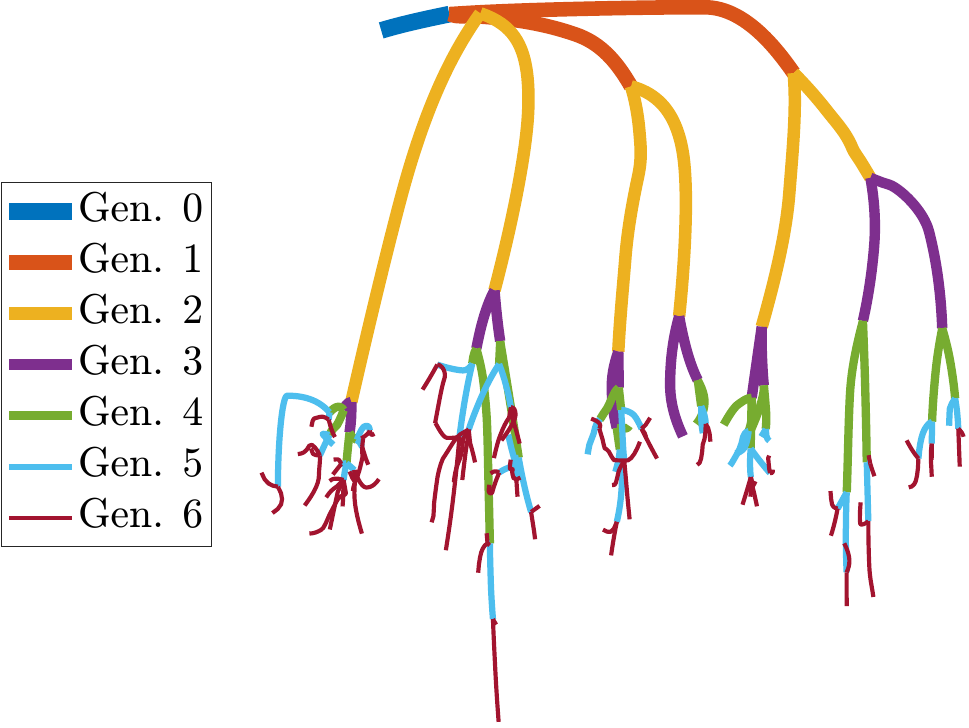}\label{fig:six}}\hspace{2cm}
  	\subfloat[]{\includegraphics[height = 5 cm]{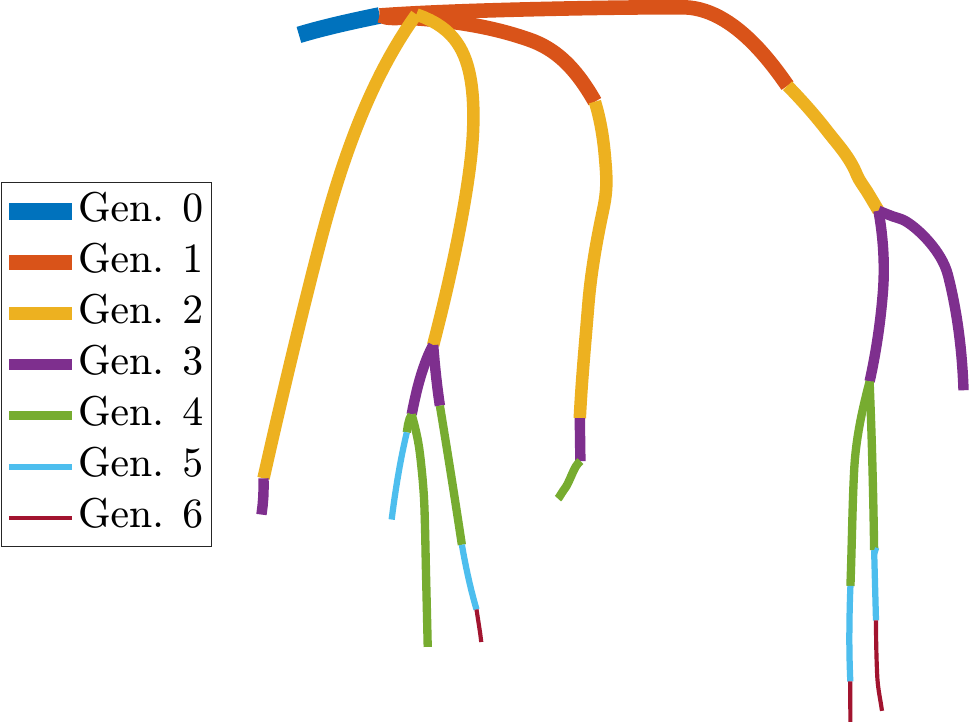}\label{fig:six:rad}}\\
  	\subfloat[]{\includegraphics[height = 5 cm]{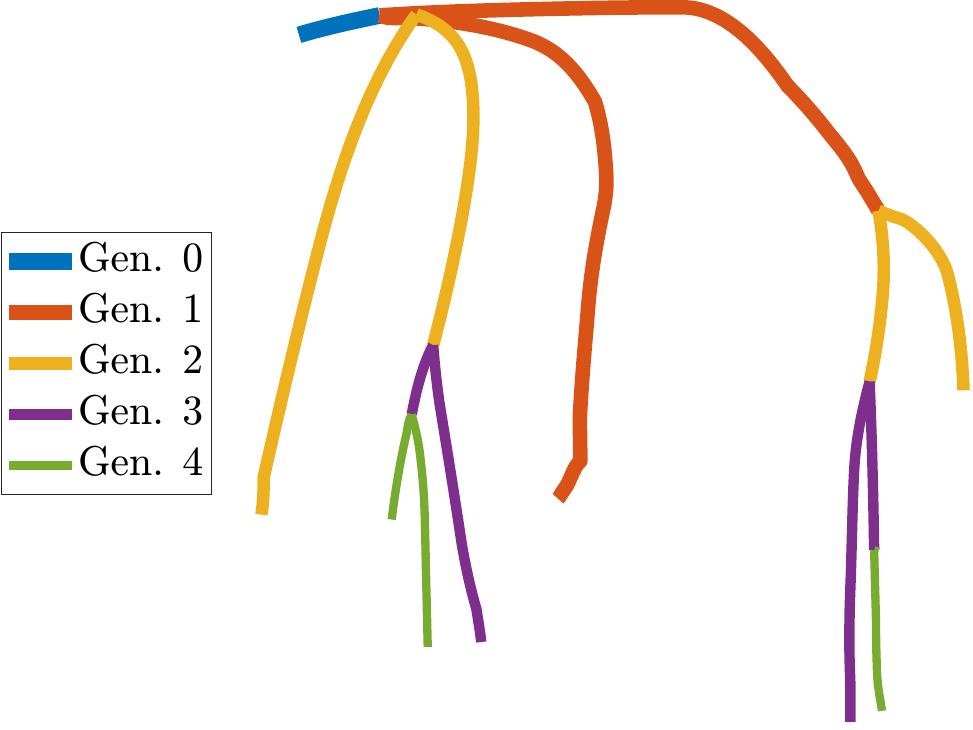}\label{fig:six:rad:series}}\hspace{2cm}
  	\subfloat[]{\includegraphics[height = 5 cm]{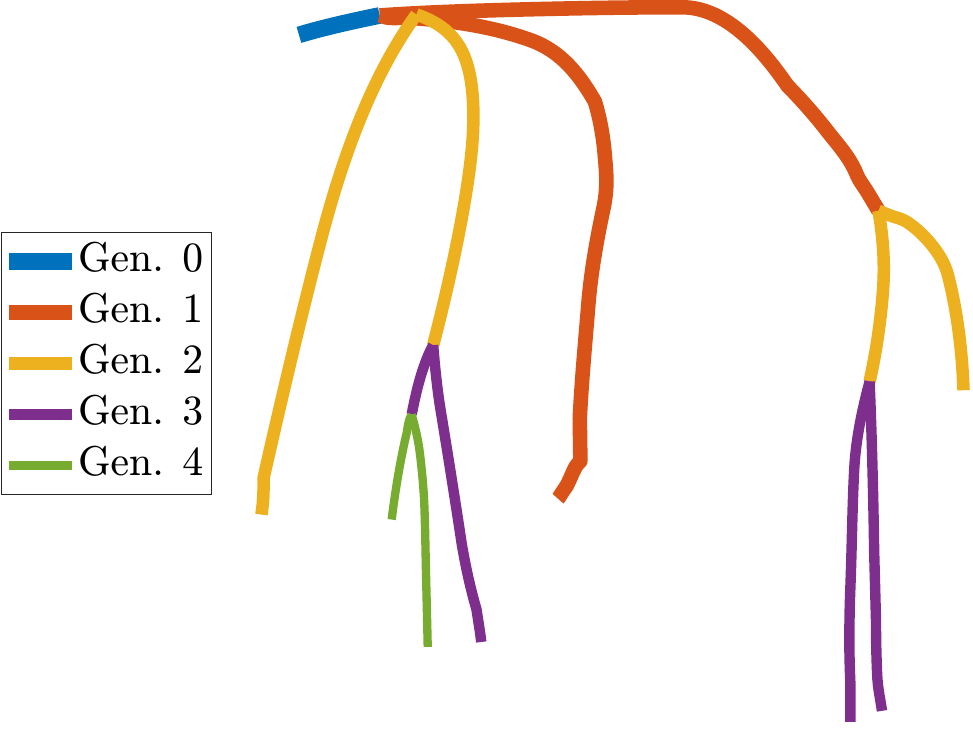}\label{fig:six:rad:series:twice}}
\caption{The evolution of a tree as it passes through some of the filters discussed here; line weight and colour indicate generation affiliation; the Mercator projection preserves the tree structure but not segment length. (a) Generations 0 -- 6 of the tree without pseudotrifurcations. (b) Segments from (a) that additionally have mean radius of at least 0.6 mm. (c) The segments from (b) have been joined in series. (d) Short terminal segments from (c) have been removed and the series join has been repeated.
}\label{fig:firstSix}
\end{figure}

\begin{figure}\centering
 	\subfloat[]{\includegraphics[height = 5 cm]{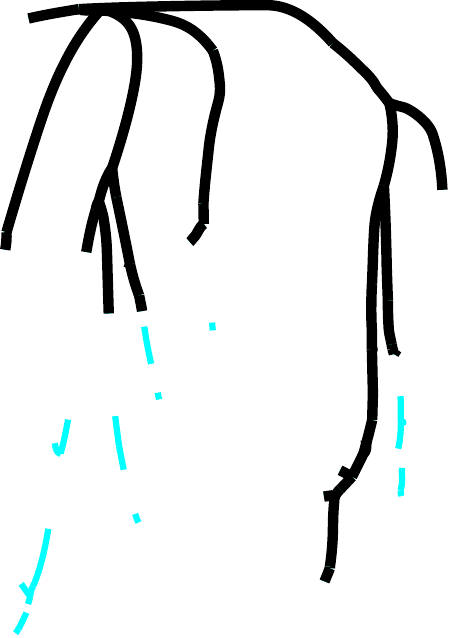}\label{fig:rad:mean}}\hfill
  	\subfloat[]{\includegraphics[height = 5 cm]{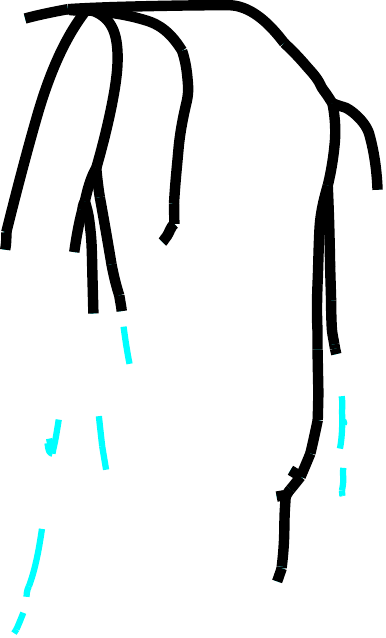}\label{fig:rad:pct}}\hfill
  	\subfloat[]{\includegraphics[height = 5 cm]{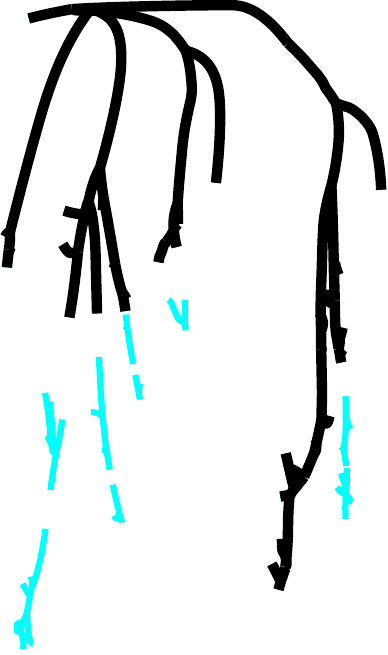}\label{fig:rad:single}}
\caption{Radius filtered and pruned trees with a threshold set at 0.6\;mm. The largest connected subtrees are shown in solid black and disconnected segments in cyan. (a) The mean radius condition is applied; there are 47 connected segments and 21 disconnected. (b) A proportion threshold is set to 0.8 and applied; there are 26 connected segments and 21 disconnected. (c) A single point threshold is applied to yield a connected tree of 75 connected segments and 53 disconnected.}\label{fig:rad}
\end{figure}

\begin{itemize}
    \item[] \textit{Mean Radius:} In a mean radius condition, each segment is considered individually, and if the mean radius of the nodes in that segment exceeds a given threshold, then the segment is eligible for inclusion in the final tree. This method may be sensitive to spurious radius values that skew the distribution of a vessel. Comparing, for example, the mean and median radius in each segment may highlight segments in which this is an issue. A tree filtered by mean radius can be seen in Fig.~\ref{fig:rad:mean}.
    \item[] \textit{Proportional Threshold:} If a certain proportion of nodes in a segment have a radius greater than the given threshold, then that segment is eligible for inclusion in the final tree. This condition may be ill suited to dealing with segments that contain few nodes.
    \item[] \textit{Single Node:} If a single node in a segment has an associated radius that exceeds the threshold, then it is eligible for inclusion in the final tree. Including any segment with at least one node with a radius that exceeds the threshold will lead to the inclusion of more vessels than the above-described mean radius condition. However, it will be sensitive to the existence of spuriously small radii values. A single node-filtered tree can be seen in Fig.~\ref{fig:rad:single}.
\end{itemize}
Other radius-based conditions exist, and it is certainly possible to apply multiple conditions within a single tree based, for example, on segment length. A proportional-threshold filtered tree can be seen in Fig.~\ref{fig:rad:pct}. In general, we use a \textit{proportional threshold} and require that at least 75\% of the nodes in a given segment exceed the lower radius bound to be eligible for inclusion. 
\textbf{This proportion was chosen after some experimentation to avoid setting an overly strict or overly lenient inclusion criterion. An overly strict criterion may limit the variety of trees that could be produced and an overly lenient criterion would produce trees that do not necessarily conform to the desired constraints.}

Figure \ref{fig:firstSix} shows the Mercator projection of the first six generations of segments. In particular, Fig.~\ref{fig:six:rad} shows the connected tree of segments in the first six generations that have a mean radius greater than 0.6\;mm. Generations 0 through 6 contain 162 segments, 28 of which meet the radius condition.

Rendering 3D spatial data in a 2D medium may lead to difficulty in the interpretation of the figures. To render the 3D trees here, we use a Mercator projection \cite{miller1942projections} that preserves the branching structure of the tree but not the length of the segments.

\subsubsection{Removing Disconnected Segments}
As radius is not monotonically decreasing with distance from the root node, some segments give rise to others in which the radius in the daughter segment exceeds that of the parent. As such, radius-based filtration methods give a set of segments that meet the given criterion for generation affiliation and radius but are not necessarily accessible from the inlet of the tree since intermediate segments have been filtered out. This results in a tree in which some segments are not connected to the component contained in the inlet node. A principal pathway must be a tree in which all segments are connected to the component containing the inlet node. Hence, if pruning results in segments becoming disconnected from the inlet, a principal pathway can be obtained in, at least, one of two ways: by removing the components that are disconnected from the component containing the inlet node; or by reconnecting the disconnected segments to the component containing the inlet node.

The former is straightforward if the generations of all segments have already been computed. Given a collection of segments, it is easy to construct sets that contain the initial $S_i$ and final $S_f$, nodes of all segments. In a connected tree, $S_i \setminus S_f$ should contain only the inlet node; any other node that is returned lies at the inlet of a disconnected segment that should be removed. Recompute $S_i$, $S_f$, and $S_i \setminus S_f$ while $|S_i \setminus S_f| > 1$. This is the procedure used to remove the highlighted segments in cyan seen in Figs.~\ref{fig:rad}.

\subsubsection{Segment Concatenation}
As can be seen in Fig.~\ref{fig:six:rad}, there appear to be segments that give rise to exactly one other vessel. As mentioned, much of the computational time spent during flow simulations with these networks is spent on matching the boundaries between vessels, hence it is computationally advantageous to join a parent segment and its single child to form a one longer segment. Such a join is advantageous as it reduces the complexity of the tree structure without loss of information.

Figure \ref{fig:six:rad:series} shows the tree from Fig.~\ref{fig:six:rad} in which series joins have been made. The resulting tree (Fig.~\ref{fig:six:rad:series}) contains 16 segments.

\subsubsection{Further Length Based Filtration}
Earlier, we saw that all segments containing two nodes are considered to be spuriously short. This is not the only circumstance under which it might be useful to remove a short segment. In Fig.~\ref{fig:six:rad:series} there is an instance in which a long segment (in generation 3) appears to give rise to exactly one other vessel (in generation 4) -- it can be seen in the lower right of the panel of Fig.~\ref{fig:six:rad:series}. However, this cannot be the case that such vessels have been concatenated to form a single longer vessel. It transpires that the vessel of generation 3 gives rise to two vessels containing 21 nodes (which is appreciated) and 3 nodes (which cannot be appreciated). The sibling containing 3 nodes is deemed to be short and is removed from the tree. Once segment concatenation has been performed again, we obtain the tree seen in Fig.~\ref{fig:six:rad:series:twice}.

Currently, there is no clear optimal criterion for the additional removal of short segments from a tree. More work is needed to establish this condition. At present, a segment is removed if it contains fewer than 5 nodes and gives rise to no further segments that are eligible for inclusion.

\subsubsection{Strahler Ordering}
The Strahler order of a tree is a metric of its branching complexity. The metric was first developed by Horton \cite{horton1945erosional} and later by Strahler \cite{strahler1952hypsometric, strahler1957quantitative} for applications in hydrology. Strahler ordering has been widely applied to vascular trees; see, for example, \cite{jiang1994lung, nordsletten2006kidney, schwen2012liver, vanbavel1992coronary}.
Strahler numbers are assigned from the bottom up as follows:
\begin{enumerate}
    \item if a branch has no children, it is given order 1
    \item if a branch has a child of order $n$ and all other children have order less than $n$, then give the branch order $n$
    \item if the branch has two or more children of order $n$, and no children with greater order, the Strahler order is $n+1$.
\end{enumerate}

\begin{figure}\centering
\includegraphics[width = 0.35\textwidth]{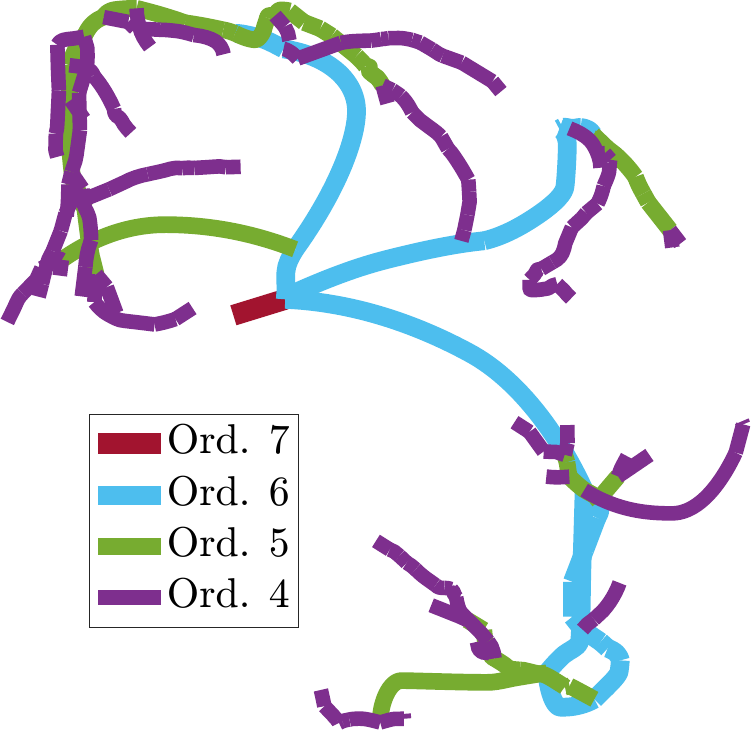}
\caption{Segments of the vascular tree with Strahler orders 4 -- 7. Line weight and color indicate the order to which a vessel belongs.}\label{fig:strahler}
\end{figure}

Figure \ref{fig:strahler} shows the largest segments of the tree that belong to orders 3 through 7. Vessels of orders 1 and 2 are omitted from the figure as they are sufficiently numerous to occlude other vessels in the tree.

Briefly, we have taken a given graph comprised of nodes, edges, and a radius measurement at every node and constructed a simplified graph by creating segments that run between two junction nodes or a junction node and a terminal node. The segments are sorted into generations, after which spuriously short segments are removed. The resulting tree can be pruned to conform to given criteria on, for example, the generation, Strahler order, or radius of a given vessel. If such filtration results in a disconnected tree, then the disconnected segments are discarded. If a terminal segment is very short, it is discarded. If a branch gives rise to a single daughter segment, these are concatenated to form a single longer segment that contains all the nodes of the two segments that were concatenated.

\subsection{Ventricular Data}
With the analysis described above, it is possible to generate coronary arterial networks that adhere to certain conditions. These networks have terminal segments, which in turn give rise to a tree of small vessels. Such small vessels, which perfuse the wall of the left ventricle, determine the regions of the myocardium that are perfused by a given coronary artery. We construct a physiologically motivated division of the left ventricle.

\begin{figure}[ht]\centering
 	\subfloat[The set $T$ with boundary $\d T$ and region $\r T$.]
  {\includegraphics[width = 0.3 \textwidth]{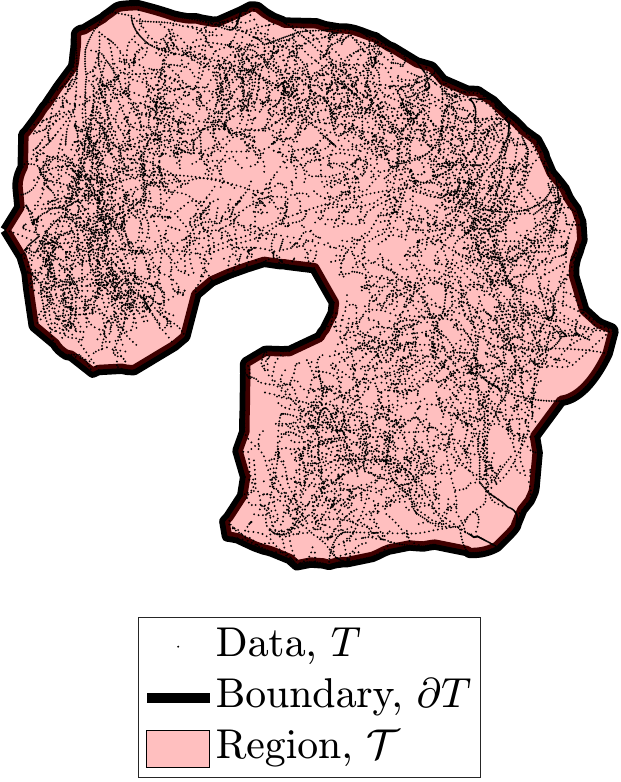}
            \label{fig:dorsal:T}}\hspace{0.1 \textwidth}
  	\subfloat[The set $V$ with boundary $\d V$ and region $\r V$.]
   {\includegraphics[width = 0.3 \textwidth]{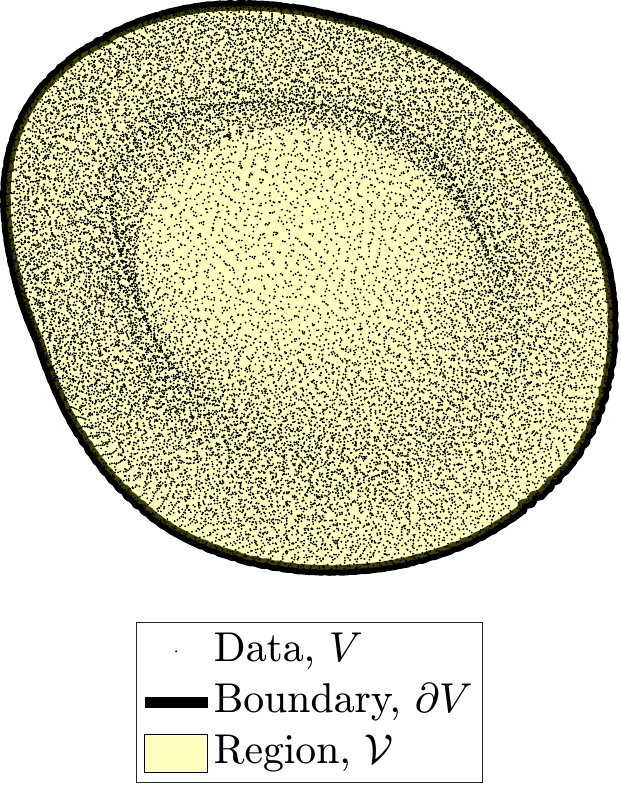}
            \label{fig:dorsal:V}}\\
  	\subfloat[The regions $\r V$ and $\r T$ are shown in cyan and yellow, respectively. Their intersection $(\r V \cap \r T)$ is shown in grey.]{\includegraphics[width = 0.3 \textwidth]{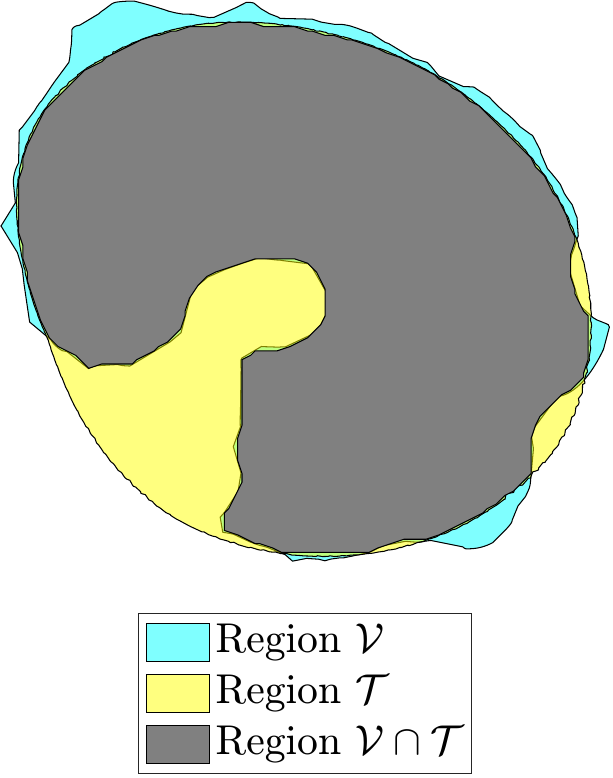}
            \label{fig:dorsal:overlap}}\hspace{0.1 \textwidth}
  	\subfloat[Regions of the ventricle that are ({$\r T \cap \r V$}) and are not ($\r V \setminus \r T$) reached by the arterial tree.]{\includegraphics[width = 0.3 \textwidth]{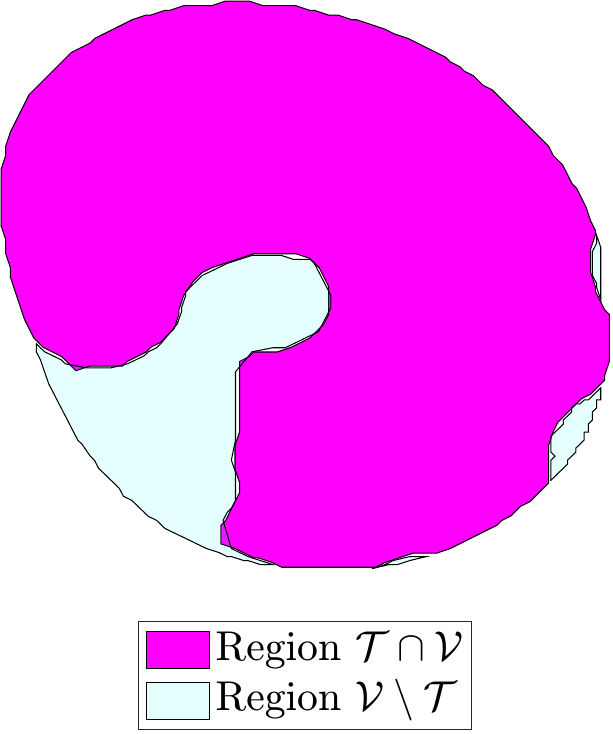}
            \label{fig:dorsal:setminus}}
   \caption{Dorsal projections of set operations on the regions and boundaries of $V$ and $T$ to isolate the unperfused regions of the ventricle $\r V \setminus \r T$.}\label{fig:dorsal}
   \end{figure}

Recall that $T \in \mathbb{R}^3$ is the collection of nodes in the vascular tree and $V \in \mathbb{R}^3$ is the collection of nodes that define the ventricular surface. Let the boundaries of $T$ and $V$ be denoted by $\d T$, and $\d V$, respectively, and the boundaries enclose the regions $\r{T}$, $\r{V}$. The data $T$ are discrete, but the region $\r T \subset \R3$ is continuous, and similarly for $V$. The regions overlap, but neither is completely contained in the other. The dorsal projections of $\R3 \to \R2$ of $T$, $\d T$ and $\r T$ can be found in Fig.~\ref{fig:dorsal}(a), and similarly for $V$ is Fig.~\ref{fig:dorsal}(b). It is obvious that there is a ventricle subregion $\r V$ that is not perfused from the left coronary arterial tree, as this defines the region $\r T$. We identify this seemingly unperfused subregion as the ventricular septum \cite{HurstHeart}. The septum is perfused from the right coronary artery tree.

\subsubsection{The Septum}
\begin{figure}\centering
 	{
  	\includegraphics[width = 0.5\textwidth]{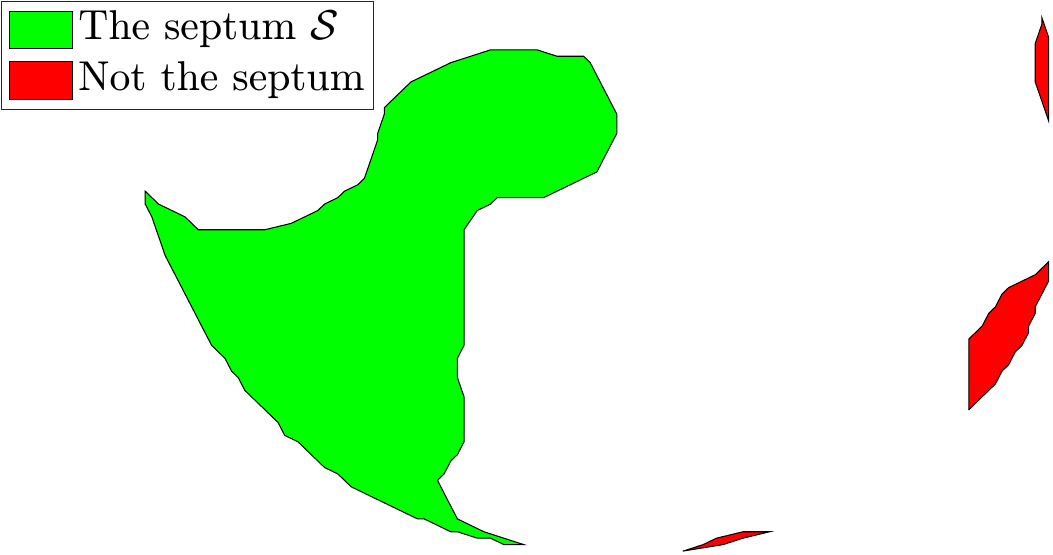}}
\caption{The separation of the septum $\r S$ from the islands of $\r V \setminus \r T$.}\label{fig:septum}
\end{figure}

By the definition of the septum here, there are nodes that belong to the septum and the ventricle, but the tree does not reach part of the ventricle. Mathematically this is $S \subset V$ and $\r S \cap \r T = \emptyset$, so $\r S \subset \r V \setminus \r T$ where $S$ is the collection of septal nodes, and $\r S$ is the septal region. Figure \ref{fig:dorsal:overlap} shows the regions $\r T$ (yellow) and $\r V$ (cyan); the intersection $\r T \cap \r V$ can be seen in grey. There are disjoint subregions that comprise the unperfused region $\r T \setminus \r V$ seen in Fig.~\ref{fig:dorsal:setminus}. The largest of these $\r S$, highlighted in green in Fig.~\ref{fig:septum}, is isolated manually using a data brushing tool. 

Although the depictions here have been in $\mathbb{R}^2$, as the regions all belong to $\mathbb{R}^3$ by finding $\r S \in \mathbb{R}^2$, $\r S \in \mathbb{R}^3$ is now also found.

\subsubsection{Initial subdivision}
Let the region of the left ventricle that is perfused from the left coronary tree be $\r L = \r V \setminus \r S$. As can be seen in Fig.~\ref{fig:septum}, there are unperfused regions that are not connected to the septum. These are assumed to be perfused and arise due to imaging constraints.

Principal pathways are generated from vascular data using the methods described in Subsections \ref{sec:correction} and \ref{sec:filter}. Let the principal pathway with $n$ terminal branches be denoted by $B_n\subset T, n \in \mathbb{Z}_{\geq 2}$. The perfused ventricular region $\r L$ should be divided into subdomains $n$. Here, this division is informed by the vasculature. The set of terminal nodes of $B_n$ is $\d B_n$, and the principal pathway is the region $\r {B}_n$.

Let the $n$ subregions of $\r L$ be denoted $\r S_i$, $i = 1, \dots, n$. Here, the $i^{\rm th}$ subregion is perfused from the $i^{\rm th}$ terminal artery of $B_n$, which is the element $i^{\rm th}$ of $\d{B}_n$ is ${\d{B}_n}_i$. By construction, the ventricle $\r V$ is the union of the subdomains
\[
\r V = \left(\bigcap_{i=0}^n \r S_i\right)
\]
where the septum $S$ is the subdomain $0^{\rm th}$ $S_0$. The node ${\d{B}_n}_0$ is the inlet node.

To generate the regions $\r S_i$, $i = 1, \dots n$ based on the vascular tree, the nodes in $T$ downstream of ${\d{B}_n}_i$ must be found for each value of $i = 1, \dots, n$. The set $U = T \setminus B_n$ contains all nodes not in the principal pathway and, therefore, contains all nodes downstream of the principal pathway. The set $U$ also contains nodes that cannot be reached from the terminii of the principal pathways.

\begin{figure}[ht]\centering
 	\subfloat[Mercator projection of the results of the terminal node search for the principal pathway $B_6$.]{
    \includegraphics[height = .2\textheight]{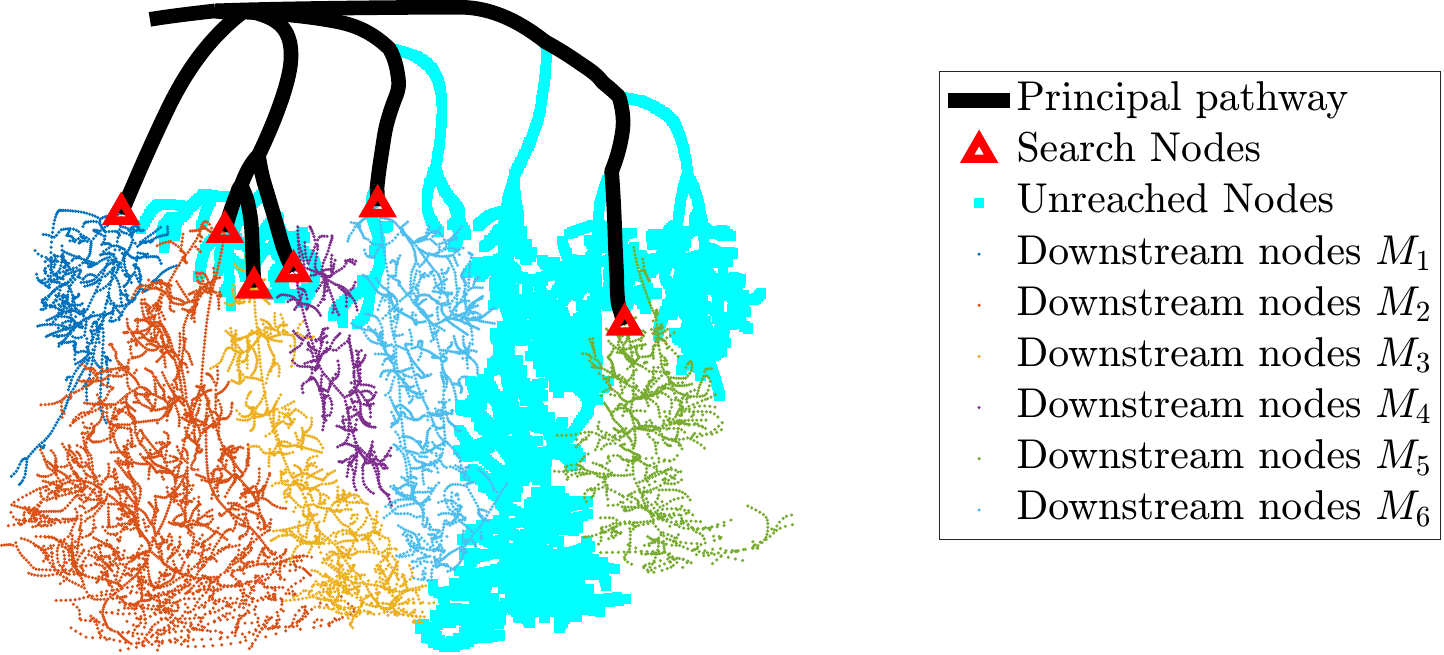}\label{fig:From:Show}}\\
    \subfloat[Mercator projection of the results of the initial node search for the principal pathway $B_6$.]{
    \includegraphics[height = .2\textheight]{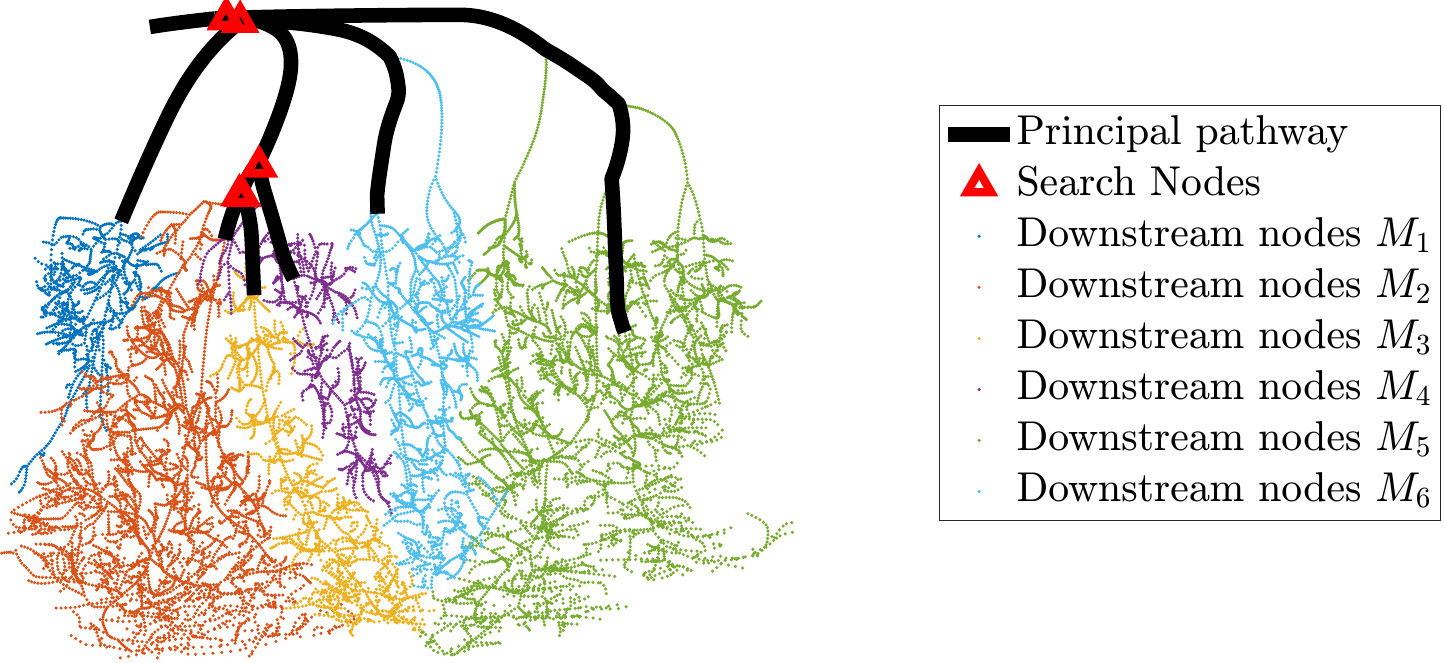}\label{fig:From:Search}}\\
    \subfloat[
    Transverse (left), sagittal (middle), and coronal (right) projections of the results of the initial node search for the principal pathway $B_6$.
    ]{
    \includegraphics[height = .2\textheight]{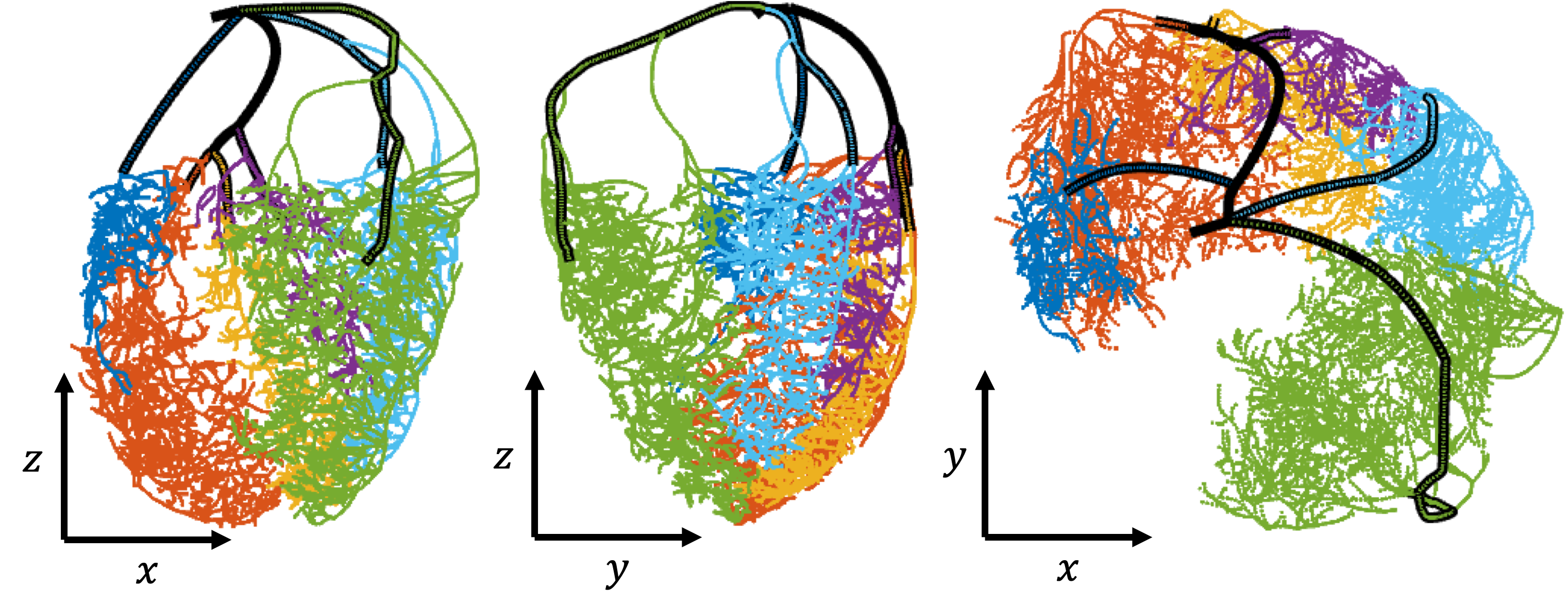}\label{fig:From:3D}}
   \caption{
   Mercator projections of $B_6$ (solid black lines) showing $M_i$ for $i \in [1, 6]$ in coloured points. The nodes contained within $U$ -- those that have not been assigned to a downstream tree -- are shown as large cyan dots. The nodes from which downstream nodes were searched are shown as red triangles.}
\end{figure}

It is straightforward to determine which nodes in $U$ lie downstream of the terminal nodes in a given principal pathway. We do this considering the path between the initial node and a given node in $U$. If the path contains the index of a terminal node of the principal pathway, the given node is downstream of that terminal node and hence is perfused from the terminal branch to which it belongs. 
If the given node lies downstream of the terminal node of the $i^{\rm th}$ terminal segment of the principal pathway, sort it into a set $M_i$ and remove it from $U$, otherwise it is left in $U$. Repeat this for all nodes in $U$ until they have all been considered. No node need be considered twice. This is called the \textit{terminal node search}. The subsets $M_i \subset U$, $i = 1, 2, ..., n$, $n=6$, contain discrete spatial nodes and are shown in Fig.~\ref{fig:From:Show}. After constructing the sets $M_i$, there may still be nodes in $U$, now considered to be the unreachable nodes; these are shown as cyan dots in Fig.~\ref{fig:From:Show}. The terminii of the principal pathway from which the sets $M_i$ are sought are shown as red triangles in Fig.~\ref{fig:From:Show}. The sets $M_i$, $i \in [1, n]$, $n=6$ are also shown in Fig.~\ref{fig:From:3D} together with the principal pathway.

In the illustrative case $B_6$, there are 4821 nodes remaining in $U$ of an original 17448 shown as cyan dots in Fig.~\ref{fig:From:Show}. It seems reasonable that a collection of nodes that branch directly from a terminal segment (but not the terminus) should be considered to be perfused by that segment. This is the rationale for moving the seed points from which $U$ and $M_i$ are generated to the initial nodes in each terminal branch. Denote the resulting vascular compartments as $M_i'$ and note that in the case of $B_6$, there are no unassigned nodes in $U'$ when looking for downstream nodes from the terminal branches. It is not always true that $|U'|=0$ (as in the case of $B_8$), but it is guaranteed that $|U'| \leq |U|$. This is called the \textit{initial node search}.

\begin{figure}[ht]\centering
 	\subfloat[]{
    \includegraphics[width = .75\textwidth]{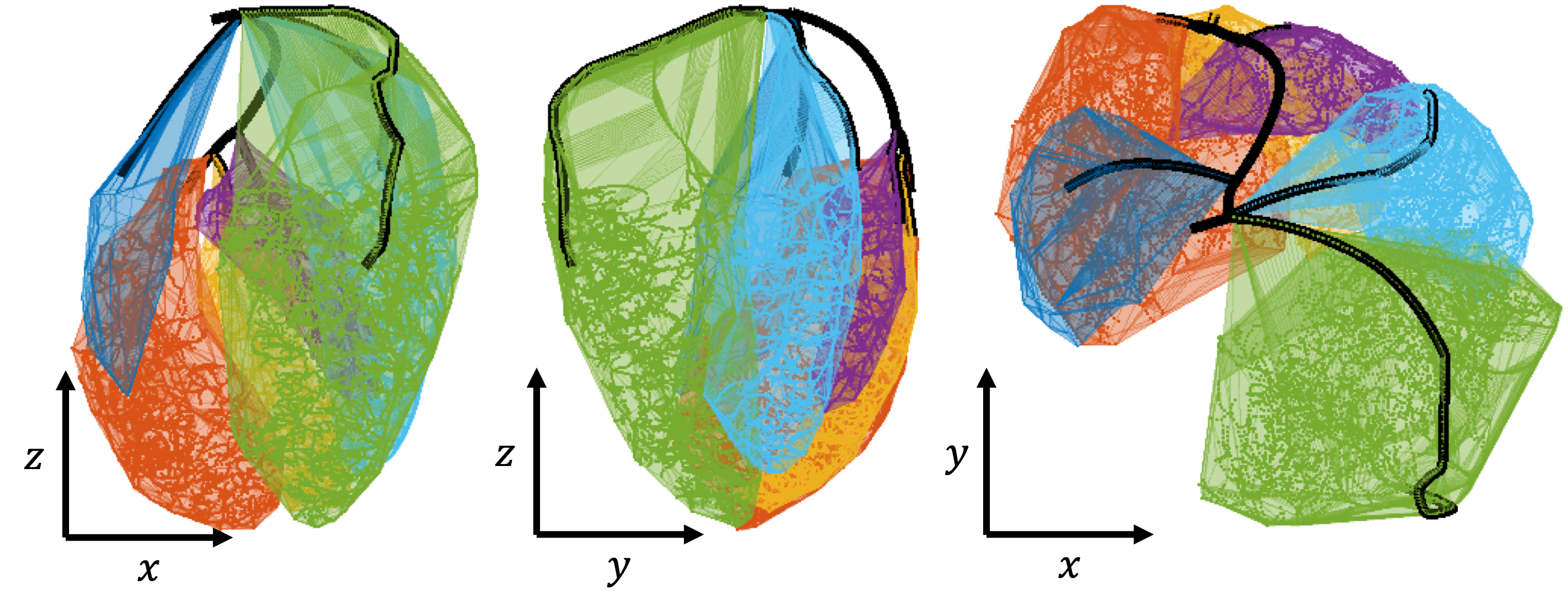}\label{fig:vasc_region}}\\
    \subfloat[]{\includegraphics[width = .75\textwidth]{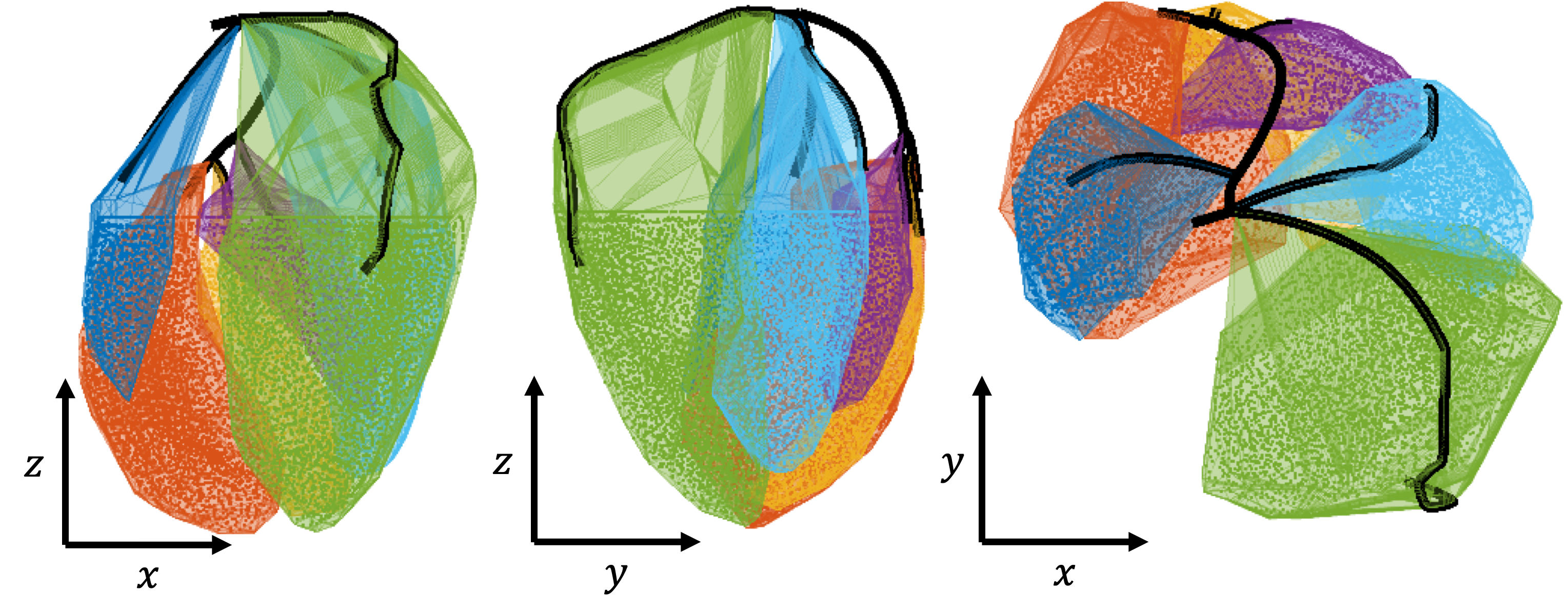}\label{fig:intersection}}\\
    \subfloat[]{\includegraphics[width = .75\textwidth]{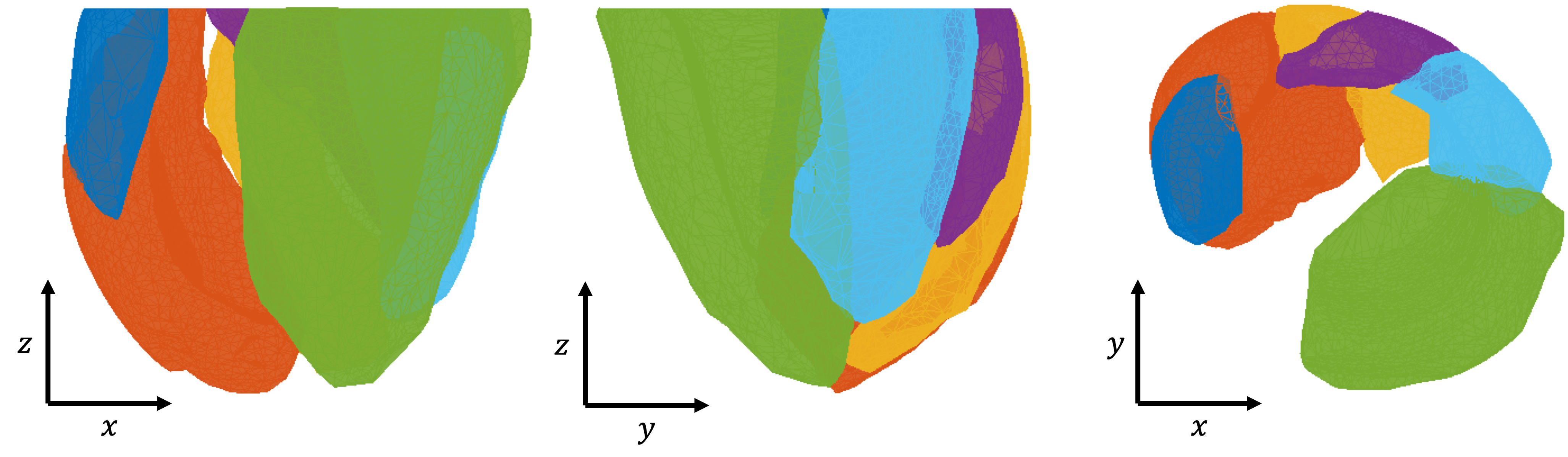}\label{fig:init_subdomain}}
   \caption{Transverse (left), sagittal (middle), and coronal (right) projections are shown of the (a) boundaries $\partial M'_i$ surrounding the point clouds $M'_i$ that are downstream of the terminal segments of the principal pathway (shown in black) (b) the nodes of $V$ that lie within each of $\r M_i'$ that form $S_i$ shown within $\r M_i'$ and (c) the regions $\r S_i$ formed from $S_i$.
   }\label{fig:From}
\end{figure}

The discrete nodes of the sets $M_i'$ can be used to define the regions that these nodes occupy called $\r M_i'$. These are the regions of the vasculature that are perfused from a given artery of the principal pathway. These regions and the principal pathway are shown in Fig.~\ref{fig:vasc_region}.
Each of the regions $\r M_i$ intersects with the ventricular region $\r V$. 
We call this intersection $\r S_i = \r V \cap \r M_i$. This is the region within the ventricle that is perfused from the $i^{\rm th}$ terminal segment of the principal pathway. In practice, this is done by constructing $\d M_i$ and finding the nodes $S$ that lie within $\d M_i$\cite{inpolyhedron} and assigning these to $\r S_i$.
The elements of $S_i = V \cap \r M_i$ within each region $\r M_i$ can be seen in Fig.~\ref{fig:intersection}. The boundaries $\partial S_i$ are shown in Fig.~\ref{fig:init_subdomain}.

\subsubsection{Creating Disjoint Subdomains}
With the modelling assumption that no point in the ventricle can be perfused from two different terminal arteries, then the restriction that the regions $\r{S}_i$ are disjoint must be imposed.
Hence, for each $i \in [1, n]$, redefine
\[
\r S_i = \r S_i \setminus \left(\bigcup_{j \in [1, n]\setminus i}^n \r S_j\right).
\]
Now, by construction the ventricle has been divided into $n+2$ subregions, so
\[
\r V = \left(\bigcap_{i = 0}^n \r S_i \right) \cap \r S_*
\]
where $\r S_*$ is a disjoint region comprised of nodes that have yet to be assigned to their final subdomain in order that the whole ventricle may be perfused. In the case of $n=6$, there are 665 of the 17,440 nodes in the volume filling point cloud that have been double assigned and 3134 that have not been assigned in this initial stage. Hence, $|S_*|= 3799$ which must be assigned to one of $S_i$, $i \in [0, 6]$.

\subsubsection{Expanding the subdomains beyond this assignation}
The set $S_*$ contains nodes that belong to a volume filling point cloud that describes the ventricle but have not been assigned to a subdomain $S_i$, $i \in [0, n]$. All nodes of the volume-filling cloud must be assigned to one of $S_i$, $i \in [0, n]$. In order to assign the nodes of $S_*$ to a subdomain, we consider the neighbouring nodes according to the volume-filling tetrahedral mesh.

The region $\r V$ is represented by a cloud of volume-filling points and associated tetrahedral. Let $E = \{a, b, c, d\}$ be a mesh element that joins nodes with indices $a$, $b$, $c$ and $d$ in a tetrahedron. Suppose that element $E$ is associated with the subdomains by $N = \{F, -1, F, F\}$, then we say that $a, ~ c, ~ d$ lies in the subdomain $F$ and the node $b$ has not yet been assigned. As three neighbours of $b$ belong to $F$, assign $b$ to $F$. There is a subdomain list $N$ for each element $N$, so when $b$ is assigned to $F$ all lists $N$ must be updated to reflect that $b$ now lies in $F$. After repeating reassignment for each element, the subdomains have collectively gained nodes, so there are more elements $E$ with lists $N$ that meet the criterion for reassignment. Repeat this process for the list of elements that contain unassigned nodes until no elements that meet the assignment criteria remain.

\begin{figure}[ht]\centering
    \includegraphics[width = .75\textwidth]{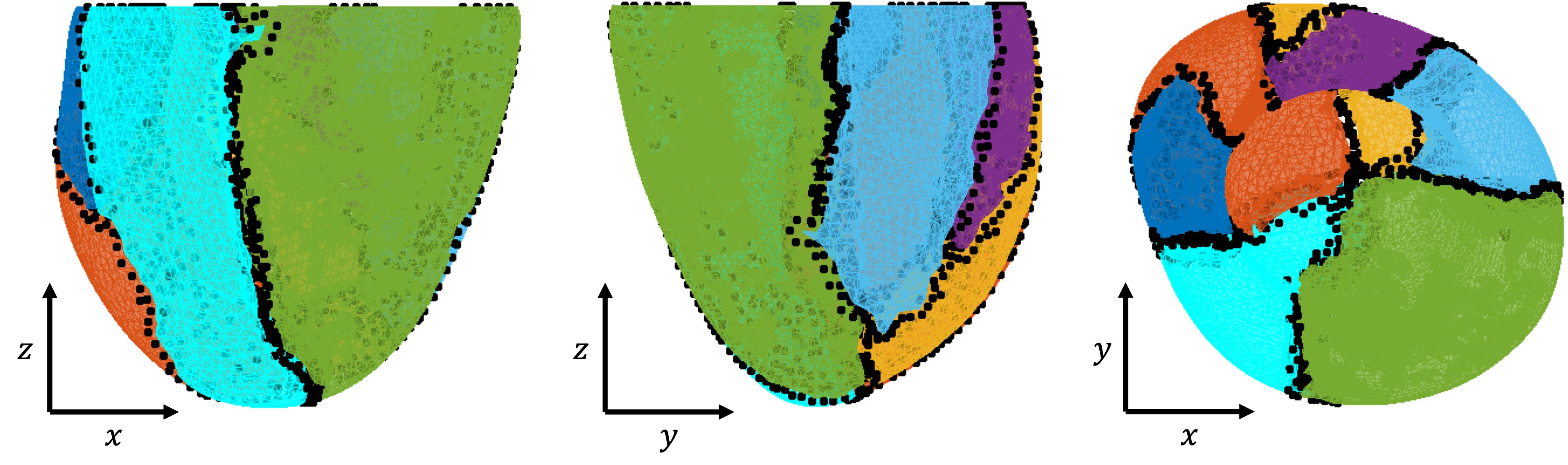}
   \caption{Transverse (left), sagittal (middle), and coronal (right) projections are shown of the (a) boundaries of the subdomains $\d S_i$, $i \in [0, 6]$ where $S_0$ is the subdomain shown in cyan. The nodes that are still to be assigned are shown as black squares. There are 1326 of these.
   }\label{fig:grow}
\end{figure}

In the case of $6 + 1$ subdomains, the first run of the algorithm described above leads to the assignment of 2,099 nodes. The second run assigns 317 nodes, the third assigns 59 nodes, and the fourth assigns 7 nodes. No nodes are assigned in the 5$^{\rm th}$ run. There are 1,326 of 17,440 nodes remaining to assign. The result of this process can be seen in Fig.\ref{fig:grow}.

\subsubsection{Assigning the remaining nodes}
If there are still unassigned nodes, these must be assigned to a subdomain. However, these lie on or near the boundaries between subdomains. In the case that an unassigned node only has neighbours that are unassigned or assigned to the same subdomain, assign the node to that subdomain. In the case that the neighbours of an unassigned node are either also unassigned or assigned to two or more subdomains, the node should be assigned to the modal subdomain of its neighbours. In the case where a node has no unique mode, the volumes of the modal neighbouring subdomains should be computed and the unassigned node is assigned to the smallest of these. This issue has not yet been encountered.

\subsubsection{Element order}
This method of subdomain growth has the flaw that the final subdomain assignment is affected by the order in which the elements are considered. The robustness of the algorithm can be assessed by randomly permuting the list of elements prior to the subdomain assignment.{ In 1000 samples for the subdivision of $B_6$, 88.96 \% (15514 of 17440) nodes are always assigned to the same subdomain. Of the 1924 nodes that are not always assigned to the same subdomain, almost all (1914) are assigned to the same subdomain at least 50\% of the time and 1178 are assigned to the same subdomain in 80\% of the trials. There are 24 nodes that are eligible for assignation to one of 4 subdomains depending on the initial order in which the elements of the mesh are considered. Subdomain 4 has a volume between 2.36 and 3.28 ml (2.85 $\pm$ 0.24 ml); these number of nodes assigned to subdomain 4 is an integer in the interval $[797, 841]$.
This is the smallest subdomain. Subdomain 5 is the largest with volume in the range 20.21 to 22.93 ml (21.75 $\pm$ 0.66 ml); this subdomain is assigned between 5172 and 5,268 nodes. Subdomain volumes are calculated by taking the computed volume assignations and constructing a concave boundary around each subdomain using the system-specific \texttt{boundary} function (MATLAB R2022b; The MathWorks Inc., Natick, MA, United States) with a \textit{shrink factor} of 1 . The \textit{shrink factor} is an optional argument of the \texttt{boundary} function that controls how tightly to the data the boundary is constructed. It lies in the interval $[0, 1]$ with a default value of 0.5. As there is some separation between the spatial locations of the nodes that belong to subdomains, the cumulative computed subdomain volume always falls short of the volume of the whole domain. A denser mesh will give a smaller shortfall with the drawback of increased computational time. The correct subdomains are not known \textit{a priori}, so there is no way to assess the absolute correctness of the grown subdomains. It can be expected that the certainty of the assignations decreases with increasing subdomain count. 
}

\begin{figure}[ht]\centering
  \includegraphics[scale = 0.5]{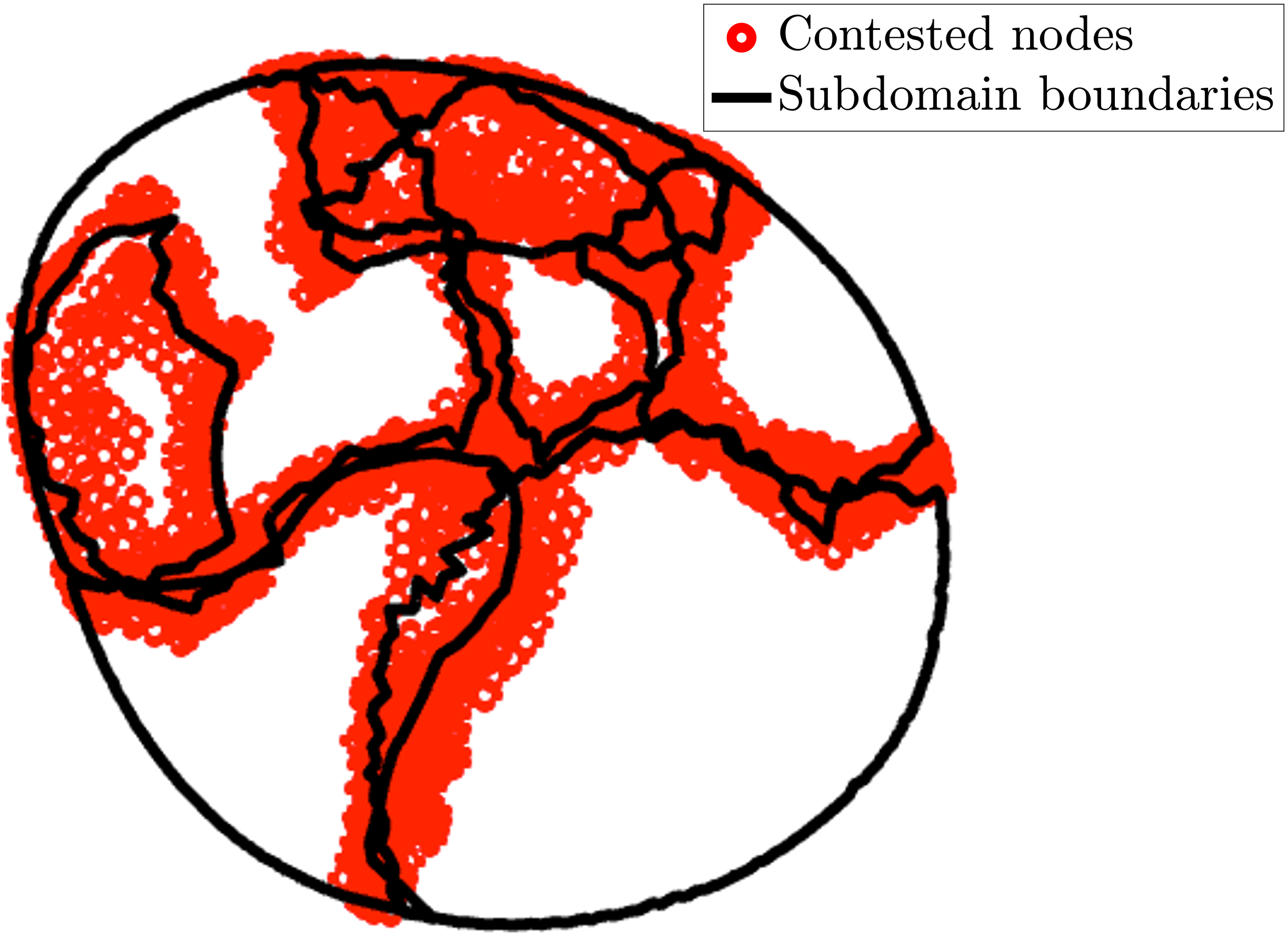}
   \caption{A dorsal projection of the subdomain boundaries for $B_6$ with the nodes of the ventricular set $V$ that are not always assigned to the same subdomain shown in red. These nodes can clearly be seen to cluster along the subdomain boundaries.}\label{fig:contested}
\end{figure}

As can reasonably be expected, the contested nodes tend to lie on the interface between subdomains as can be seen in Fig.~\ref{fig:contested}.

\subsubsection{Final subdivision}
\begin{figure}\centering
 	\subfloat[]{\includegraphics[width = 0.6\textwidth]{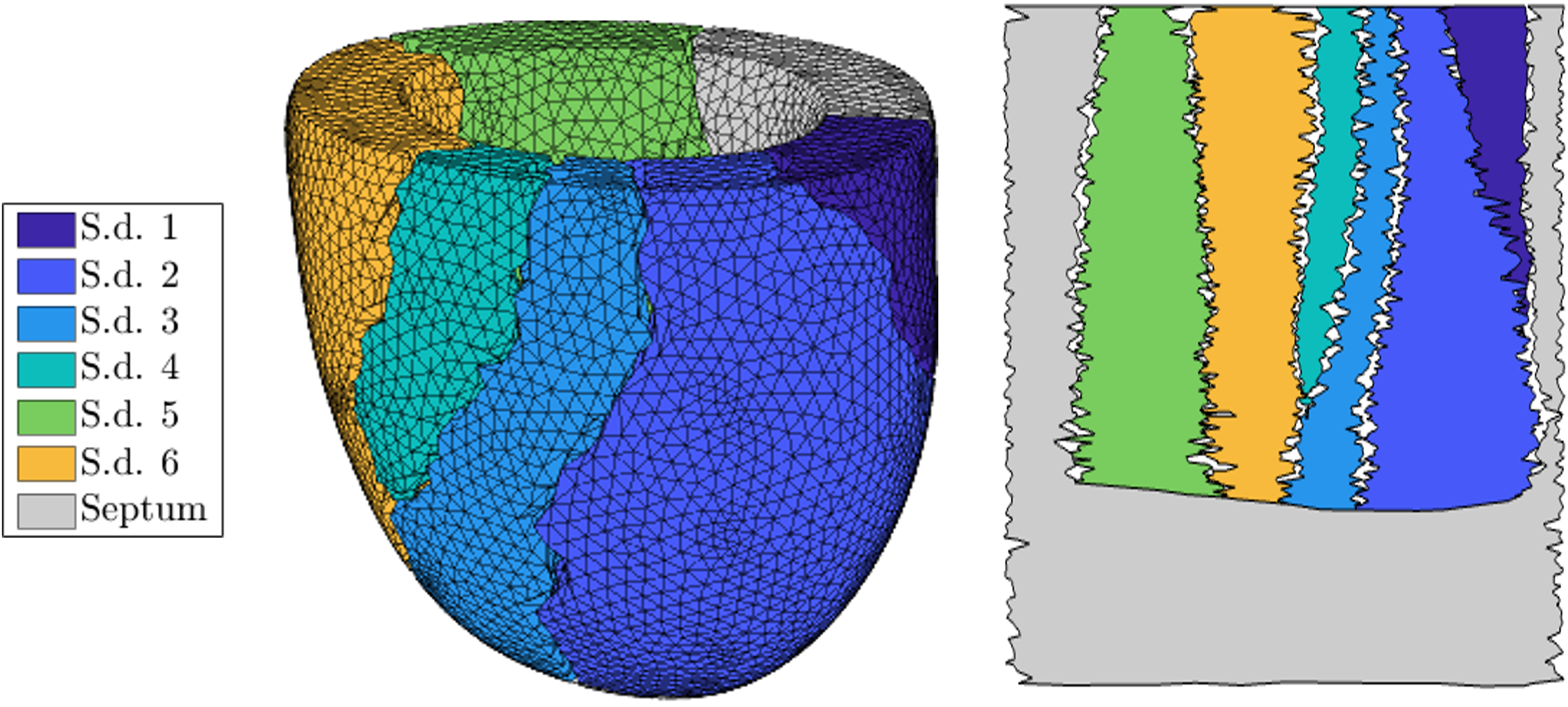}\label{fig:final:B6:show}}\\
  	\subfloat[]{\includegraphics[width = 0.6\textwidth]{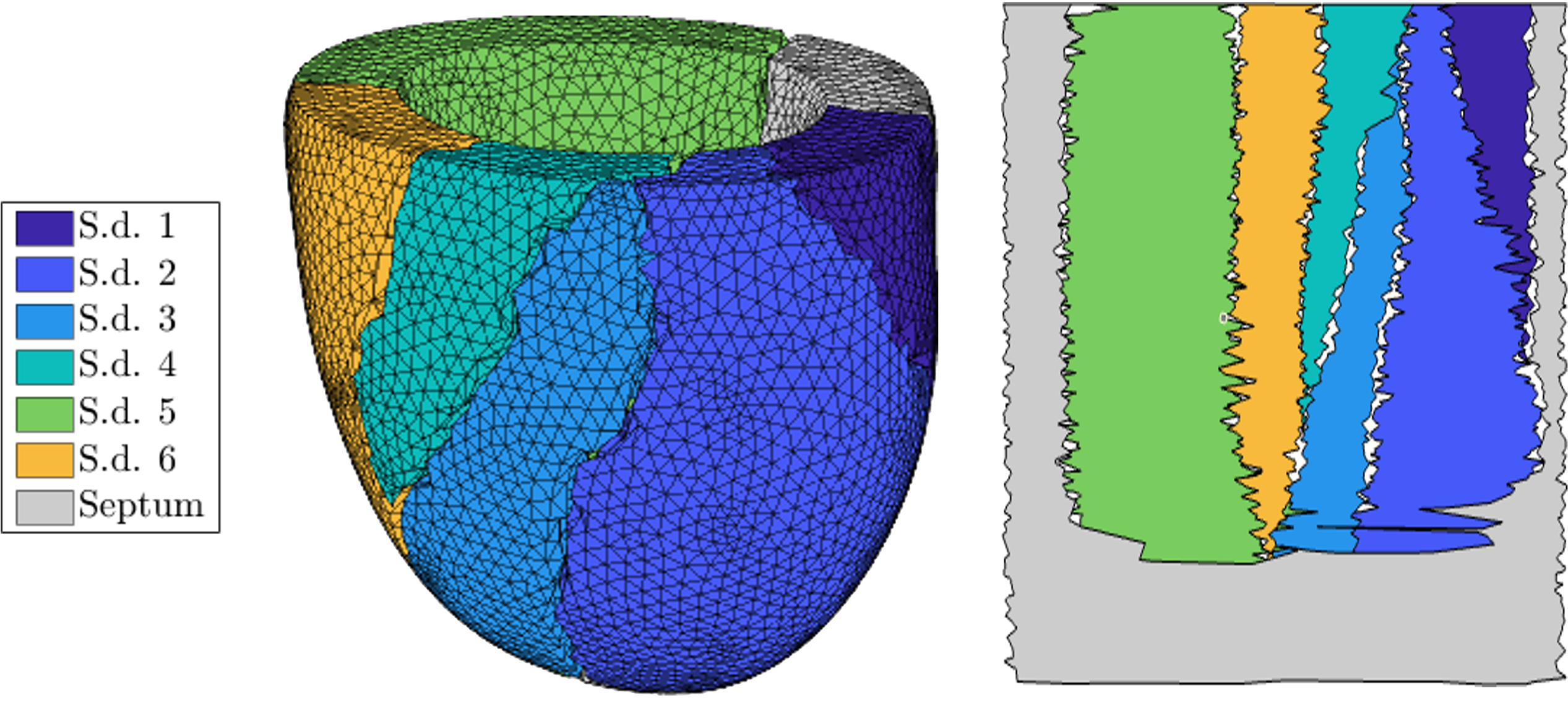}\label{fig:final:B6:search}}
   \caption{The final division of the ventricle into 6+1 subdomains using the (a) final and (b) terminal node search from the principal pathway $B_6$, i.e. by constructing hulls surrounding each of the sets of downstream nodes shown in Fig.~\ref{fig:From:Show} and Fig.~\ref{fig:From:Search}, respectively before finding the region of the ventricle that lies in each hull and growing the subdomains until all nodes are assigned. The right hand panels show the Mercator projections. The legend lists subdomain number (abbreviated to "S.d.") for the 6 subdomains of the perfused region and the septum (shown in grey). The colourings of the subdomains is preserved in the Mercator projection.
   }\label{fig:final}
\end{figure}

The final subdivision for $B_6$ from the terminal and initial nodes of the terminal branches of the principal pathway are shown in Figs.~\ref{fig:final:B6:show} and \ref{fig:final:B6:search}, respectively. Subdomain 5 can be seen to be much larger when constructed using an initial node search, rather than a terminal node search. Expected since most of the cyan nodes shown in Fig.~\ref{fig:From:Show} are absorbed into the downstream tree from which the subdomain is grown. The optimal selection of nodes from which to search for downstream trees in order to grow subdomains is currently unclear. This is an area that should be investigated in future research.

The final subdivision of the ventricle with the tree $B_{20}$ features more subdomains than that for $B_6$. The division with the terminal node search results in a subdomain with 0 volume, where the initial node search results in no subdomains of 0 volume. Given that there is a subdomain of 0 volumes, the assumption that there is a bijective map between the terminal branches of the principal pathway and the subdomains breaks down. In this case, two terminal branches would perfuse a single subdomain. Given this, if the bijective assumption is made, care must be taken when constructing the principal pathway to ensure that there are sufficiently few terminal branches so that no subdomain has 0 volume.

\begin{figure}\centering
        \includegraphics[width = 0.7\textwidth]{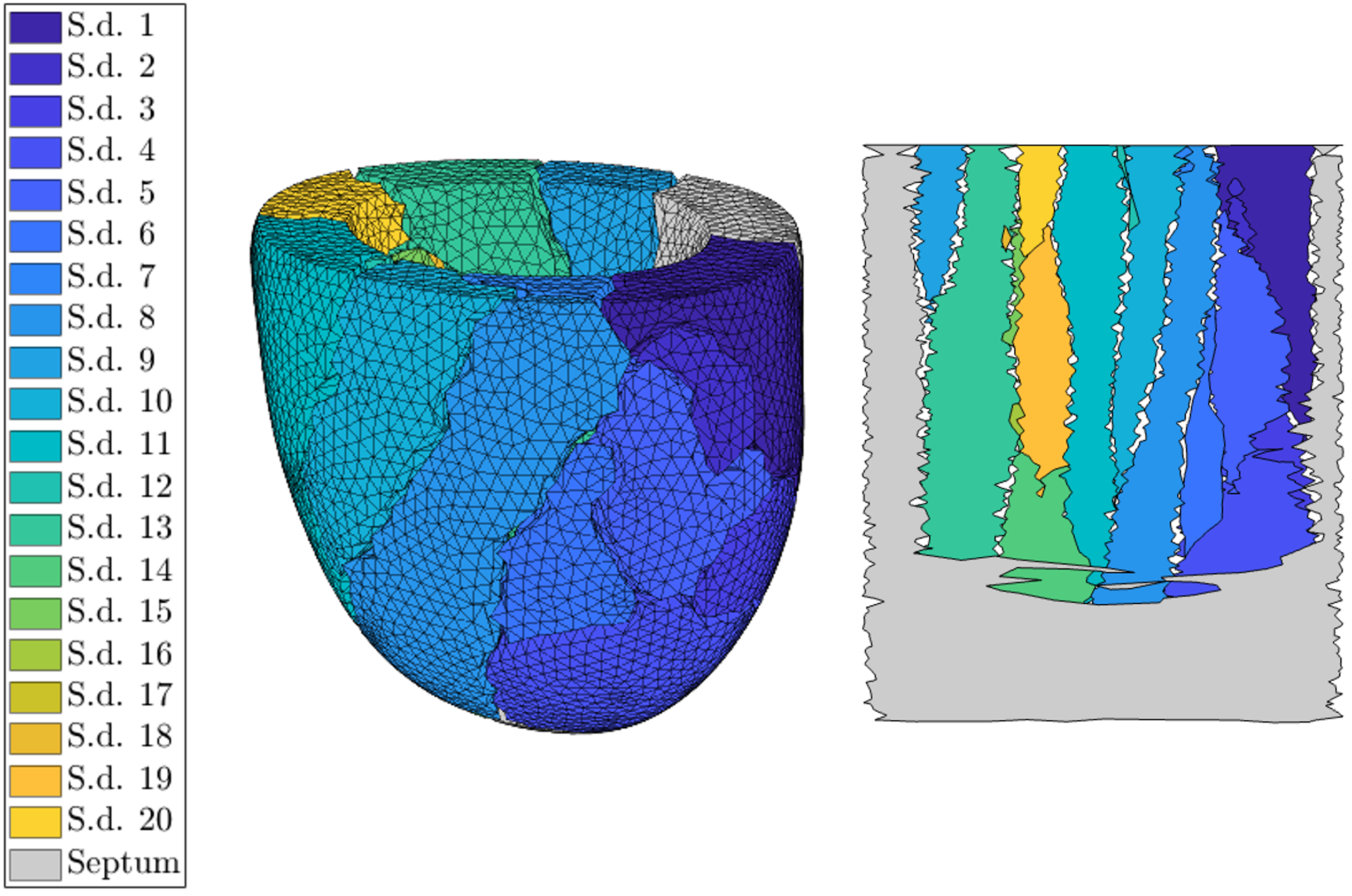}
   \caption{
   Three divisions of the ventricle into 20+1 subdomains from the terminal nodes of the terminal branches of the principal pathway $B_{20}$. The legend lists subdomain number (abbreviated to "S.d.") for the 20 subdomains of the perfused region and the septum (shown in grey). The colourings of the subdomains is preserved in the Mercator projection shown to the right.
   }\label{fig:final:B20:search}
\end{figure}

\paragraph{Final subdivision}
The final subdivisions for $B_6$ and $B_{20}$ can be found in Fig.~\ref{fig:final} for the inlet and terminal node searches for $B_6$, and an initial node search for $B_{20}$. These are shown in $\R 3$ and as Mercator projections. For $B_6$, subdomain 5 can be seen to be much larger when constructed using an initial node search, rather than a terminal node search. This is expected since most of the cyan nodes shown in Fig.~\ref{fig:From:Show} are absorbed into the downstream tree from which the subdomain is grown. The optimal selection of nodes from which to search for downstream trees in order to grow subdomains is currently unclear. This is an area that should be investigated in future research.

Unsurprisingly, the final subdivision of the ventricle with tree $B_{20}$ features more subdomains than that of $B_6$. The division with the terminal node search results in a subdomain with 0 volume, where the inlet node search results in no subdomains of 0 volume.

\subsection{Euclidean distance based division}
Other than the physiologically motivated subdivision augmented with what is essentially a nearest-neighbor algorithm based on a 3D analogue of a taxicab metric, it is possible to divide the ventricle with other metrics. Perhaps the easiest to implement is a Euclidean distance-based algorithm. 

\begin{figure}[ht]\centering
 	\includegraphics[width = 0.4\textwidth]{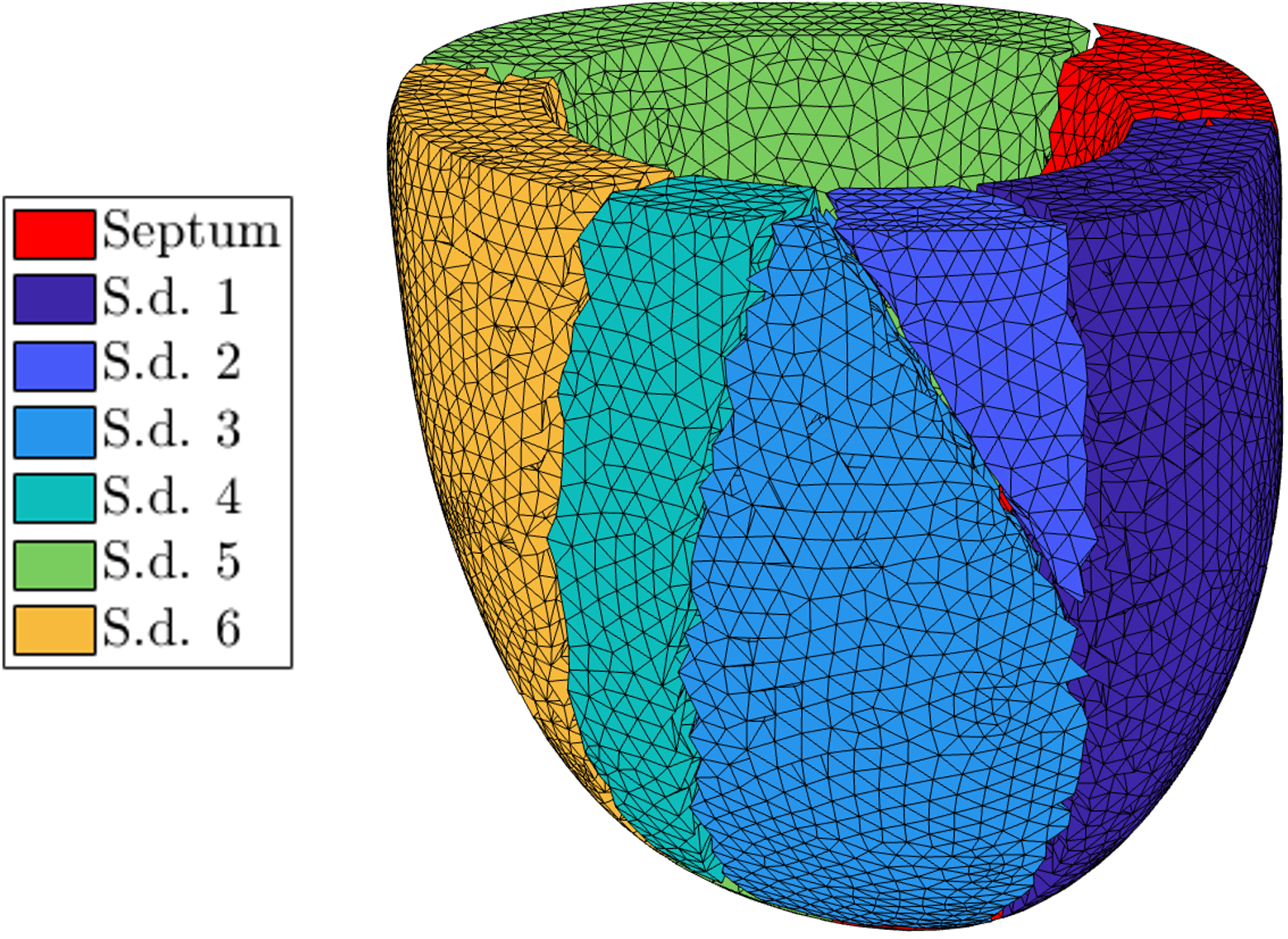}
        \caption{Division of the ventricle into 6+1 subdomains with  the Euclidean distance from the terminii of the terminal branches of the principal pathway.}\label{fig:E6}
\end{figure}

A nearest-neighbour algorithm with Euclidean distance as the metric requires nodes from which the Euclidean distance is computed. Again, we wish to divide the ventricle into $n+1$ subdomains for a tree with $n$ terminal branches $B_n$. For each of the terminal nodes of the terminal branches of $B_n$, the Euclidean distance between all the nodes in $\r V \setminus S_0$. The single nearest node to each of the $n$ search nodes is then used as the seed for ventricle division with the Euclidean distance. The division of the ventricle into 6+1 subdomains with a Euclidean approach is shown in Fig.~\ref{fig:E6}. The main differences between this division and those shown in Fig.~\ref{fig:final:B6:show} or Fig.~\ref{fig:final:B6:search} are that the edges of the subdomains are very straight and that (in the case of subdomain 2 for example), the sizes can widely vary.

\subsection{American Heart Association's left ventricular regions}
The American Heart Association division of the left ventricle \cite{manuel2002standardized} is well established and was developed to standardise the nomenclature of the various subregions of the left ventricle. For consistency with the figures here, the longitudinal axis is taken to run up the centre of the ventricular cavity, and the heart to be oriented such that the apex is at the bottom. Essentially, the ventricle is divided into four layers; one of these is the apex, which extends from the lowest point in the ventricular cavity to the lowest point on the outside wall. The remaining portion of the ventricle is divided into three layers of equal height. The upper most two layers are divided into sextants of equal angle, and the third layer into equal quarters. The apical region is not divided further.

Given that this division is geometrically based, and the one discussed here is physiologically based, an exact correspondence should not be expected between the AHA subdivision and any seen here. However, the AHA division is physiologically motivated and is essentially intended as a tool to standardise and clarify discussions pertaining to the ventricle. Hence, it is useful to compare the divisions both to facilitate comparisons with other models and as verification that the division here is reasonable.

\begin{figure}[ht]\centering
 	\subfloat[]{\includegraphics[height = 6cm]{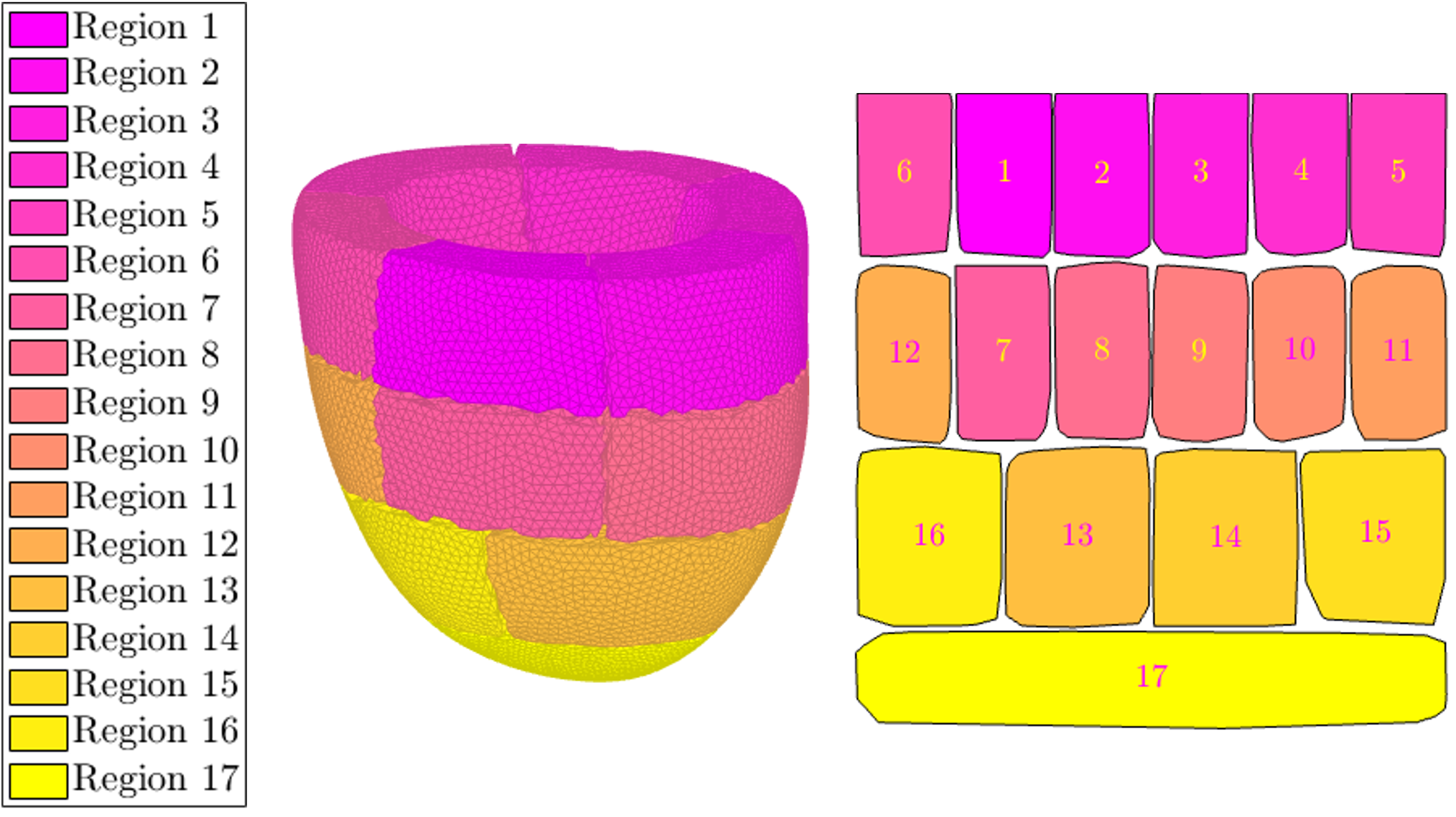}\label{fig:AHA:regions}}\\
        \subfloat[]{\includegraphics[width = 0.4\textwidth]{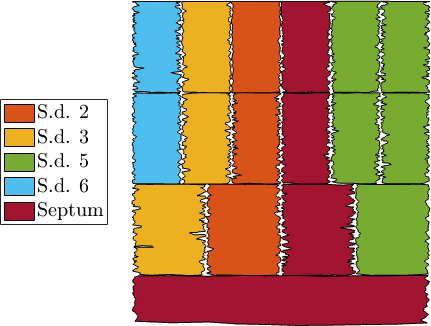}\label{fig:AHA:B6}}
        \caption{(a): The AHA regions of the left ventricle in $\R 3$ and $\R 2$ as a Mercator projection and (b) the regions of $B_6$ mapped to the AHA regions.}
\end{figure}

To compare the divisions, we must make the AHA regions. This is straightforward with the definitions given by Manuel \textit{et al.} \cite{manuel2002standardized}. The 17 subregions can be seen in Fig.~\ref{fig:AHA:regions} in $\R 3$ with the same view as is seen in Figs.~\ref{fig:final} and \ref{fig:E6}.  It is straightforward to construct a map between the AHA regions and the generated subdomains, as the assignation of all nodes is known. In general, this is neither surjective nor bijective. In the case of $B_6$, there are 7 subdomains, so the subdomains will be, on average, larger than the AHA regions. The subdomains may span many AHA regions, or vice versa. As the ventricular data do not align perfectly with the AHA regions, the map between them can get messy. In order to simplify the map,  choose the mapping criteria that the subdomain with the largest overlap (by node count) with an AHA region maps to that region. The map from the subdomains of $B_6$ to the AHA regions is shown in Fig.~\ref{fig:AHA:B6}. Here, the subdomains and 4 were located on the boundary between two regions and therefore were not mapped to any region with the chosen mapping criteria. 

Such a map is useful as it could ease knowledge exchange between modellers and stakeholders in these models, such as interested clinicians.
In the specification of the AHA regions, the septum corresponds to regions 3, 4, 9, 10, \& 15. Here, the septum corresponds to regions 3, 9, 14, \& 17. The alignment between the subdivision here and the AHA regions is imperfect, however this may be due to a misalignment of the axis of the ventricular data. However, the map in its current form may be sufficient as a tool to communicate information about the ventricle with various stakeholders.

\section{Discussion \& Development}

The algorithms developed for the processing of the vascular data would benefit from optimisation, although they are currently sufficiently fast as the data sets are relatively small and each principal pathway need only be generated once. The procedures described here should be applicable to any data set of comparable size.
The radius data are noisy, which negatively impacts the accuracy of any radius-based filtering methods; Bartolo et al.~\cite{bartolo2024computational} devise methods to reduce the estimated radius noise by using statistical change points to determine the centre-line that best represents the vessel. Bartolo et al.~\cite{bartolo2024computational} also show that the lengths and radii of vessels in a principal pathway has a strong impact on the haemodynamic predictions. Hence, developing smoothing algorithms for vessel radii in datasets that have already been segmented is worthwhile. Similarly, \cite{mackenzie2021thesis} shows that the total volume of the tree also has a strong influence on the outcome of the haemodynamic simulations.

The vascular data discussed above constitute a partial left coronary artery tree. The Strahler order of the tree shown in Fig.~\ref{fig:segs} is 7.
Kassab \textit{et al.} \cite{kassab1993morphometry} find the Strahler order of the left porcine coronary arterial tree to be 11. In their corrosion casting study, Kassab \textit{et al.} captured images of vessels with radii in the range 4.23\;$\si{\micro\metre}$ -- 1.7\;$\si{\milli\metre}$. The radius range in the data discussed here is 103.5\;$\si{\micro\metre}$ -- 1.7\;$\si{\milli\metre}$. We conclude that the discrepancy in the Strahler order of the trees arises from the lower resolution of the data discussed here; indeed they record the length and diameter of more than 8000 segments in a summary table of the porcine left anterior descending artery, not including the left circumflex artery as a principal pathway. As the tree is extensively pruned in order to generate principal pathways suitable for use in computational domains, the lower resolution does not represent an issue at this time.

As mentioned above, it is currently unclear how to choose the optimal node(s) in the terminal branches of $B_n$. The quantification of the impact of terminal node choice for subdomain growth could be a fruitful area for future research. It seems reasonable that truncated terminal arteries with large distal radius give rise to larger distal trees than terminal arteries of smaller radius. This possible link has not been explored here. Any such relationship between the radius of the terminal artery and the volume of the downstream tree could be used to introduce a weighting to the metrics used to divide the ventricle into subdomains. That is, a node could have a higher probability of being assigned to a larger subdomain than a smaller one.
Moreover, the data discussed here seem rarely to be available. However, even without access to the data it may be desirable to subdivide the ventricle for perfusion studies. Future research will focus on the development of mappings from a given left ventricle to the one discussed here that will allow a ventricle without an accompanying detailed arterial tree to by divided into physiologically motivated subdomains.

\section{Conclusions}
Briefly, using spatial data describing coronary arterial vasculature of a single porcine heart obtained from fluorescence cryomicrotome images \cite{goyal2012vasculature} and image processing techniques, we have developed algorithms to organise and search the data in order to build the main pathways from the data. These principal pathways will be used in computational perfusion studies such as those discussed in Mackenzie \cite{mackenzie2021thesis}. It is natural to ask which regions of the ventricle are perfused from a given large artery in the generated 1D tree, and with sufficient data, this can begin to be answered by considering where the nodes in the vascular tree that are downstream of the truncation points lie. By definition, the area in which the downstream tree lies is perfused by that tree. However, this is not the only way to divide the ventricle. In principle, any distance metric can be used to divide the ventricle. As can be expected, different metrics lead to different divisions. However, while the data are available, it makes sense to use a physiologically motivated subdivision, representing an improvement of the method presented by Termeer \textit{et al.}\cite{termeer2010patient}.

The objectives of the study were to develop algorithms that robustly and reliably allow a user to generate arterial trees from a large graph representing the left coronary vascular tree and to determine the region(s) of the ventricle that are perfused from the terminal arteries of a generated principal pathway. Both goals have been achieved, and the trees generated here have been used in 1D haemodynamic studies \cite{mackenzie2021thesis}. 

The main drawback of the methods discussed here is the availability of data to which they can be applied. Such data comes from processed images. We have treated both the acquisition and initial processing of images to form a skeletonisation as a black-box. To move toward a patient-specific coronary flow model, it is important that the complete procedure of processing imaged data be as reliable and efficient as possible. A key future goal of this project is to integrate the framework developed here into a larger data collection and processing pipeline.

The data discussed here do not include either a right coronary arterial tree or a venous side, yet such networks are vital for coronary flow models. As has been seen, large data sets detailing porcine coronary arterial and cardiac venous morphometry have been presented by Kassab et al. \cite{kassab1993morphometry, kassab1994morphometry}; comparing the data contained in these with those discussed here to generate a right arterial and venous tree that would be invaluable in such studies. These data have been successfully reconstructed to form full coronary arterial trees in realistic ventricles \cite{kaimovitz2010full, kaimovitz2005large}.

\appendix
\section{Murine Pulmonary Vascular Networks} \label{appendix}

\begin{figure}[ht]\centering
    \subfloat[]{\includegraphics[width = 0.15\textwidth]{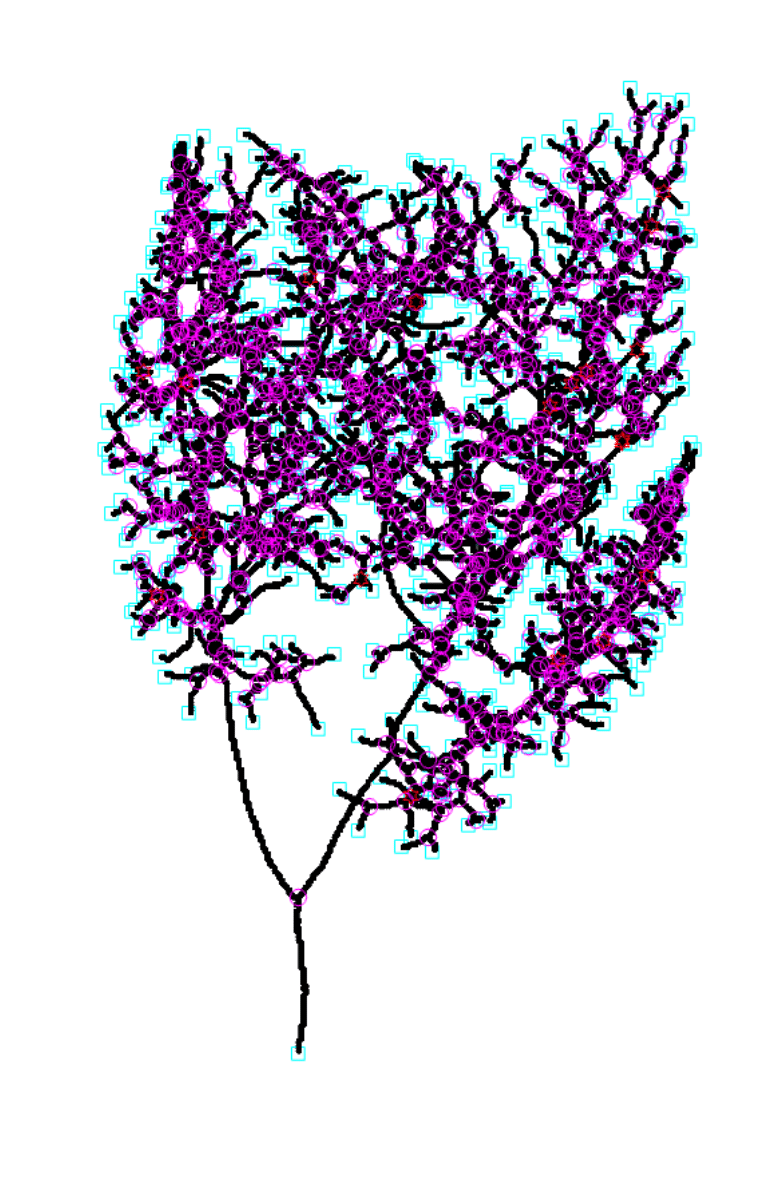}}
    \subfloat[]{\includegraphics[width =  0.15\textwidth]{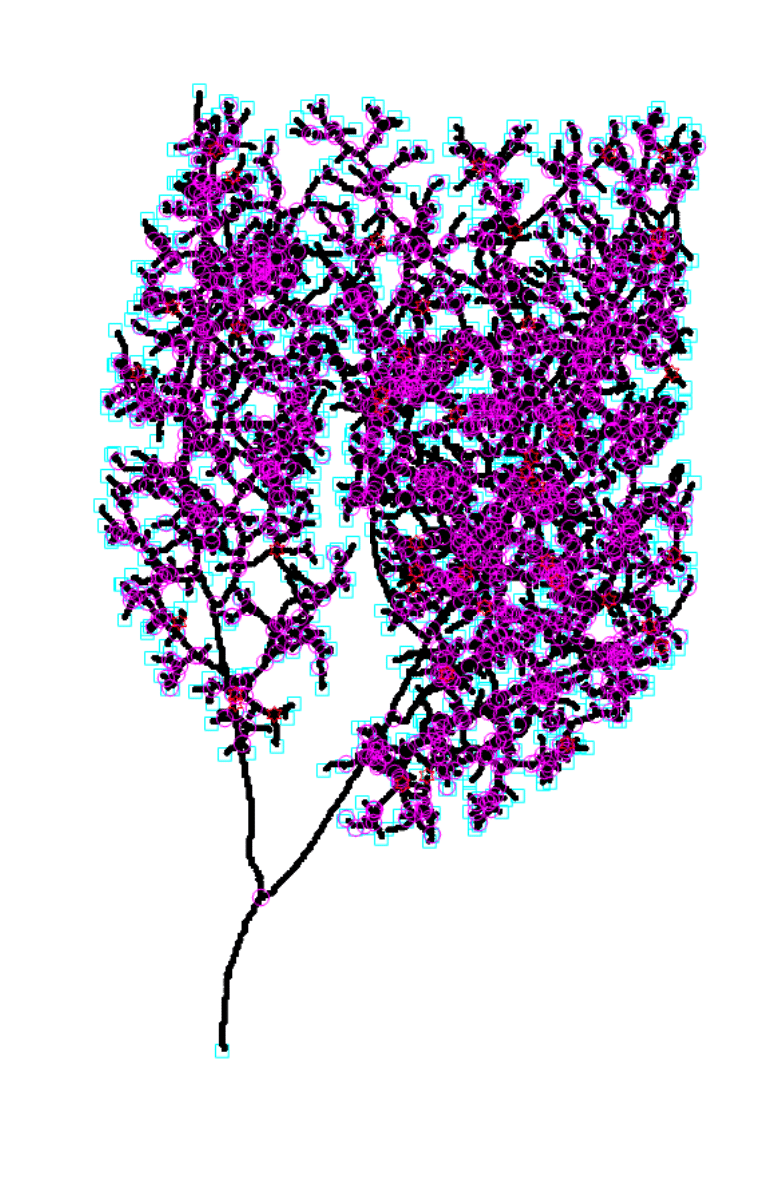}}
    \subfloat[]{\includegraphics[width =  0.15\textwidth]{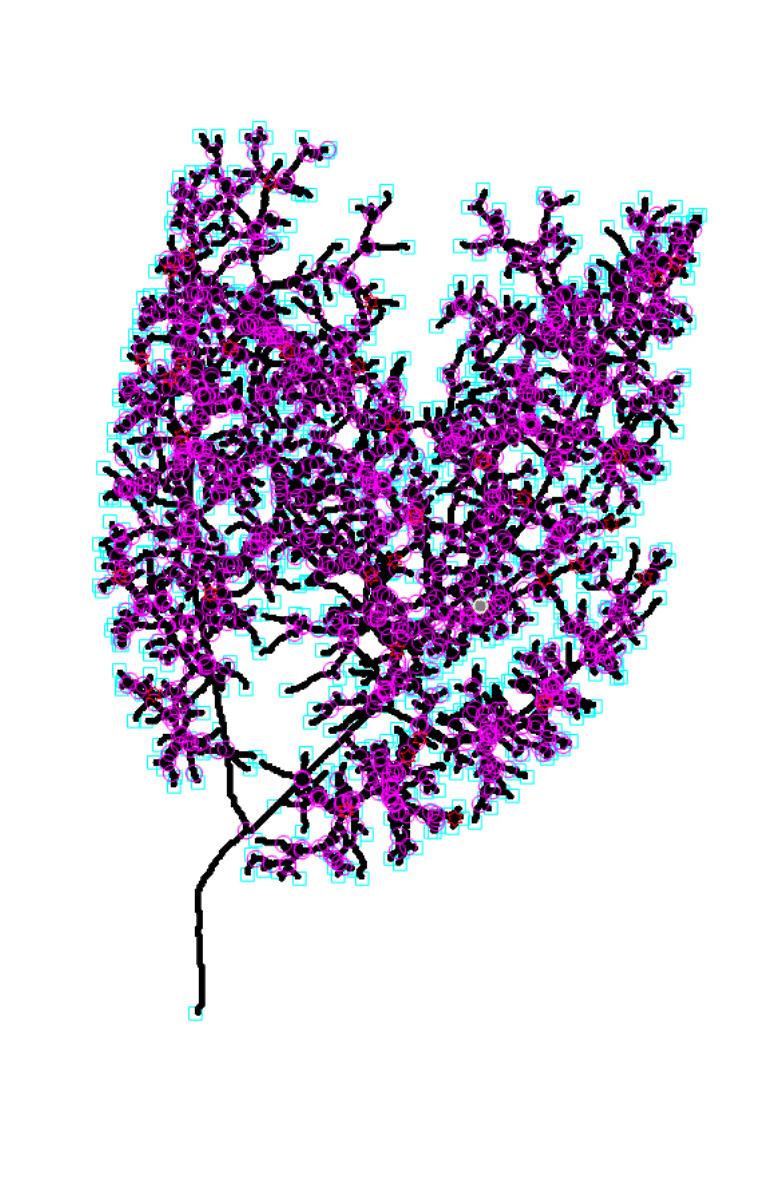}}
    \subfloat[]{\includegraphics[width =  0.15\textwidth]{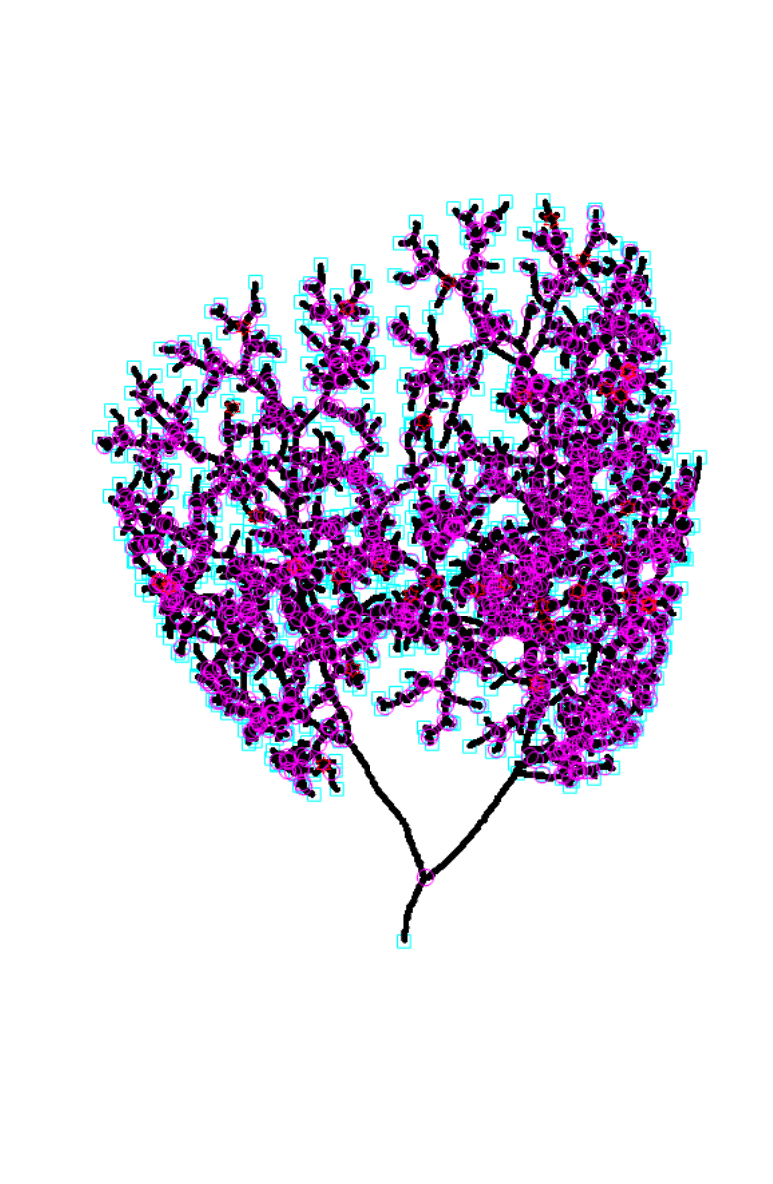}}
    \subfloat[]{\includegraphics[width =  0.15\textwidth]{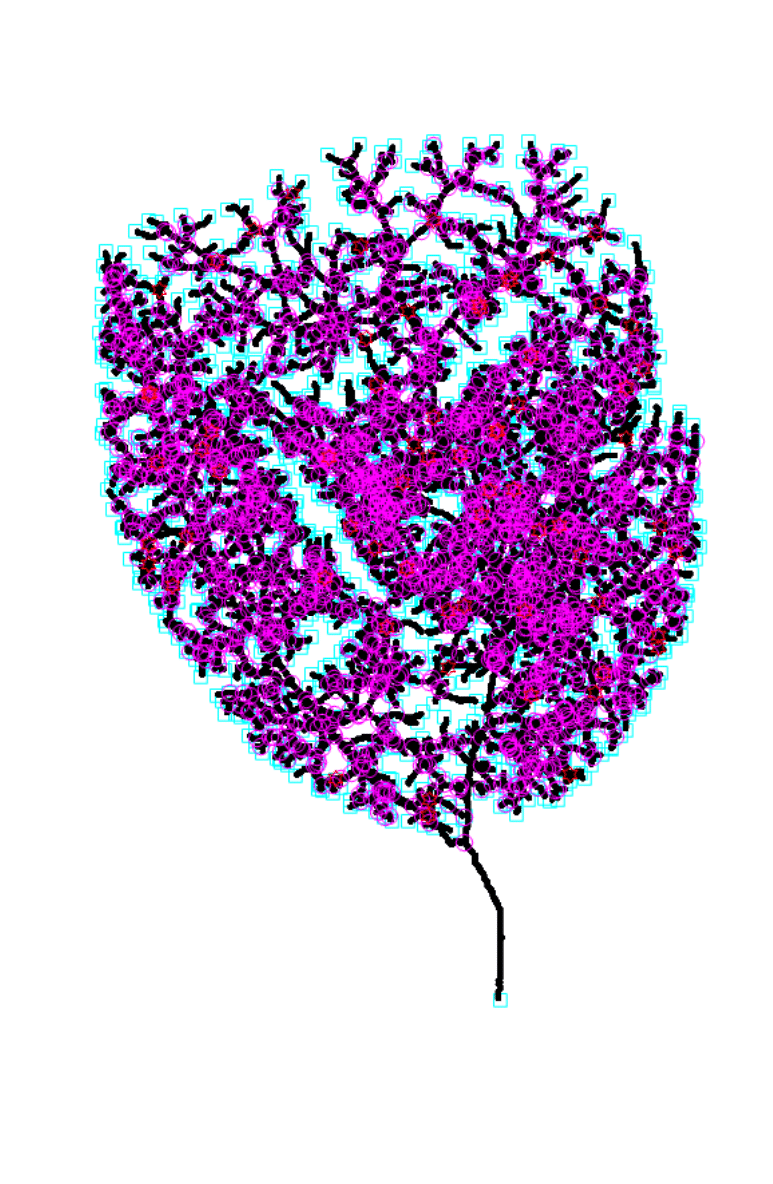}}
    \subfloat[]{\includegraphics[width =  0.15\textwidth]{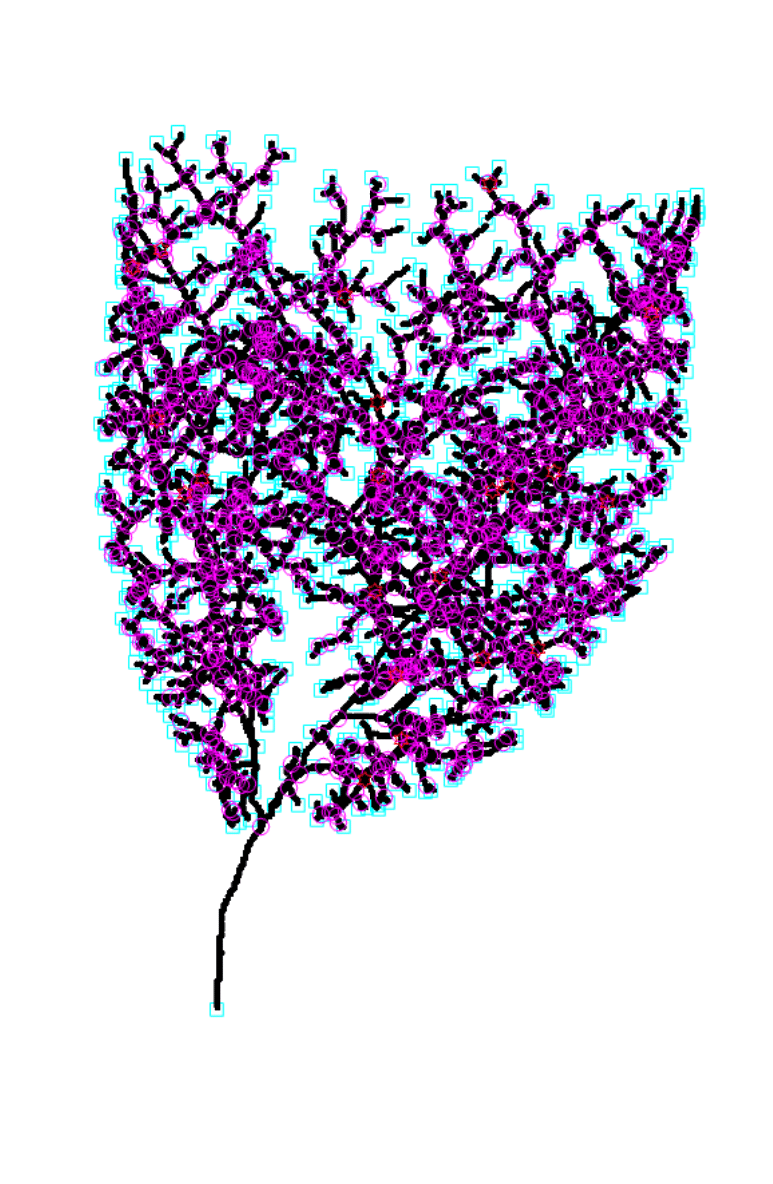}}
    \caption{The pulmonary venous networks of 6 mice. Cyan nodes are the terminal nodes, magenta nodes represent junctions, and body nodes are shown in black. The networks of the control mice are shown in the top row.} \label{fig:mouse1}
\end{figure}

Although the methods and algorithms discussed above were developed for use in the coronary arterial data set, the methods are applicable to any divergent graph. Here, we illustrate that the methods are applicable to any suitable graph by applying them to networks that represent the pulmonary venous system of six male C57BL6/J mice, three of which are controls (Fig.~\ref{fig:mouse1}(a)--(c)) and the remaining 3 have induced pulmonary hypertension (Fig.~\ref{fig:mouse1}(d)--(f)). Details of imaging and initial image processing can be found in Chambers et al.~\cite{chambers2020pulmonary} and in Chambers \cite{chambers2022morphological}.

Here, the control mice are the numbers (a) -- (c) and the hypertensive mice are (d)--(f). We are not evaluating the differences between control and hypertensive mouse pulmonary morphometries, but illustrating that the tools developed here are widely applicable.

The mouse data are in a format similar to the coronary data: there are nodes in a graph that represent the vasculature in $T \subset \mathbb{R}^3$ that are joined to form a network by edges $E$. There is a radius value associated with each node. Considering the frequency with which the index of each node appears in the list of edges, we can sort the nodes into junctions, terminal, and body nodes. This is seen in Fig. \ref{fig:mouse1}. Here, all 2D views of the networks are shown in the horizontal plane \cite{schroder2020rodent} defined by the long and the left-right axes of the mouse's body.

\begin{figure}[ht]\centering
    \subfloat[]{\includegraphics[width = 0.15\textwidth]{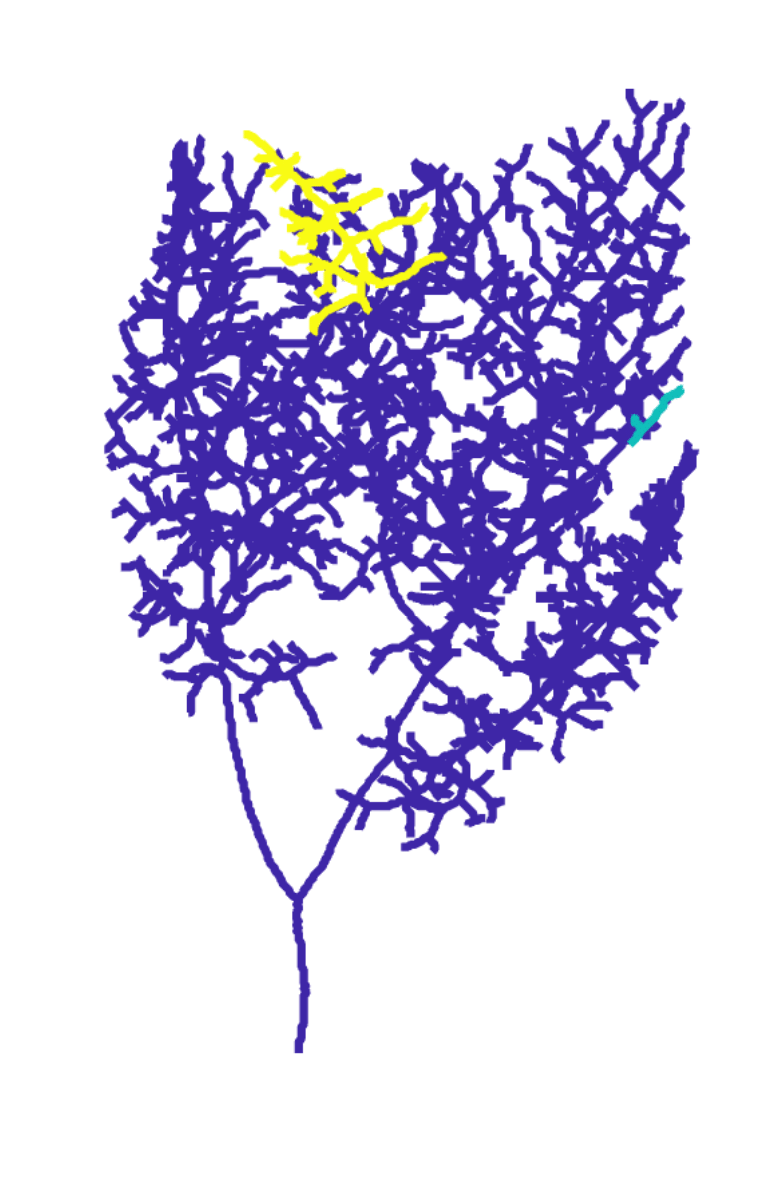}}
    \subfloat[]{\includegraphics[width =  0.15\textwidth]{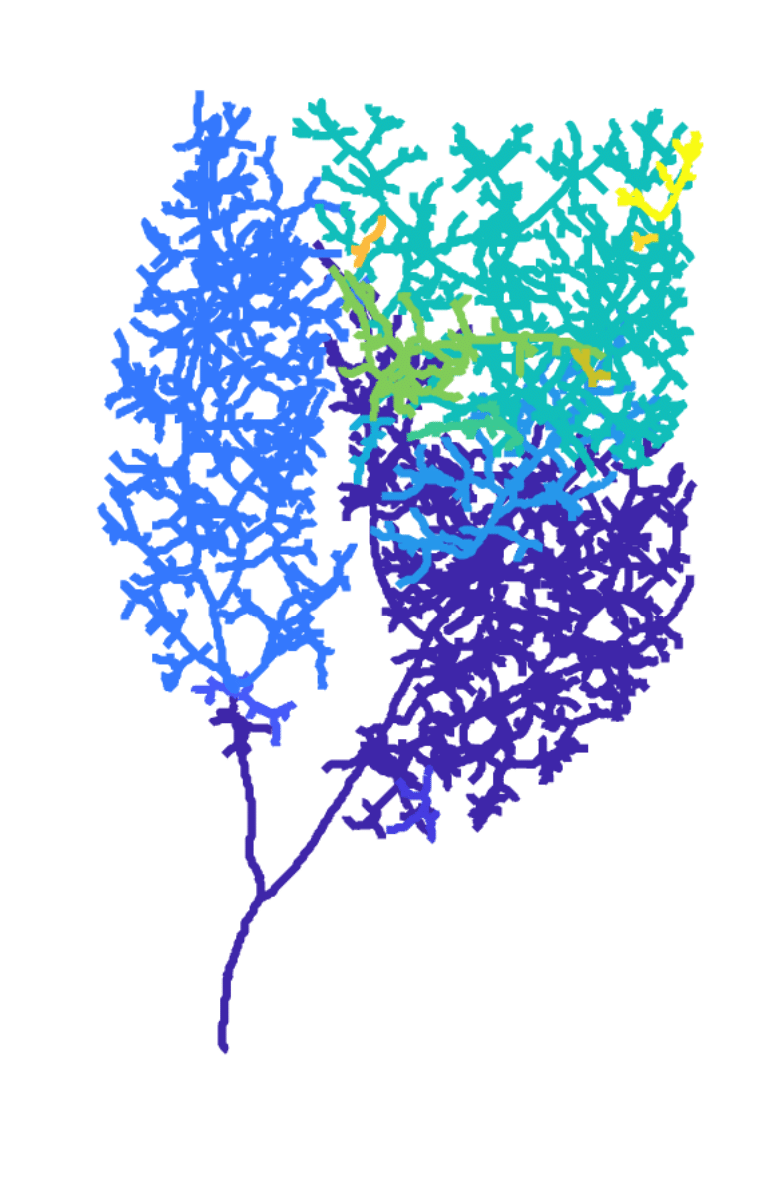}}
    \subfloat[]{\includegraphics[width =  0.15\textwidth]{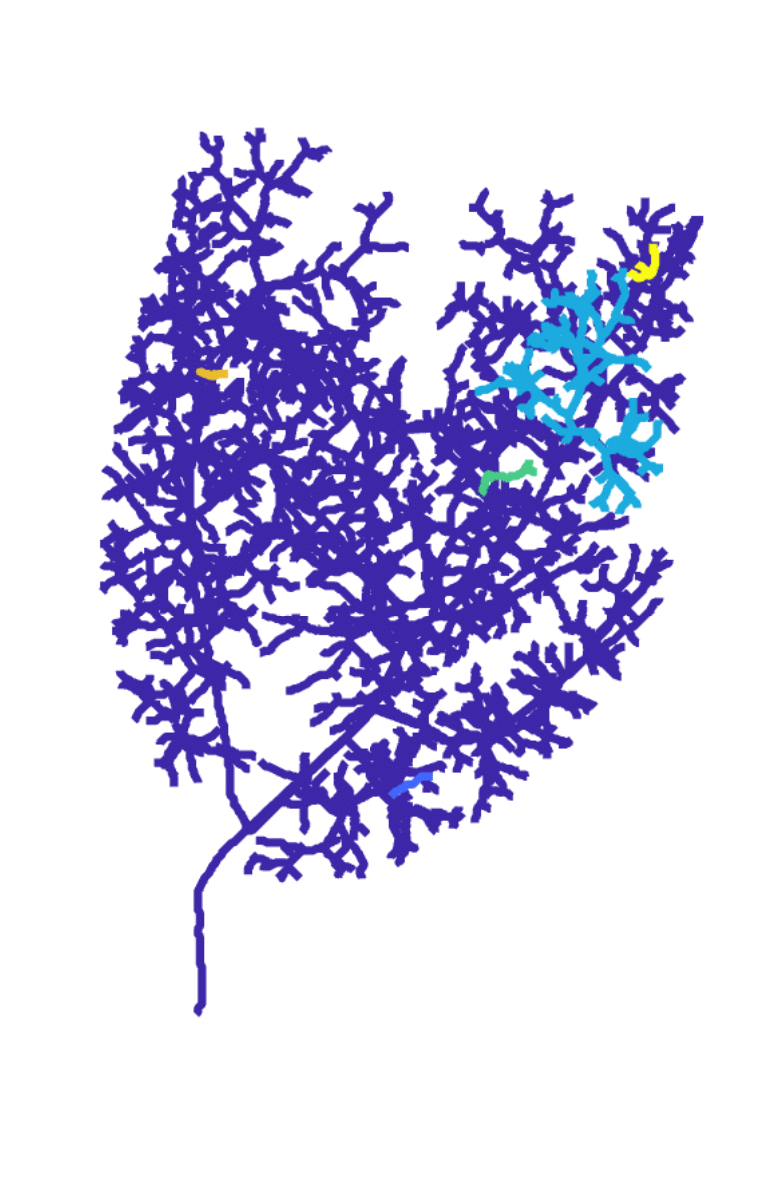}}
    \subfloat[]{\includegraphics[width =  0.15\textwidth]{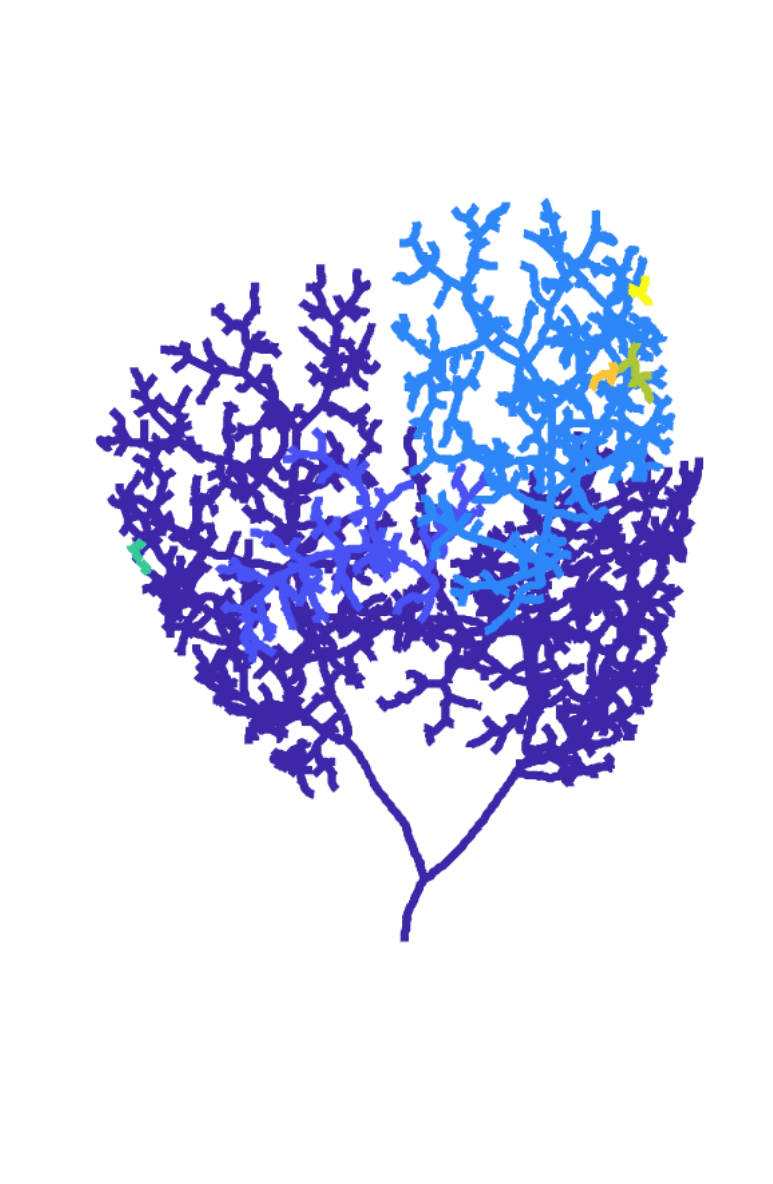}}
    \subfloat[]{\includegraphics[width =  0.15\textwidth]{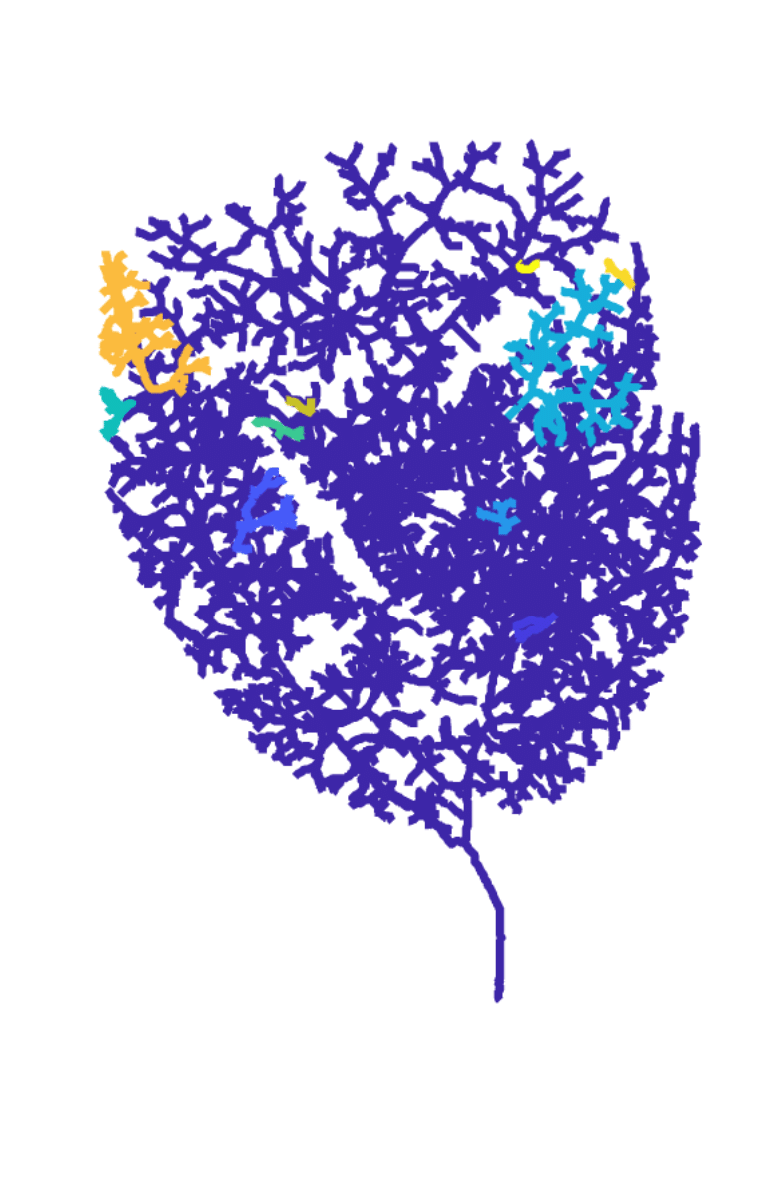}}
    \subfloat[]{\includegraphics[width =  0.15\textwidth]{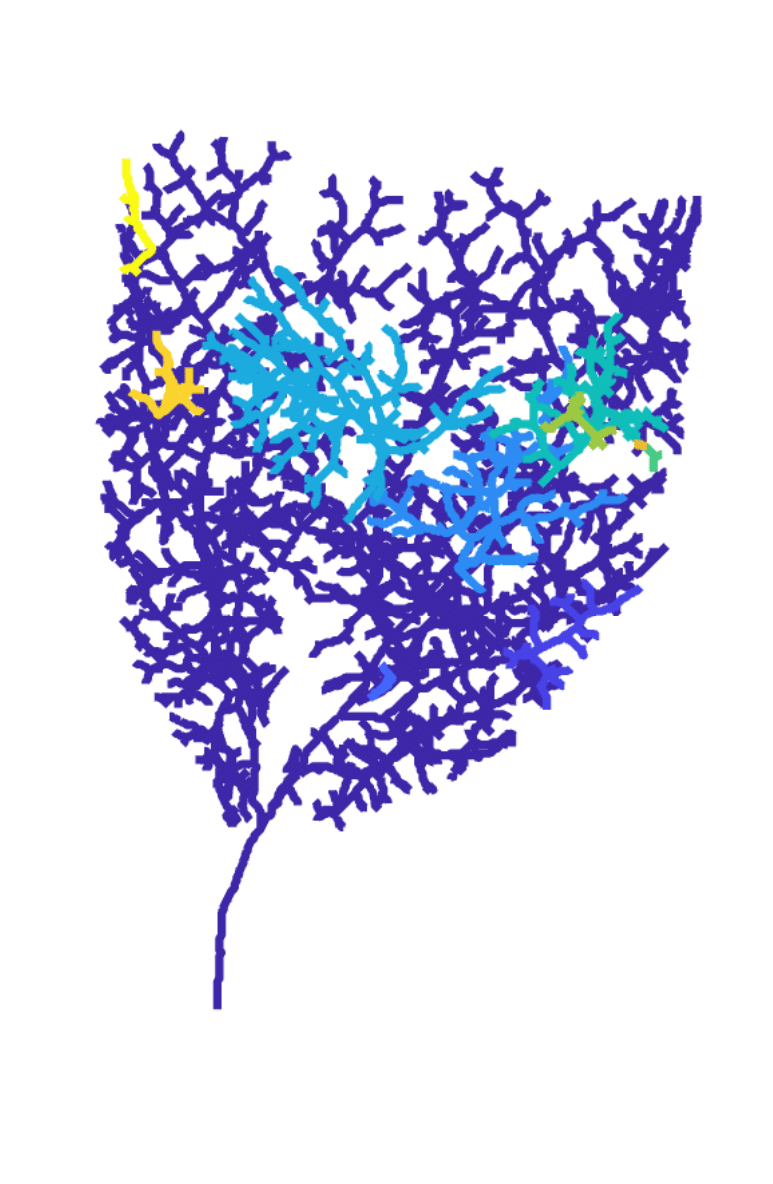}}
    \caption{The subtrees that comprise each of the pulmonary venous networks of 6 mice. In each panel, each disconnected subtree is shown in a different colour. Some subtrees are small and are hence hard to see.} \label{fig:A:discon}
\end{figure}

Initially, each of these pulmonary networks are comprised of disconnected subtrees. The subtrees are found by finding the inlet node (defined here as the terminal node with the least $z$ coordinate and finding the segments of the tree that extend from there. Subtrees that are disconnected from the inlet nodes are also found in this way, but by defining a subtree inlet node, chosen as any terminal node in the subtree. The disconnected components for each of the trees are shown in Fig.~\ref{fig:A:discon}. From this we can see that some of the trees have many disconnected components.

Network (a) (Fig.~\ref{fig:mouse1}(a)) is comprised of three disconnected subtrees. The largest of the subtrees contains 1976 (97.34\% by count) of the vessels. The other subtrees contain 51 (2.51\%, shown in yellow at the top of the network), and 3 (0.15\% shown in cyan to the right of the panel). The vast majority of the network is contained within the largest subtree. Therefore, take the largest subtree to be the whole network. It is not always the case that the vast majority of the network is contained within a single subtree, as in Fig.~\ref{fig:mouse1}(b). If the largest subtree contains less than 95\% of the vessels (by count) of a network, we wish to form a larger network from the subtrees. 

\begin{figure}[ht]\centering
    \subfloat[]{\includegraphics[width = 0.2\textwidth]{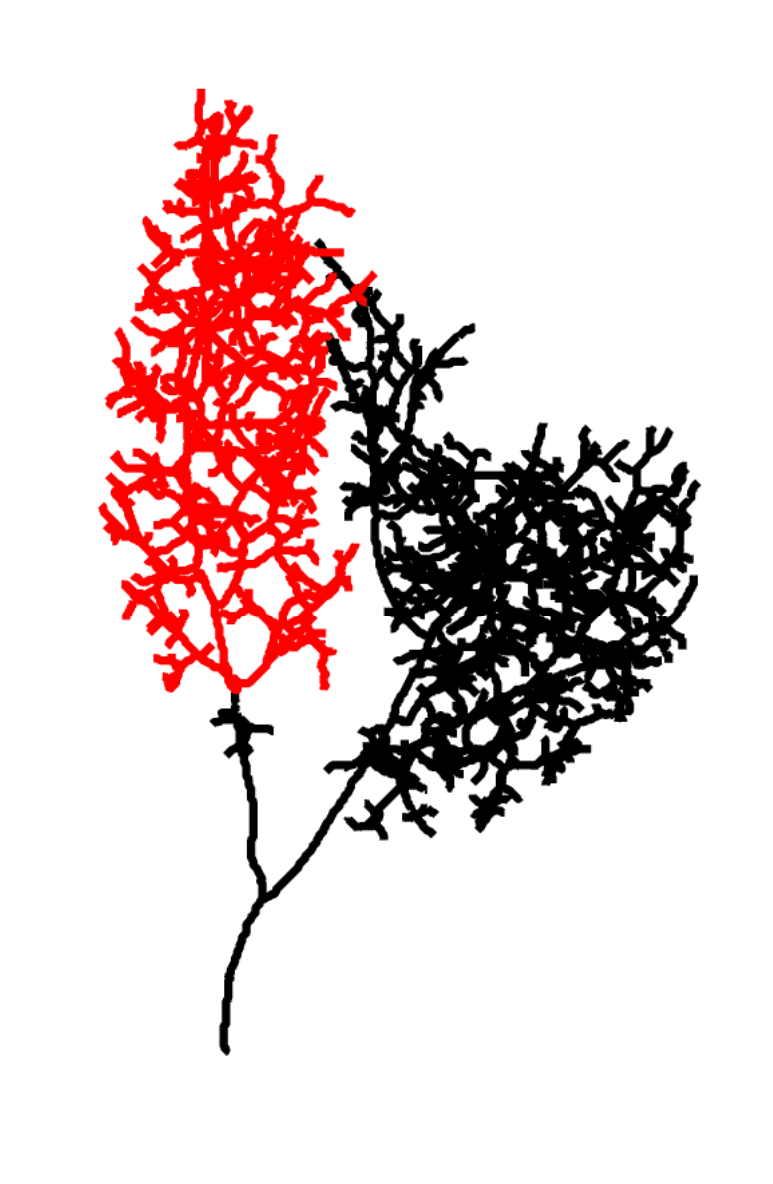}}
    \subfloat[]{\includegraphics[width = 0.2\textwidth]{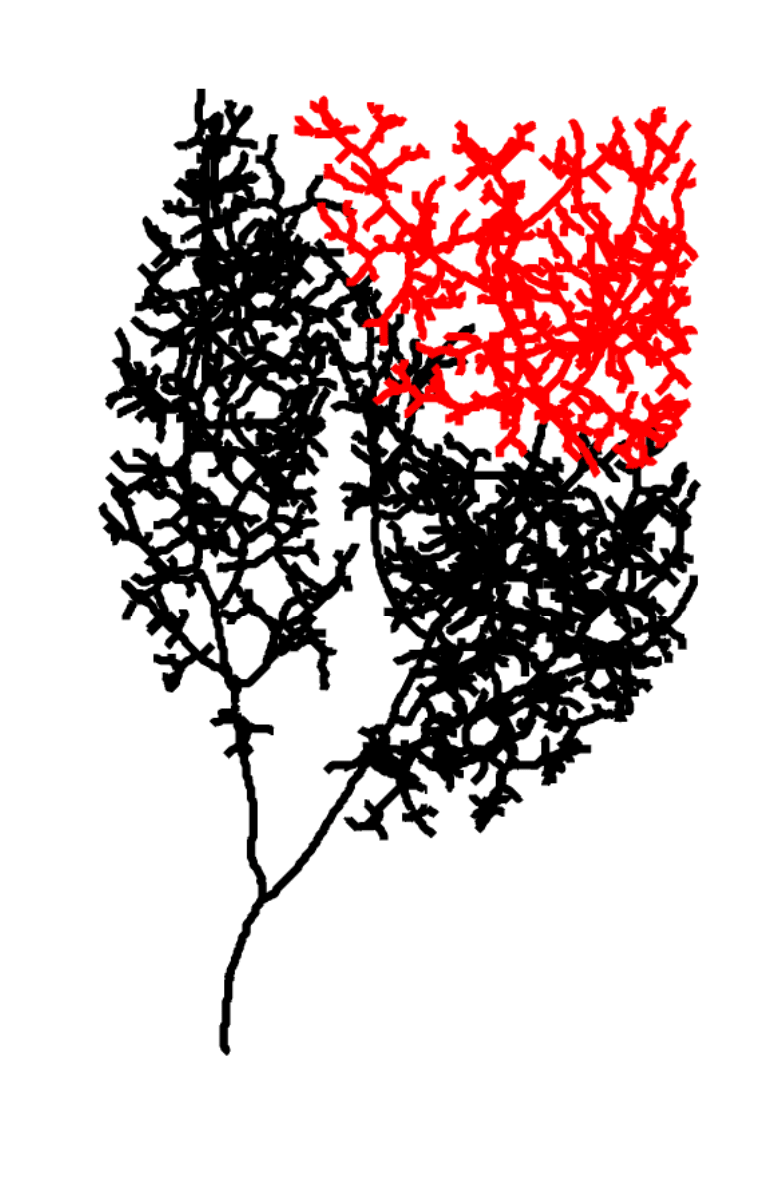}}
    \subfloat[]{\includegraphics[width =  0.2\textwidth]{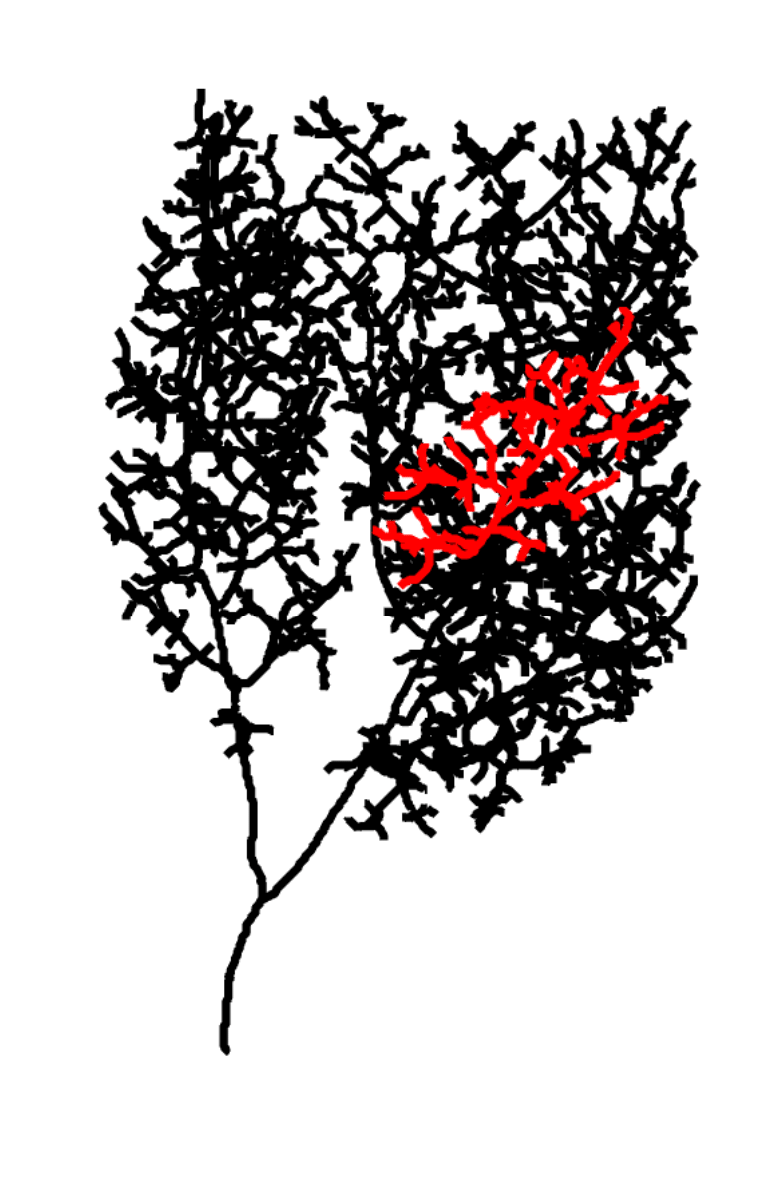}}
    \subfloat[]{\includegraphics[width =  0.2\textwidth]{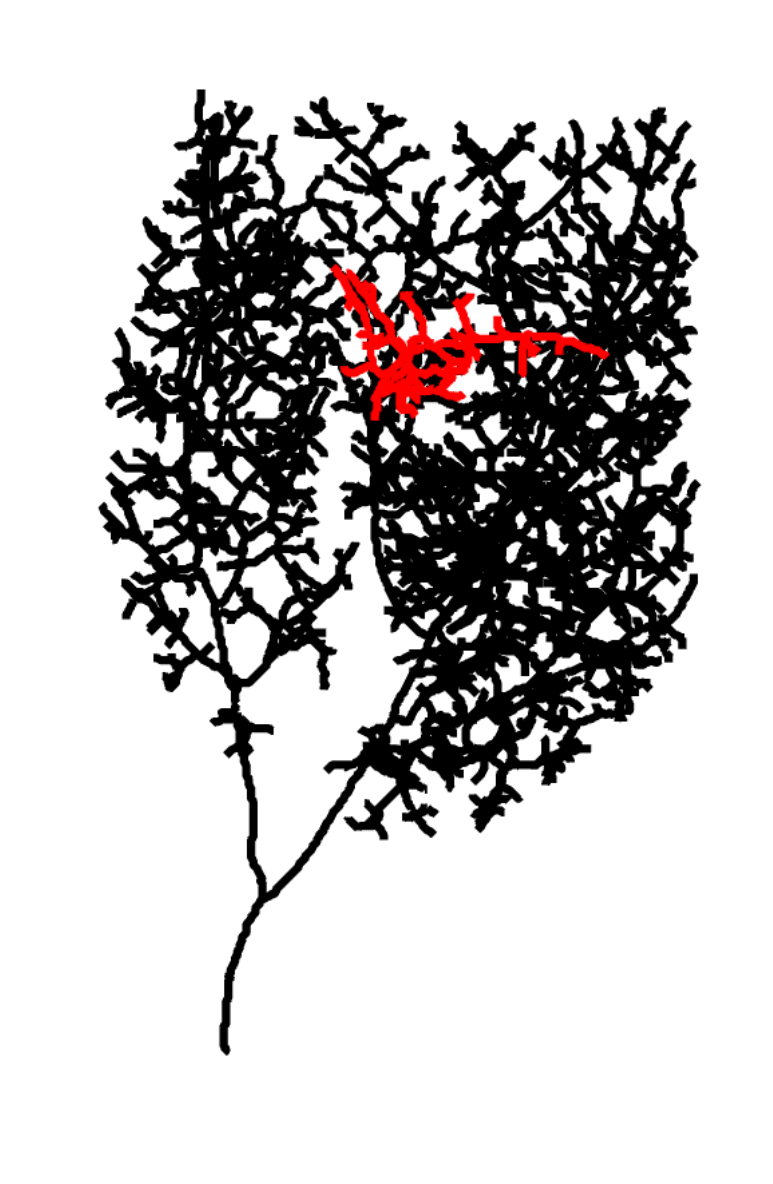}}
    \caption{The steps of joining the subtree containing the inlet node (shown in black) with the closest disconnected subtree (in red). Panel (a) represents the first join of trees of size 1104 and 872, (b) the second joining of trees of size 1976 and 710, (c) the third joining trees of size 2686 and 174 and (d) the final join of trees of size 2860 and 95.} \label{fig:A:joins}
\end{figure}

The tree shown in the 2$^{\rm nd}$ panel contains 13 subtrees with between 1104 (36.1\% by count) and 23 (0.1\% by count) of the vessels. Essentially, to join a subtree with the largest subtree, find the minimum distance between all of the terminal nodes in the largest subtree and all of the terminal nodes in each of the smallest subtrees. Create a new edge that bridges the terminal nodes in the subtree containing the inlet and the other subtrees that minimises the distance between these. There are now fewer disconnected trees, and the subtree containing the inlet is larger. Continue this process until some stopping condition is met, that is, until 95\% of the node segments belong to the subtree containing the inlet. 

For the tree shown in the second panel, the sequence of subtrees that are joined is shown in Fig.~\ref{fig:A:joins}. The first step (shown in Fig.~\ref{fig:A:joins}(a)) joins two trees of size 1104 (36.1\%) and 872 (28.52 \%) to form the largest subtree of size 1976 segments (64.62\%). The second (shown in Fig.~\ref{fig:A:joins}(b)) joins the largest subtree with one of 710 segments to form one of 2686 segments (87.84\% of the tree). The third (shown in Fig.~\ref{fig:A:joins}(c)) joins the largest subtree of 2686 with one of 174 segments to form one of 2860 segments (93.53\%) and finally, this is joined with one of 95 segments (shown in Fig.~\ref{fig:A:joins}(d)) to form a largest subtree containing 2955 segments (96.63\%). This process is performed with the other 4 trees to produce the connected subtrees that contain at least 95\% of the segments, including the inlet. As we start from the inlet, the segments that are most likely to be excluded from the final trees are at the distal ends of branches, so probably also contain fewer nodes than more proximal branches. The practical result of this is that at 95\% of the segments contain at least 95\% of the nodes.

\begin{figure}[ht]\centering
    \subfloat[]{\includegraphics[width = 0.16\textwidth]{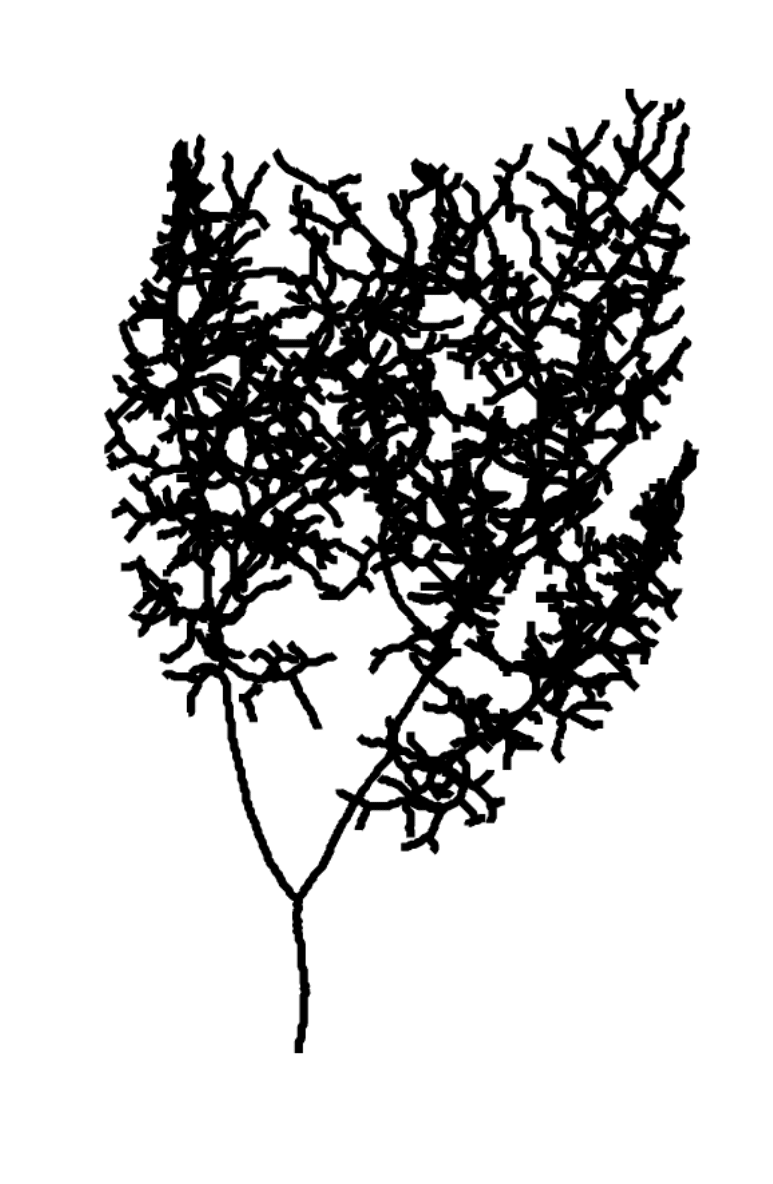}}
    \subfloat[]{\includegraphics[width =  0.16\textwidth]{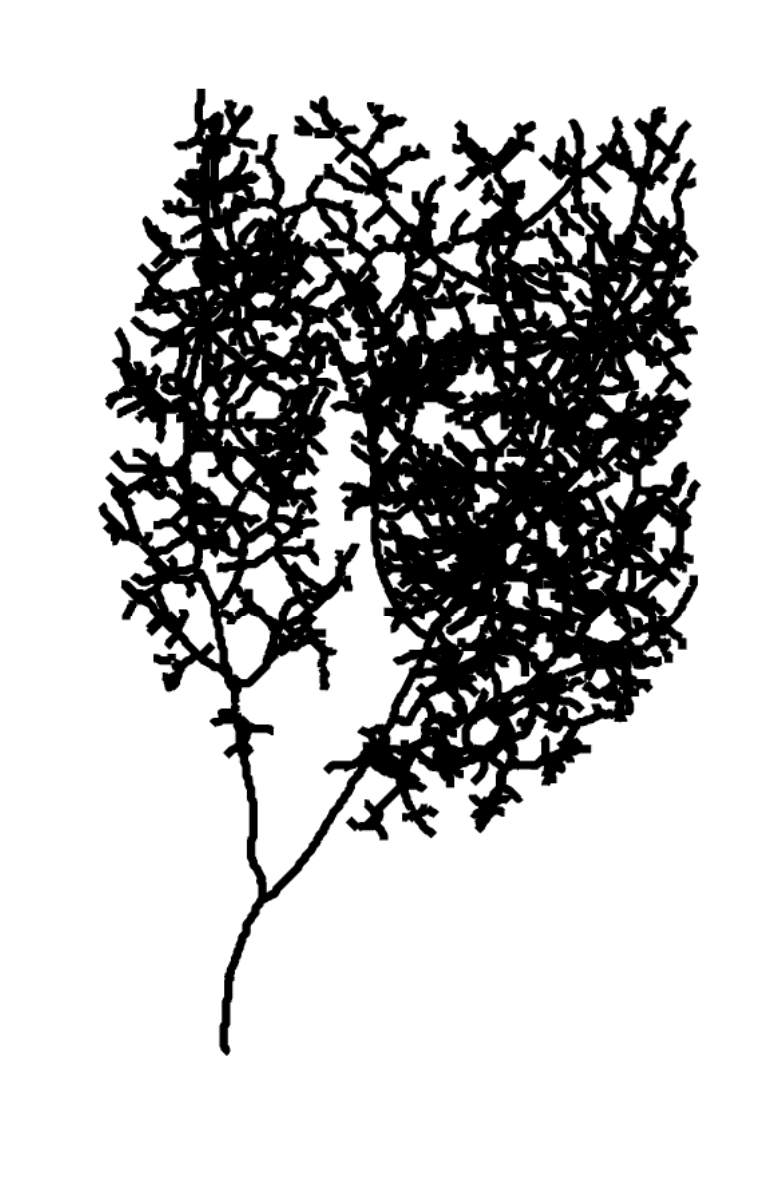}}
    \subfloat[]{\includegraphics[width =  0.16\textwidth]{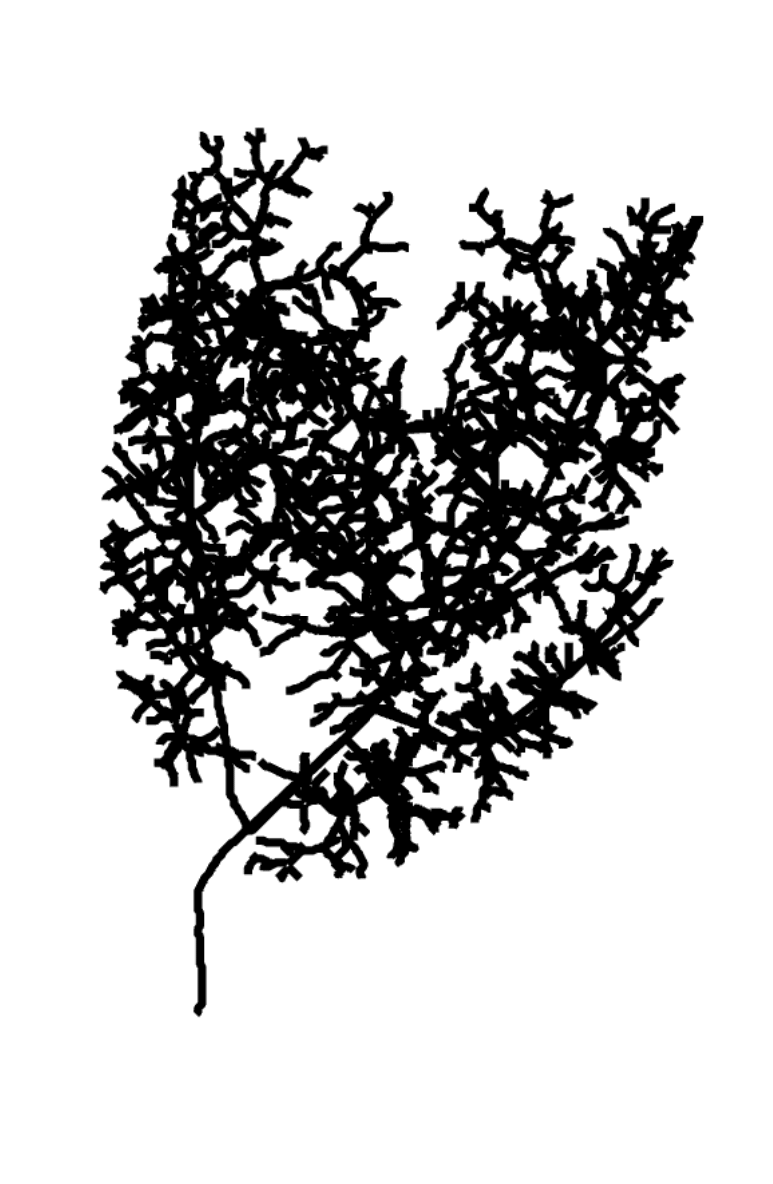}}
    \subfloat[]{\includegraphics[width =  0.16\textwidth]{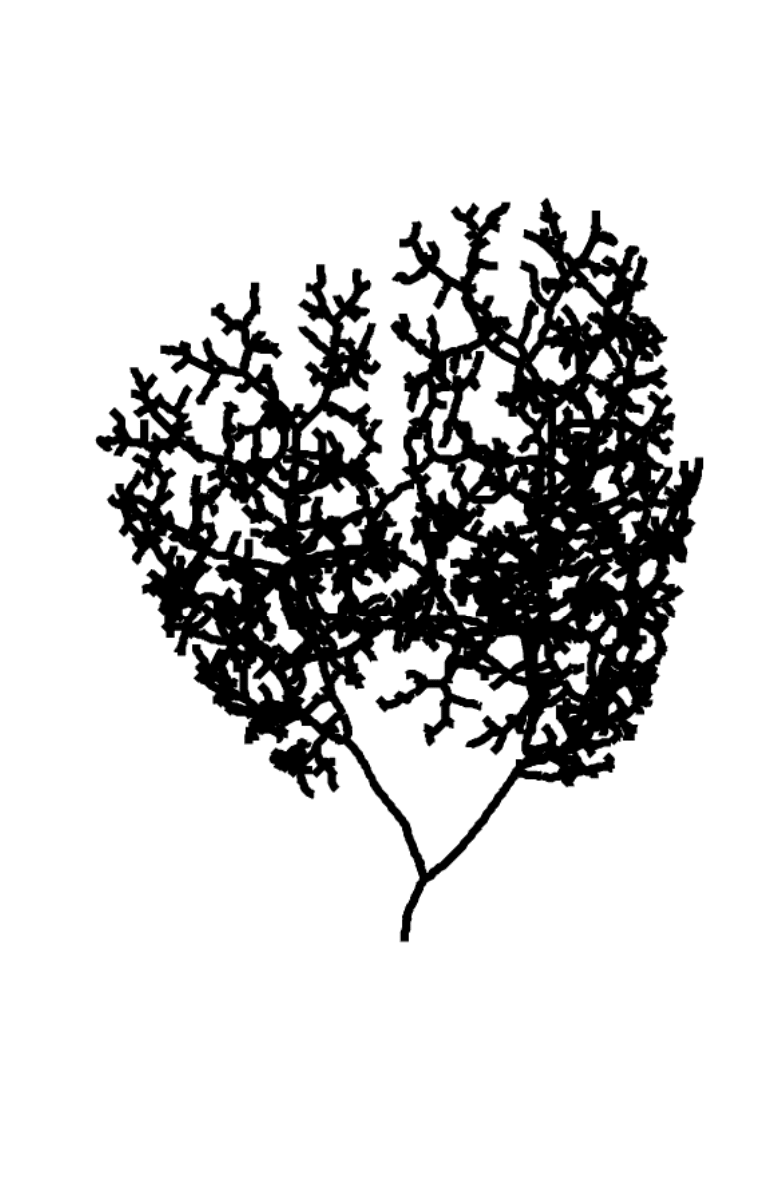}}
    \subfloat[]{\includegraphics[width =  0.16\textwidth]{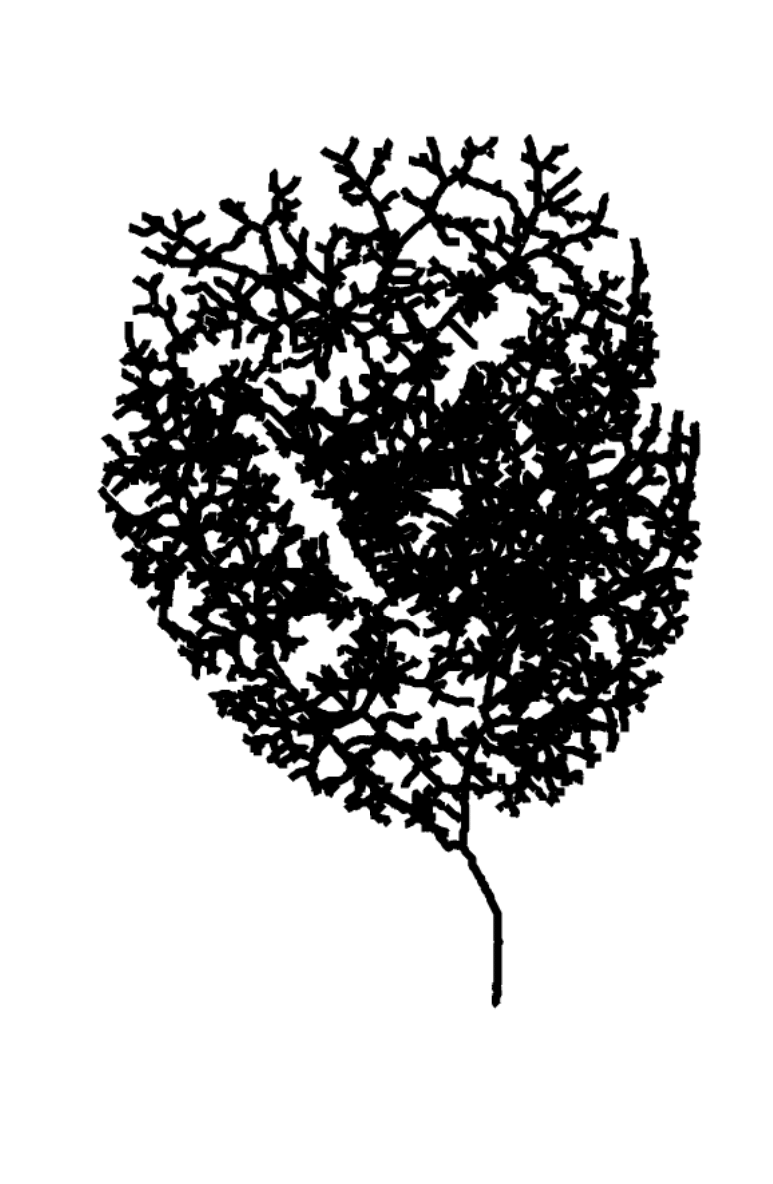}}
    \subfloat[]{\includegraphics[width =  0.16\textwidth]{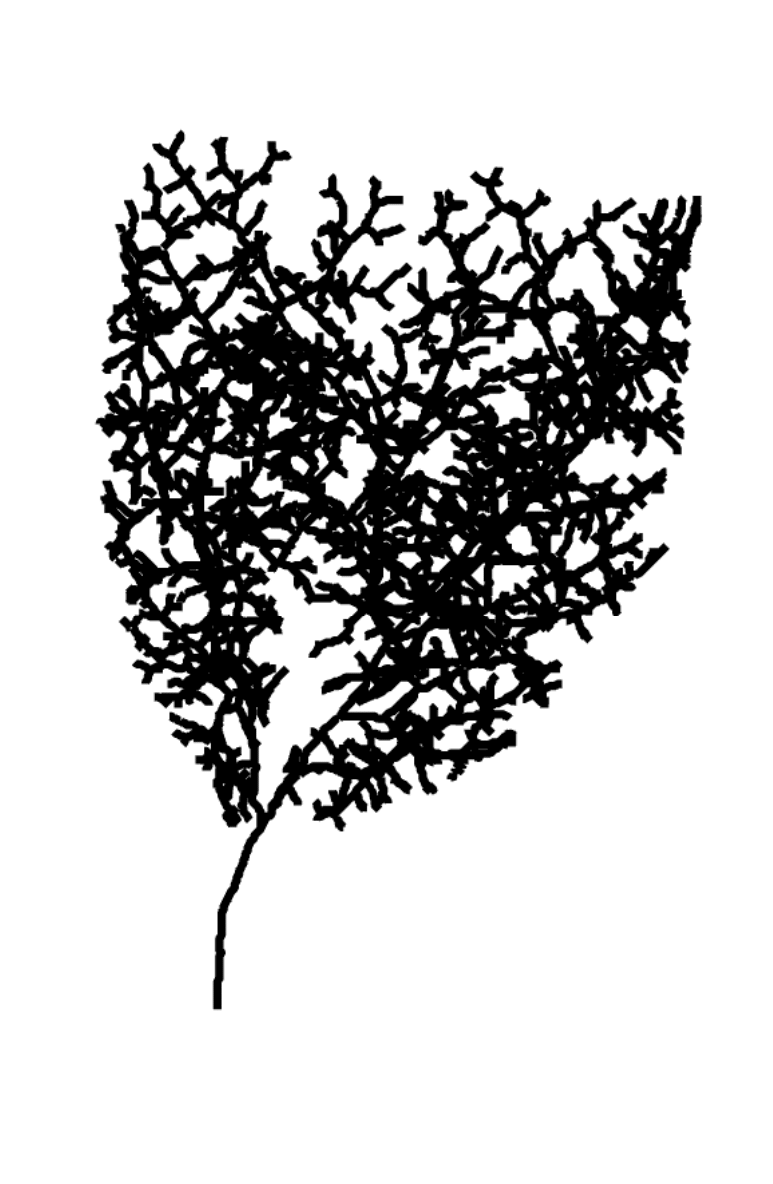}}
    \caption{The subtrees that result from joining the disconnected subtrees until one of the subtrees contains at least 95\% of the total segments.} \label{fig:A:most}
\end{figure}

We have now obtained 6 connected networks in the correct format. The segments were generated during the process of joining the subtrees. They are shown in Fig.~\ref{fig:A:most}.

\begin{figure}[ht]\centering
    \subfloat[]{\includegraphics[width = 0.16\textwidth]{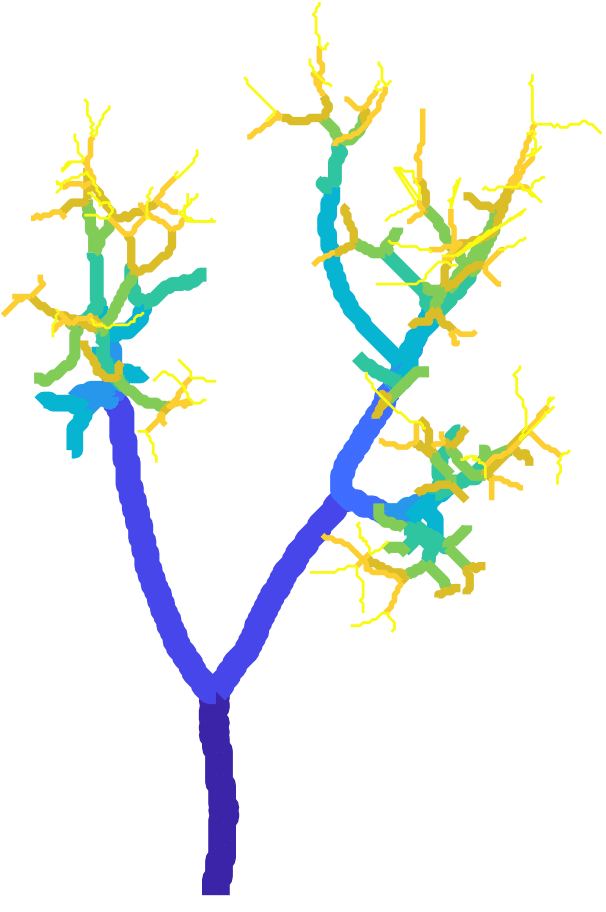}}
    \subfloat[]{\includegraphics[width =  0.16\textwidth]{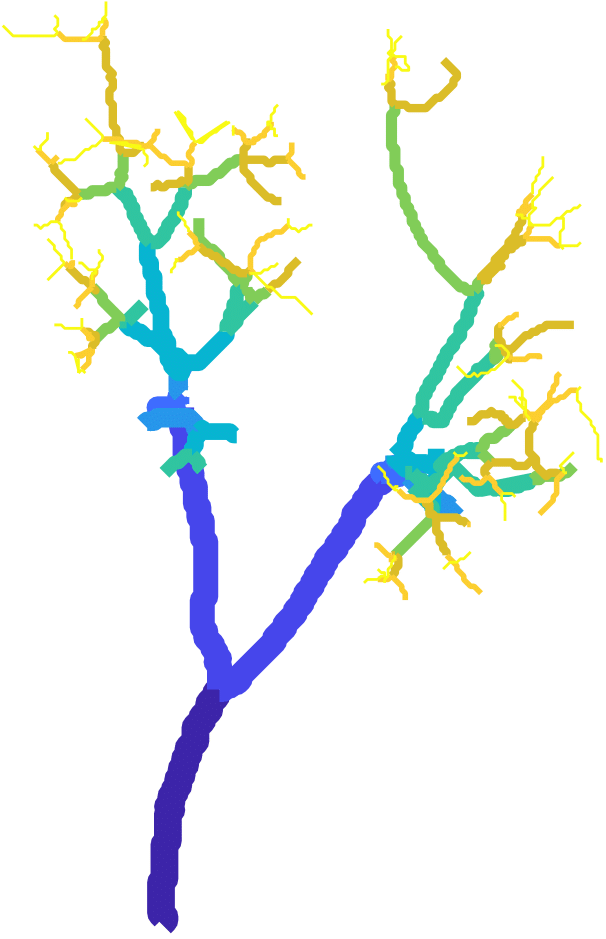}}
    \subfloat[]{\includegraphics[width =  0.16\textwidth]{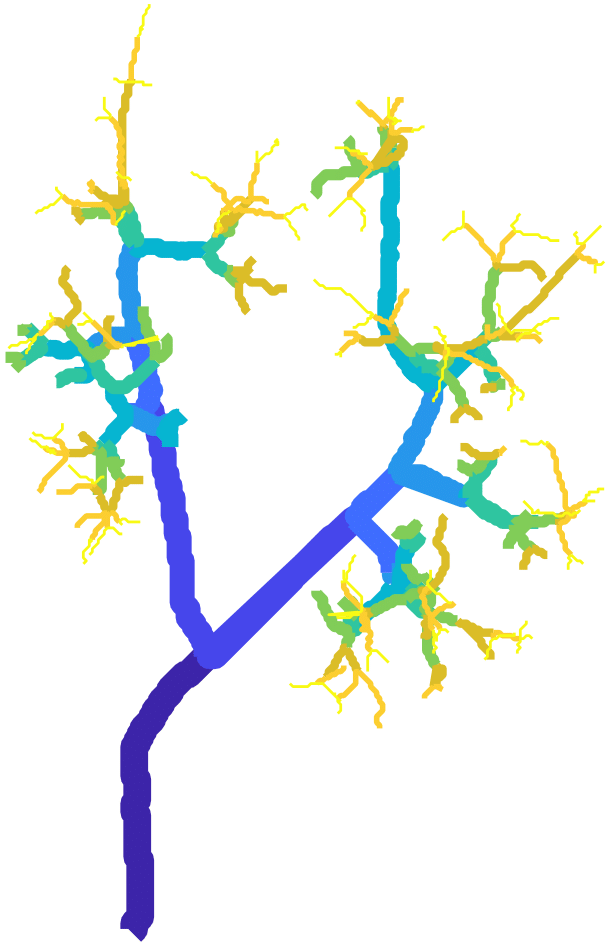}}
    \subfloat[]{\includegraphics[width =  0.16\textwidth]{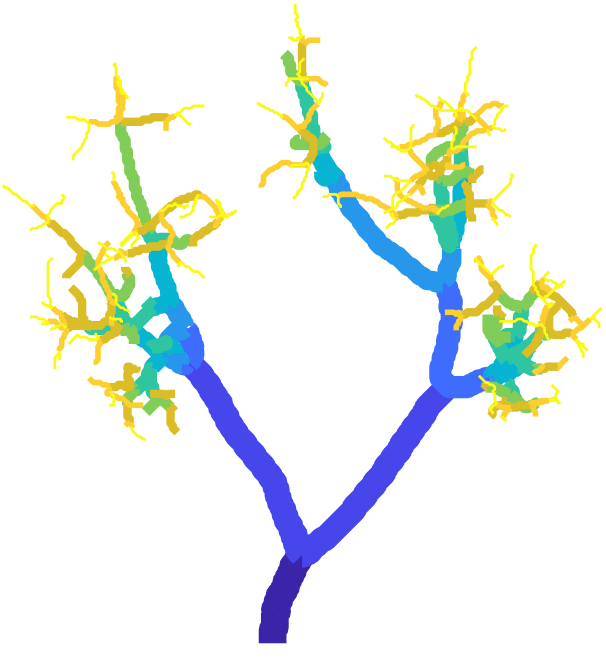}}
    \subfloat[]{\includegraphics[width =  0.16\textwidth]{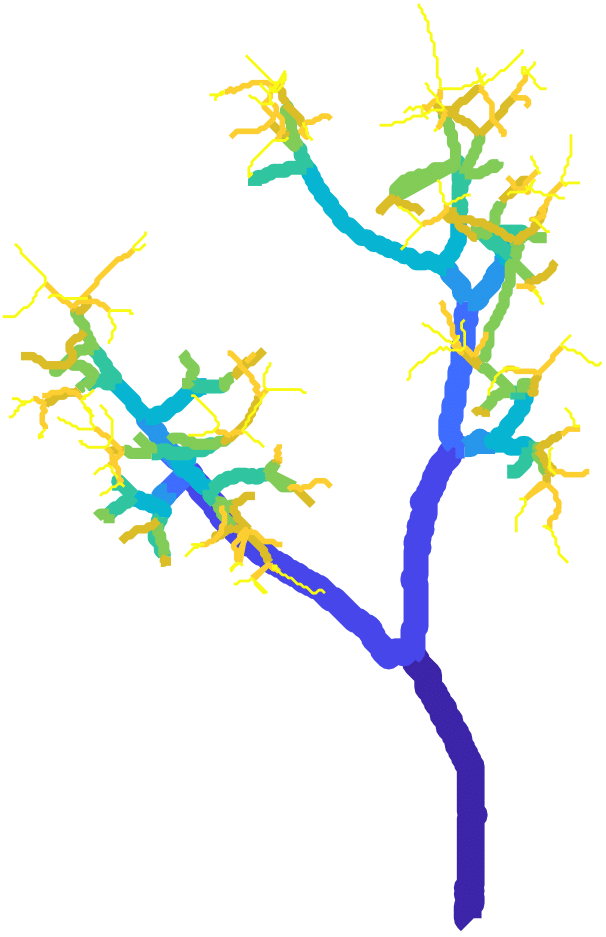}}
    \subfloat[]{\includegraphics[width =  0.16\textwidth]{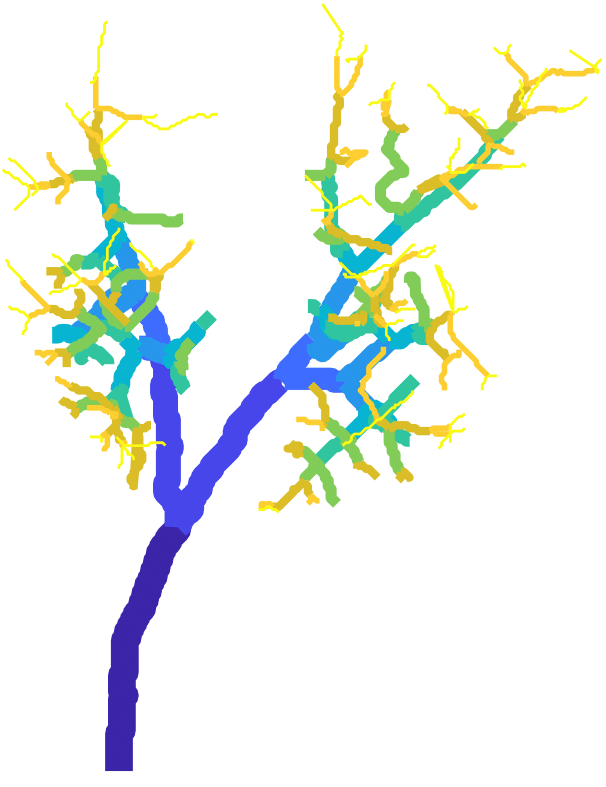}}
    \caption{The first 10 generations of vessels in each network is shown. Vessels in lower generations (closer to the inlet vessel) are shown in thicker lines.} \label{fig:A:gens}
\end{figure}

\begin{figure}[ht]\centering
    \subfloat[]{\includegraphics[width = 0.16\textwidth]{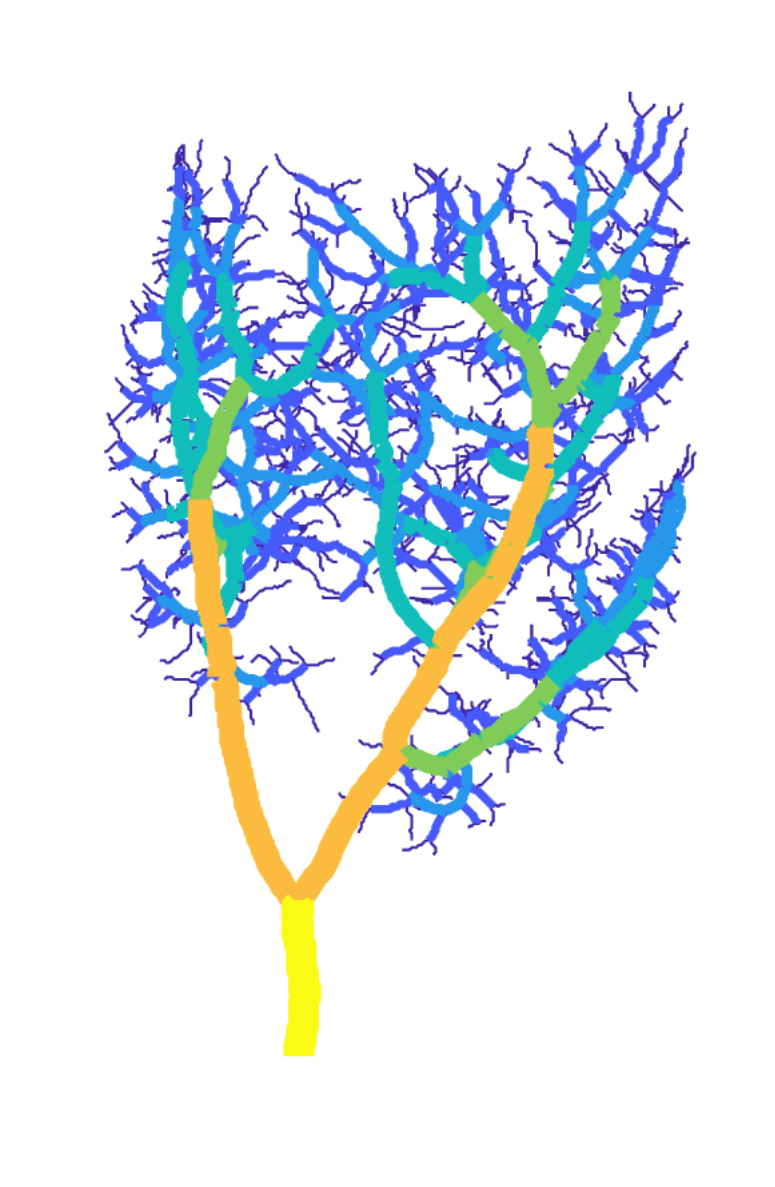}}
    \subfloat[]{\includegraphics[width =  0.16\textwidth]{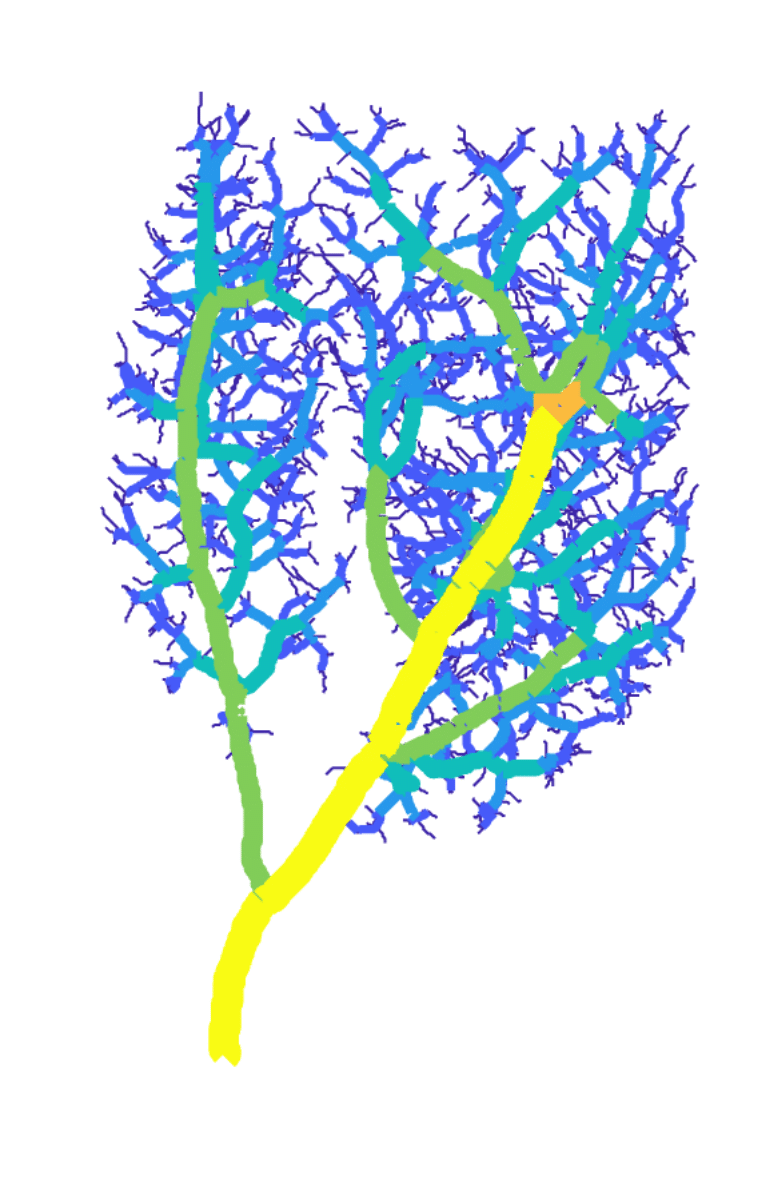}}
    \subfloat[]{\includegraphics[width =  0.16\textwidth]{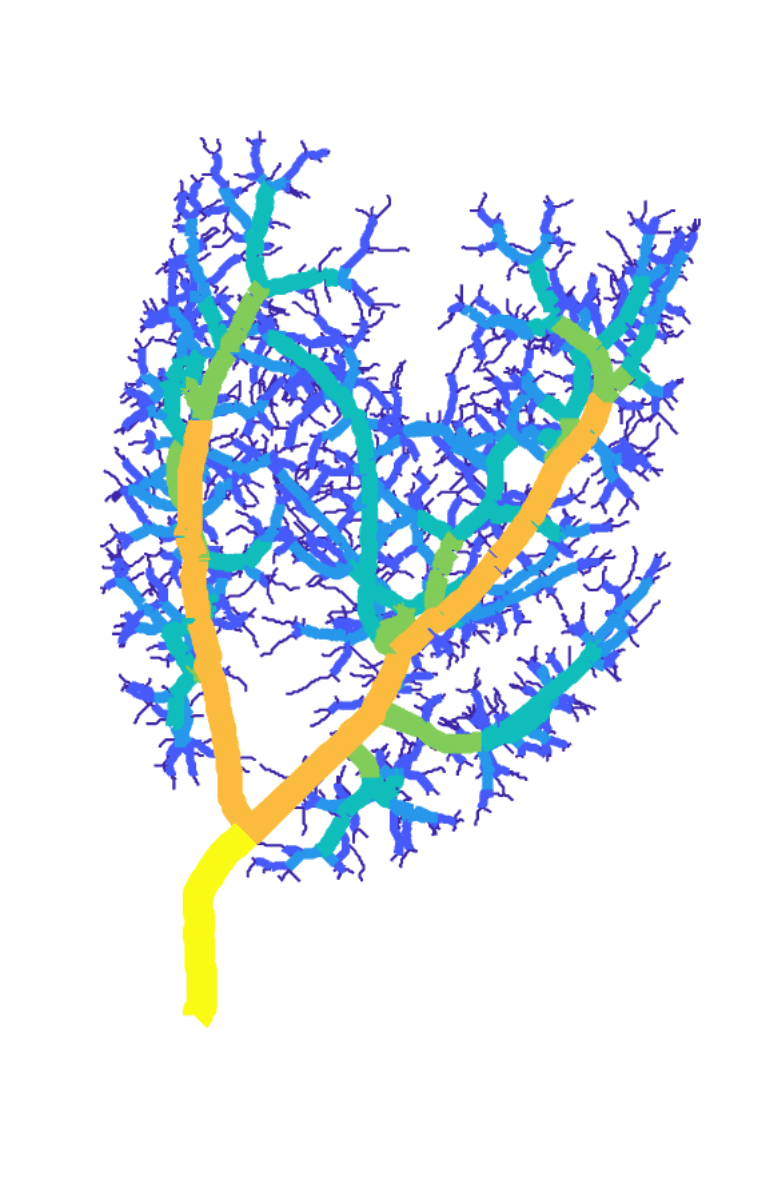}}
    \subfloat[]{\includegraphics[width =  0.16\textwidth]{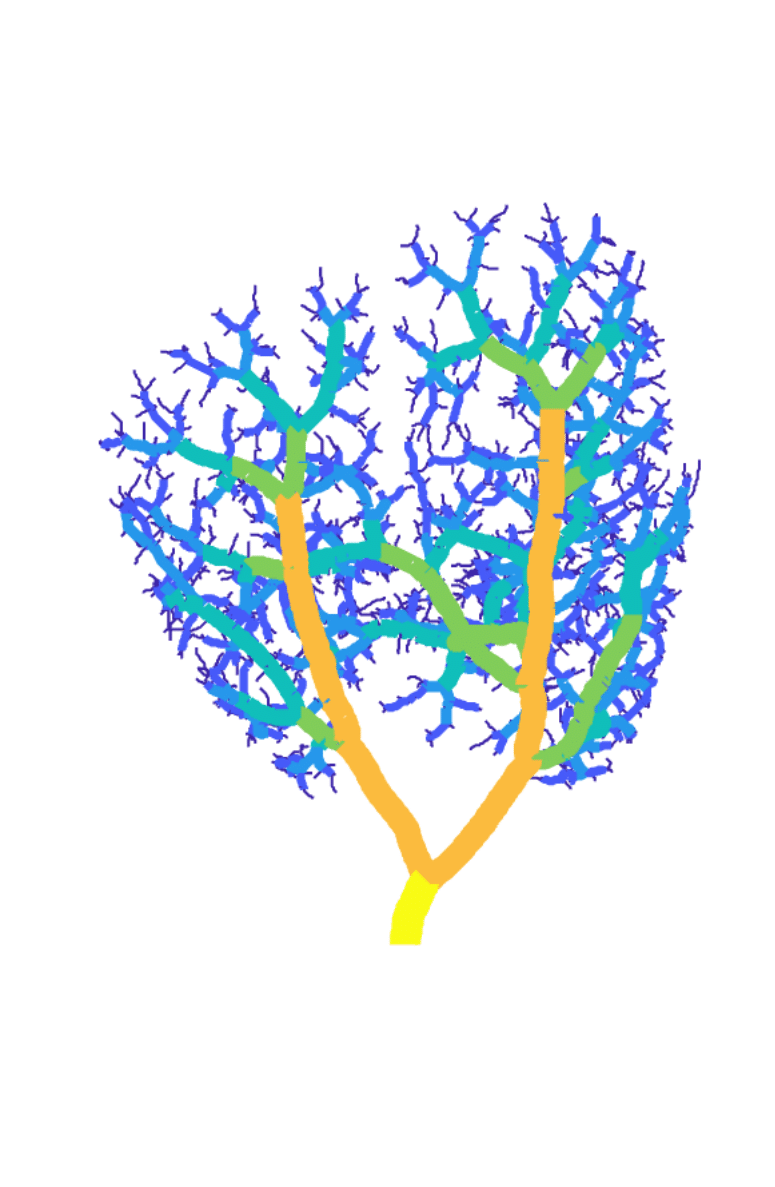}}
    \subfloat[]{\includegraphics[width =  0.16\textwidth]{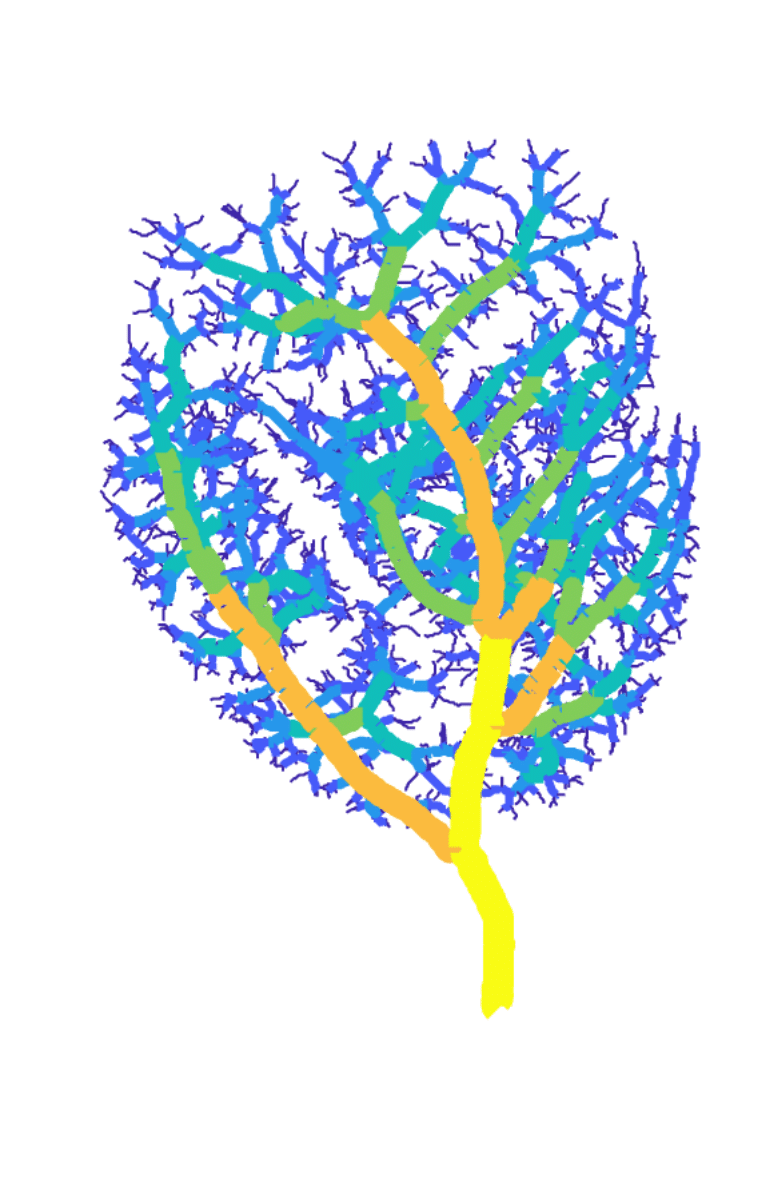}}
    \subfloat[]{\includegraphics[width =  0.16\textwidth]{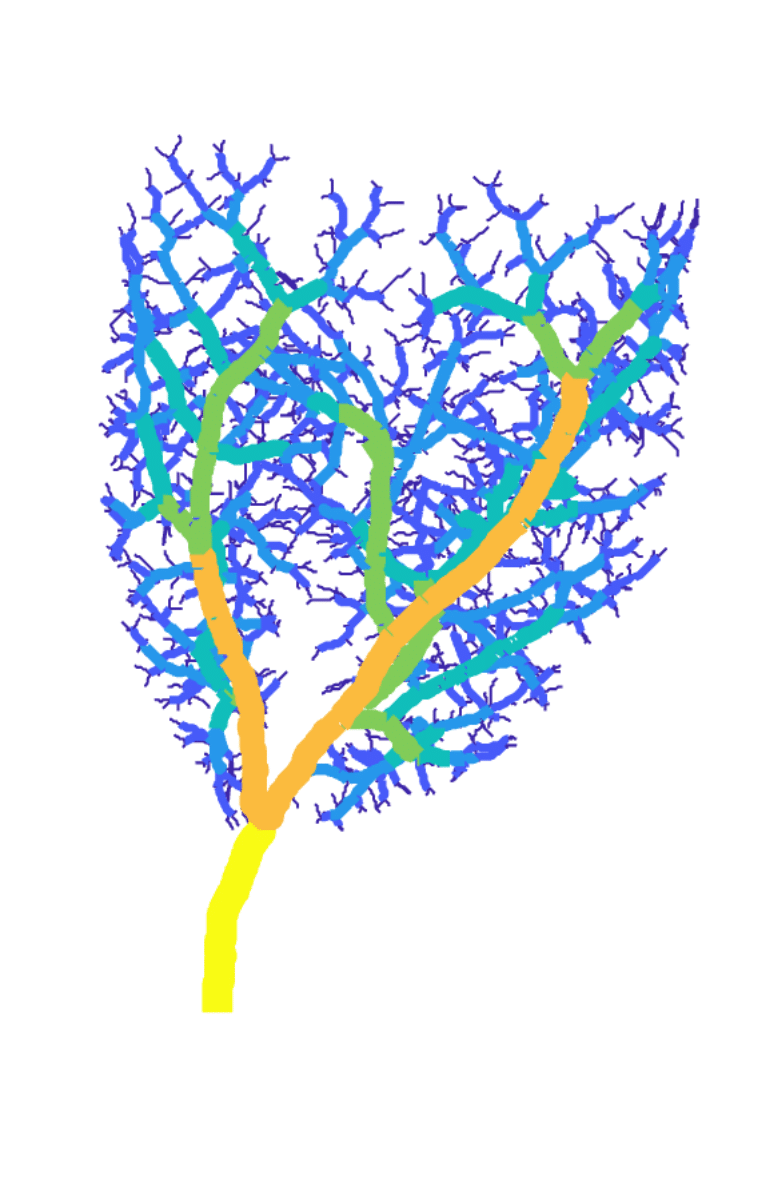}}
    \caption{The Strahler order of the segments of each network are indicated by line weight and colour. Thicker, more yellow edges correspond to a higher Strahler order than thinner, more blue edges.} \label{fig:A:Strahler}
\end{figure}

\begin{figure}[ht]\centering
    \subfloat[]{\includegraphics[width = 0.16\textwidth]{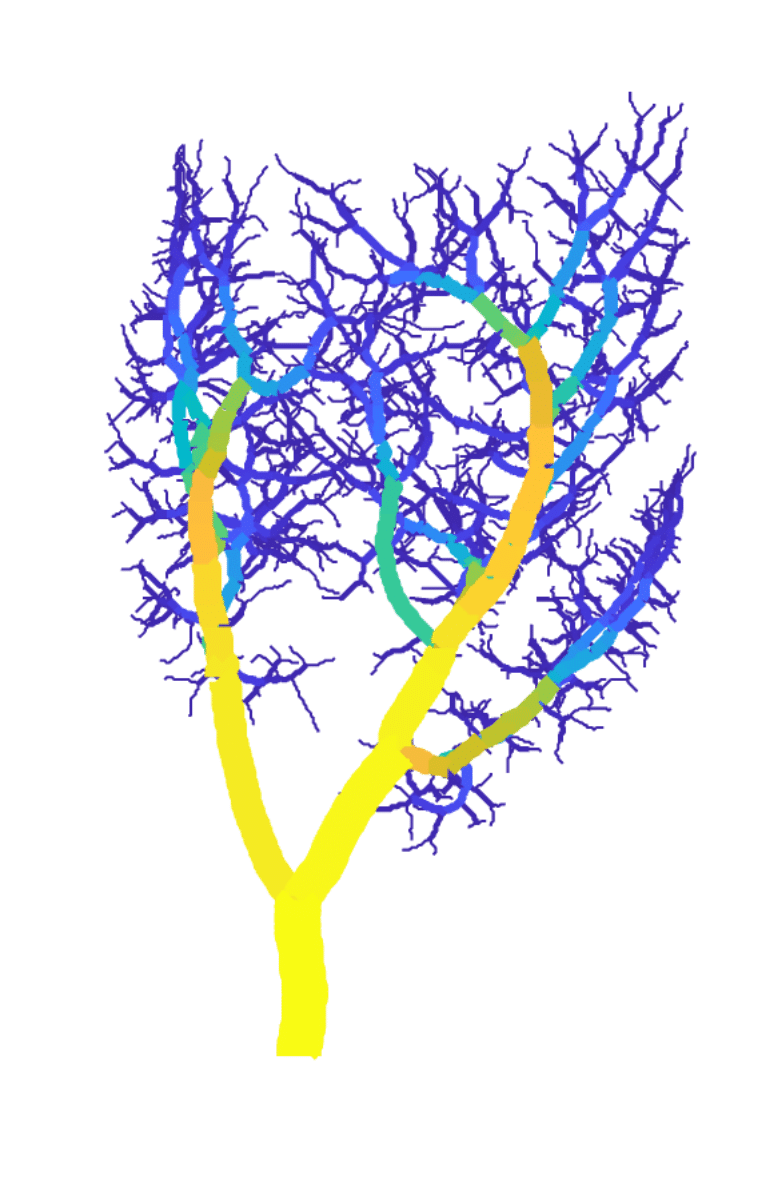}}
    \subfloat[]{\includegraphics[width =  0.16\textwidth]{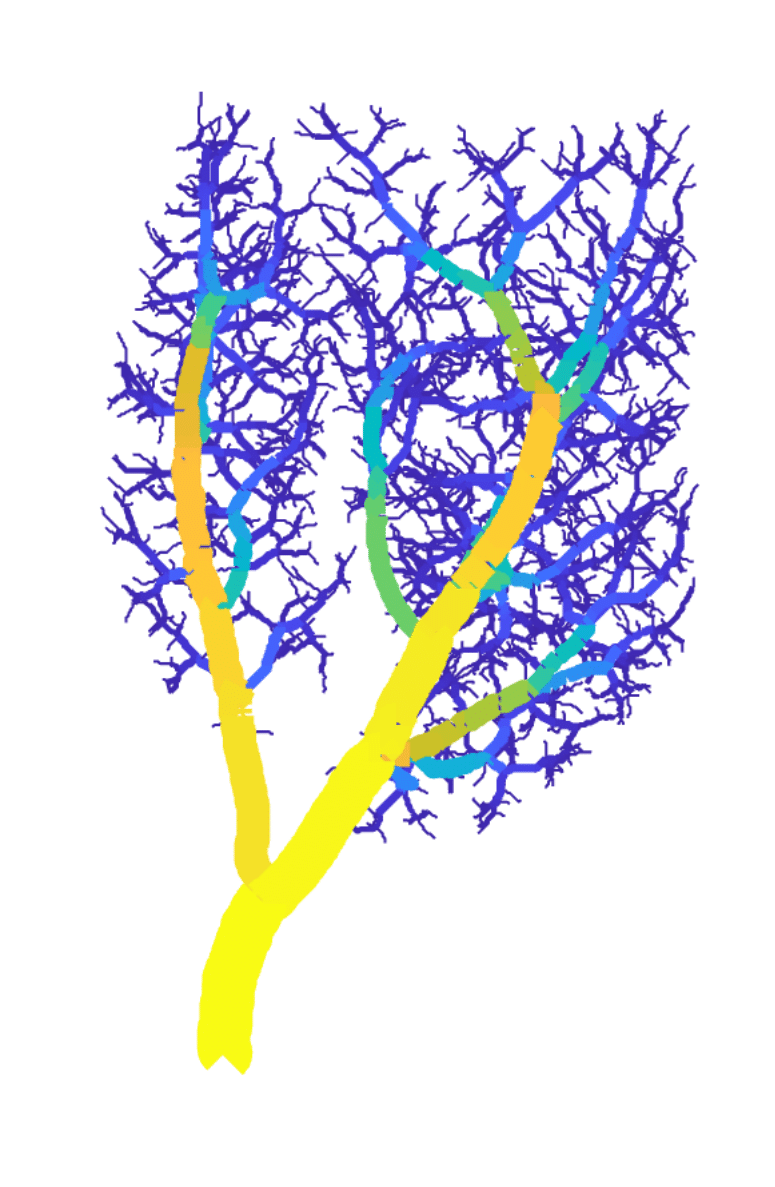}}
    \subfloat[]{\includegraphics[width =  0.16\textwidth]{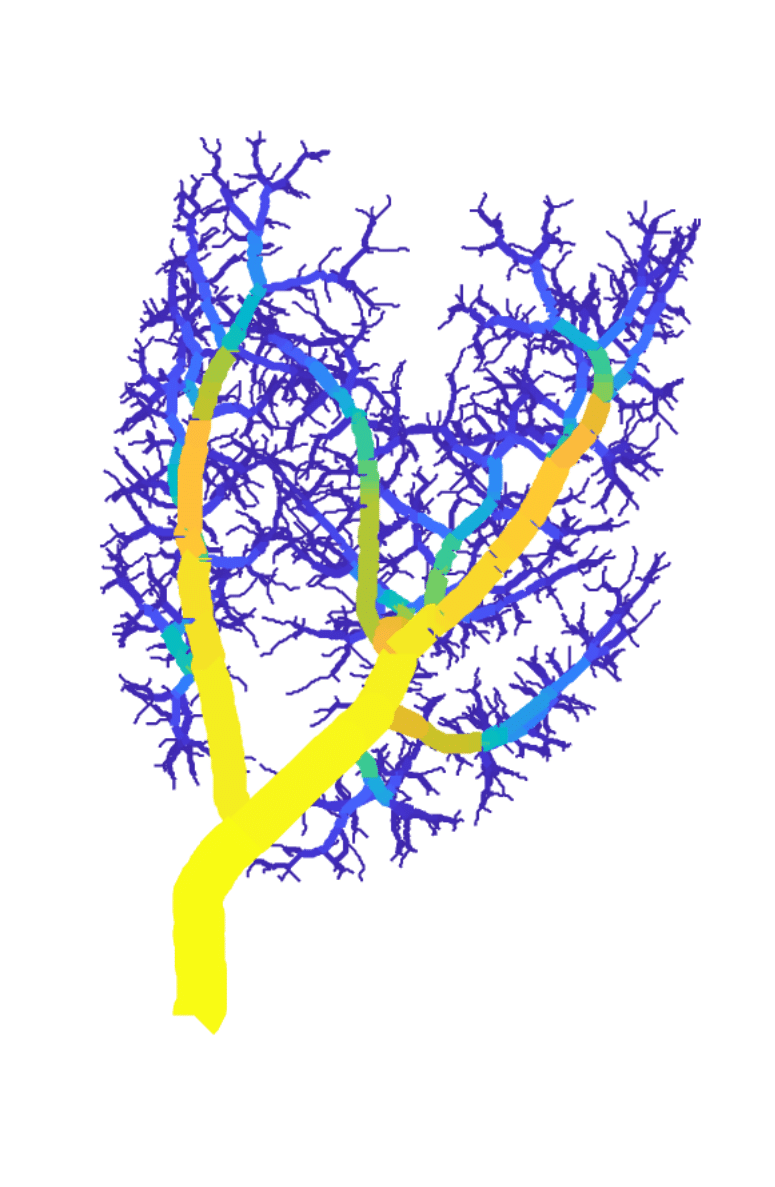}}
    \subfloat[]{\includegraphics[width =  0.16\textwidth]{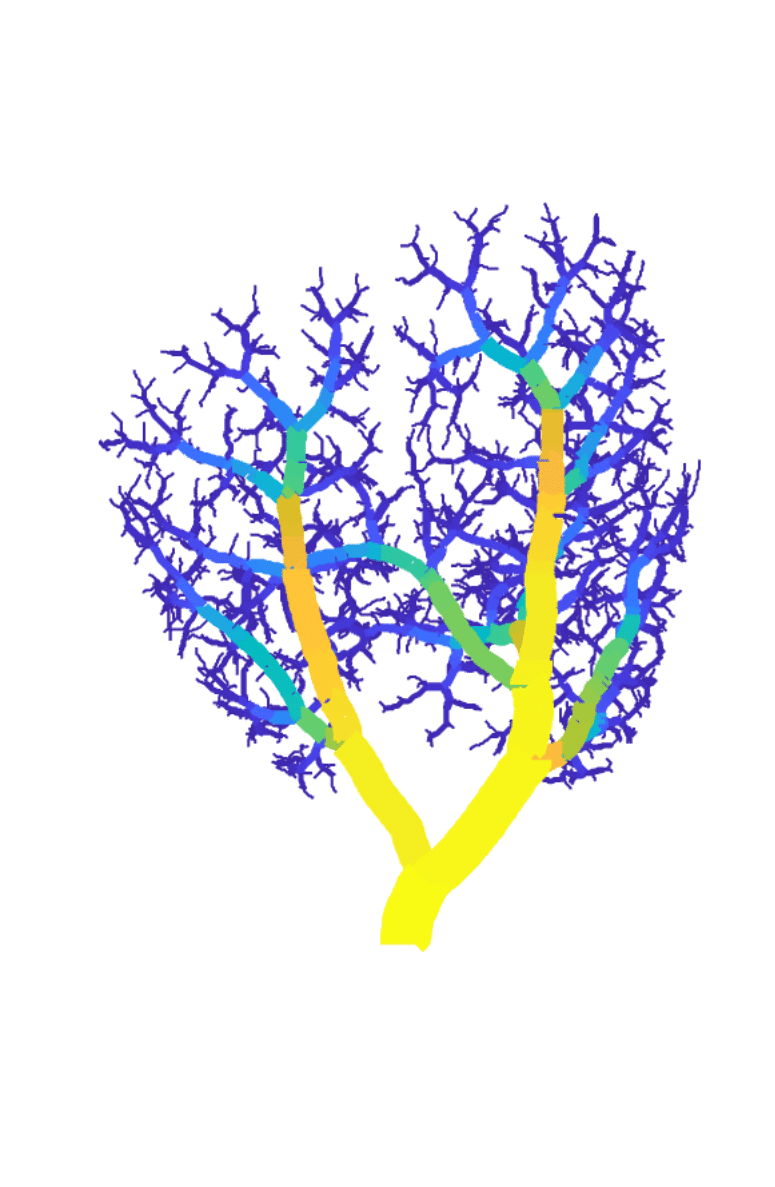}}
    \subfloat[]{\includegraphics[width =  0.16\textwidth]{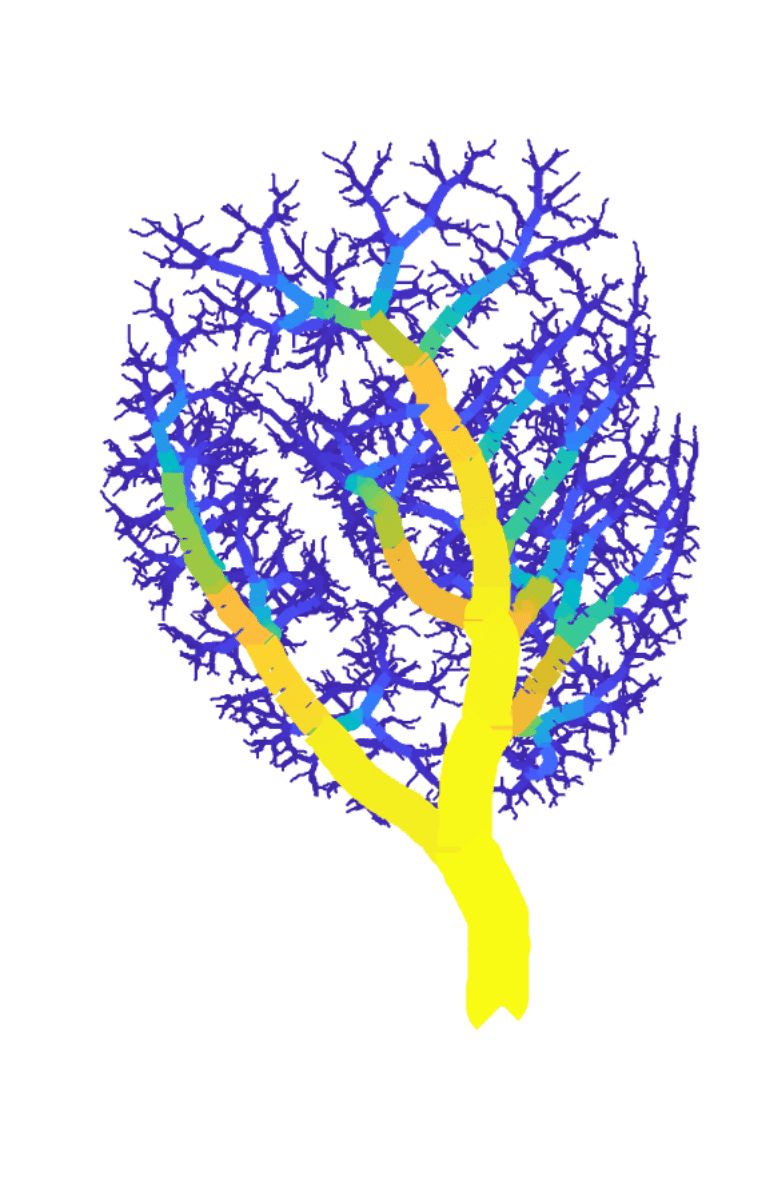}}
    \subfloat[]{\includegraphics[width =  0.16\textwidth]{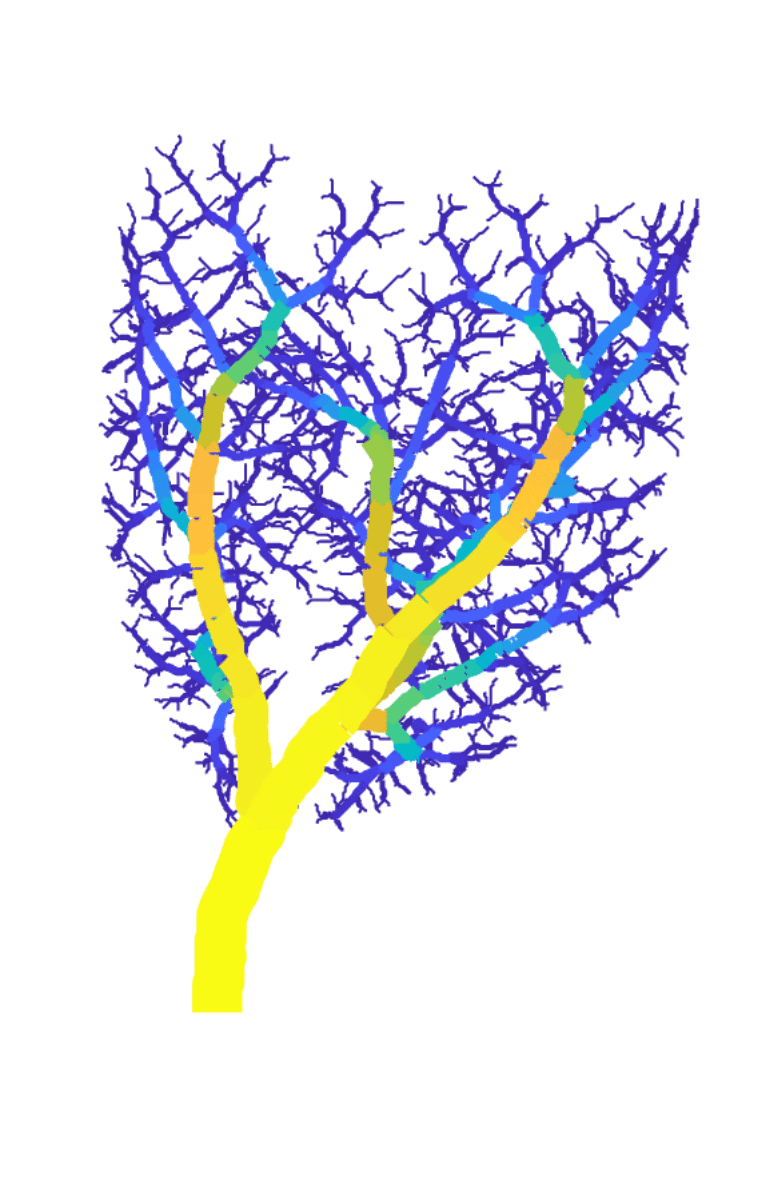}}
    \caption{The Strahler order of the segments of each network are indicated by line weight and colour. Thicker, more yellow edges correspond to a higher Strahler order than thinner, more blue edges.} \label{fig:A:Shreve}
\end{figure}

To these connected networks, we can apply various tree-metrics. The first 10 generations of the networks are shown in Fig.~\ref{fig:A:gens}. Generation is a top-down metric: vessels closer to the inlet have a lower generation number. Examples of bottom-up metrics are the Strahler \cite{strahler1952hypsometric} and Shreve \cite{shreve1966statistical} order. In both all terminal branches excluding the inlet are given an order of 1. In Shreve order, a branch that gives rise to branches of order $a$ and $b$ has order $a+b$. Strahler order is described above. They can be seen in Figs.~\ref{fig:A:Strahler} and \ref{fig:A:Shreve}, respectively.

From left to right, the trees contain 29, 38, 30, 35, 39, and 32 generations. The trees contain 1976, 2955, 2611, 2653, 4311, and 2442 branches. The maximal Strahler orders are 7 in all trees. The maximal Shreve orders are 997, 1493, 1320, 1339, 2183, and 1228. In general, trees that contain more branches also contain more generations and have a higher maximal Shreve order. This makes intuitive sense. 

\begin{figure}[ht]\centering
    \subfloat[]{\includegraphics[width = 0.2\textwidth]{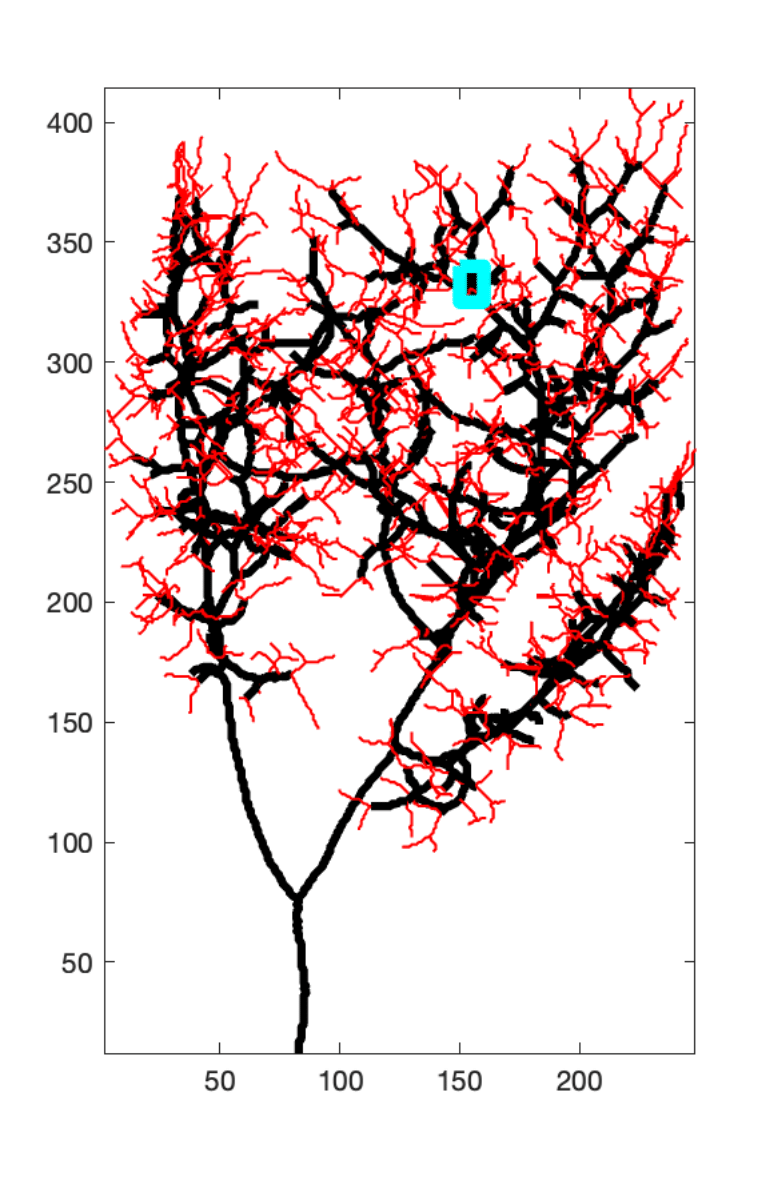}}
    \subfloat[]{\includegraphics[width =  0.2\textwidth]{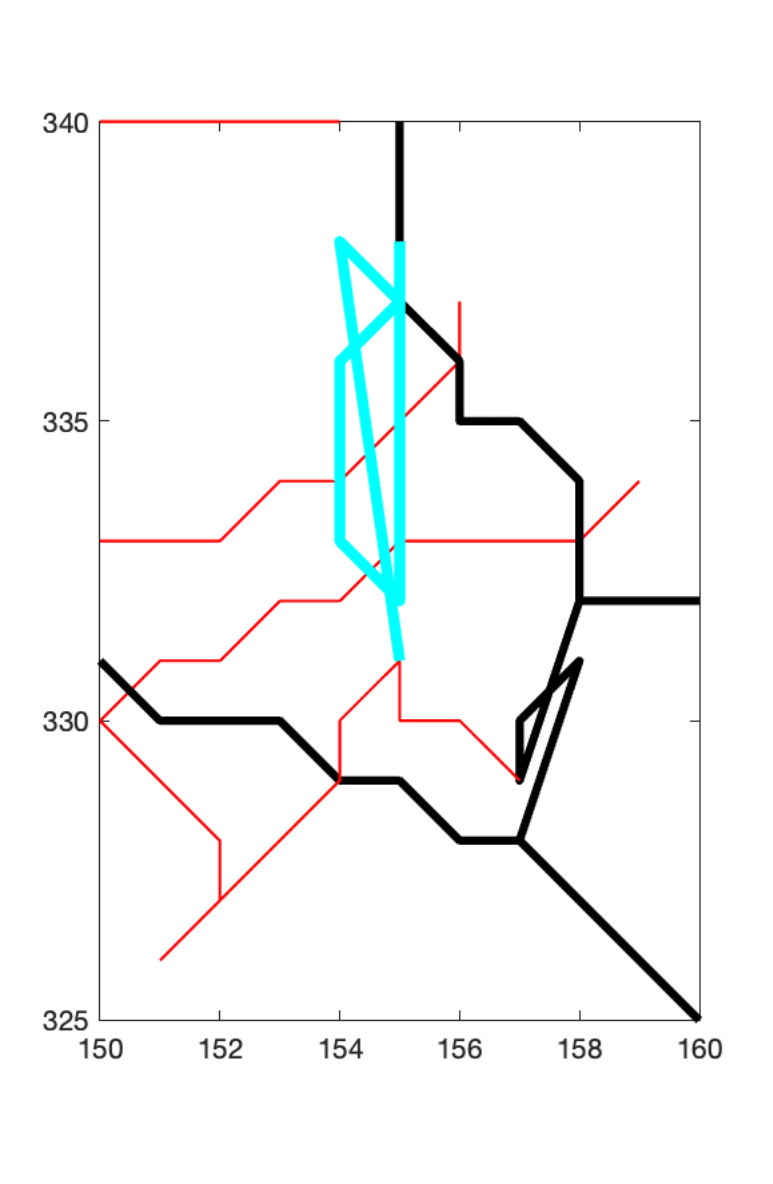}}
    \subfloat[]{\includegraphics[width =  0.2\textwidth]{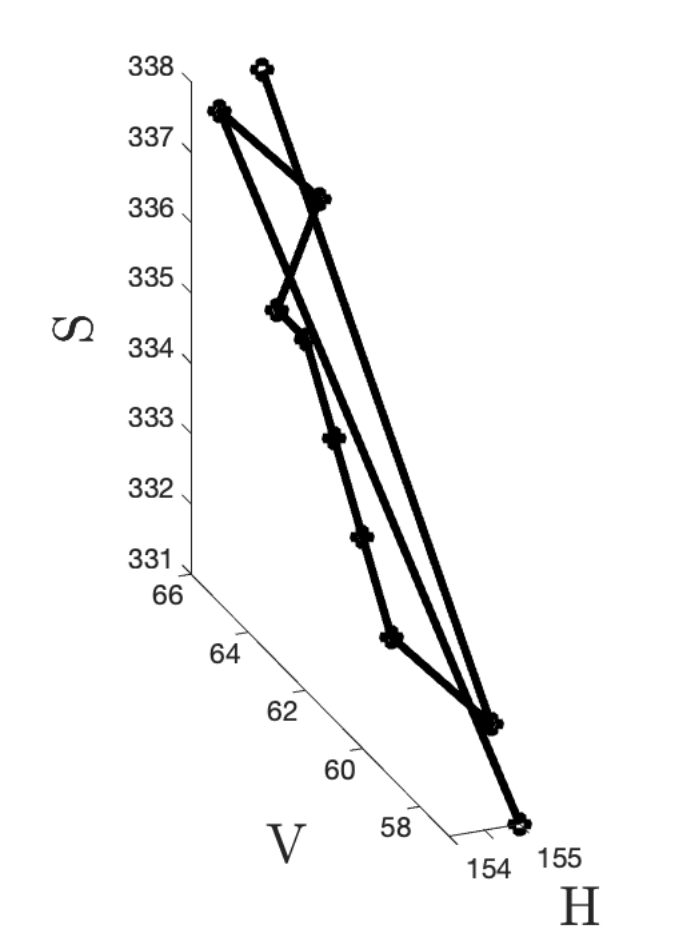}}
    \subfloat[]{\includegraphics[width =  0.2\textwidth]{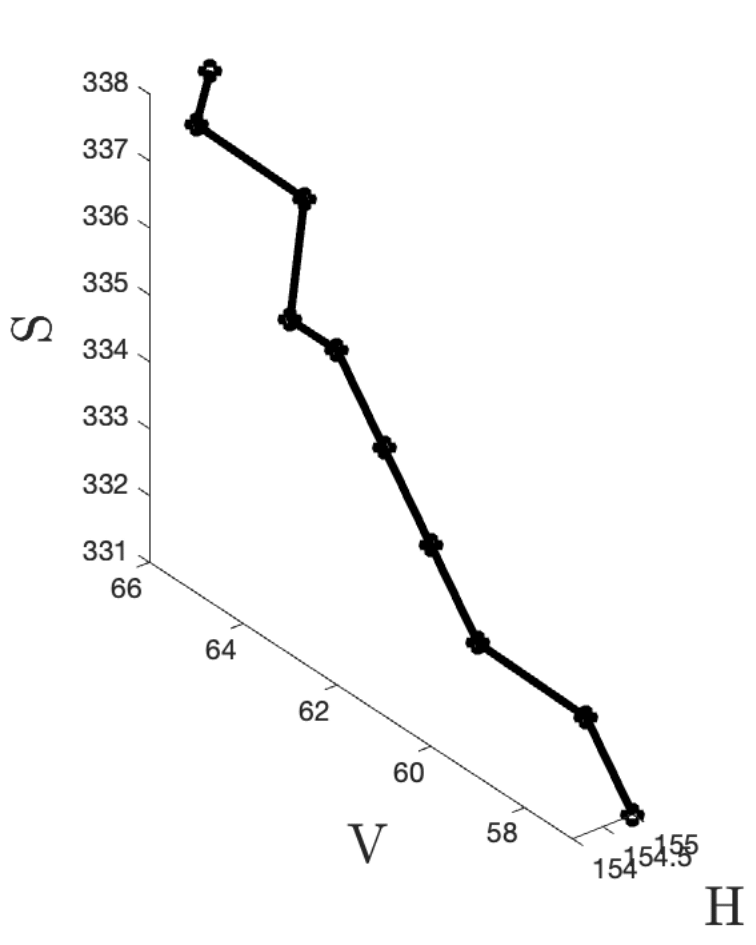}}
    \caption{(a) The network shown in Fig.~\ref{fig:A:most}(a). Red branches have mean radius less than the global mean, black branches are larger than the global mean. Panel (b) shows a zoomed-in version of the contents of the cyan box shown in panel (a). The cyan segment in this panel is spuriously long. Panel (c) shows a 3D view of the spuriously long segment alone. Here, the axis labels denote the Horizontal, Vertical, and Sagittal axes of the mouse \cite{schroder2020rodent}. Panel (d) shows the fixed segment. All dimensions are measured in microns ($\si{\micro\metre}$).} \label{fig:A:long}
\end{figure}

Due to an unknown data processing error that has lead to incorrect indexing, some segments are spuriously long. One such spuriously long segment is seen in context in Fig.~\ref{fig:A:long}(b) and alone in Fig.~\ref{fig:A:long}(c). The fixed segment in shown in Fig.~\ref{fig:A:long}(d) is achieved by reversing the order of the nodes listed to create the segment that do not include the end point. For instance, if the spuriously long segment were defined by the list $\{a, b, c, d\}$, the segment is edited to be $\{a, c, b, d\}$. The interior (i.e. away from the end points) nodes in this list have their order reversed. We detect spuriously long segments by finding the arc length of all segments and comparing the length of a segment of $n$-many nodes with the other segments of $n$-many nodes. If the length of the segment we are currently considering is two standard deviations greater than the mean, we check if the length is decreased by reversing the order of the interior nodes. The proposed segment with reversed interior nodes replaces the original segment if and only if it has a lower arc length than the original segment. If the length is decreased, it is considered to have been spuriously long and is now fixed.

The total arc length of the vessels in network (a) is 6271\;mm and a volume of 5207\;mm$^3$. The network contains 1976 segments, of which 46 were spuriously long. After removing the short segments, the total length of the arc of the network is reduced to 6184\;mm and the volume is reduced to 5151\;mm$^3$. A change of about 1\% for both metrics. The other trees follow similar patterns of changes. Larger trees that contain proportionally more segments have more spuriously long segments. Due to the size and number of spuriously long segments, they do not impact the appearance of the whole tree.

\begin{figure}[ht]\centering
    \subfloat[]{\includegraphics[width = 0.2\textwidth]{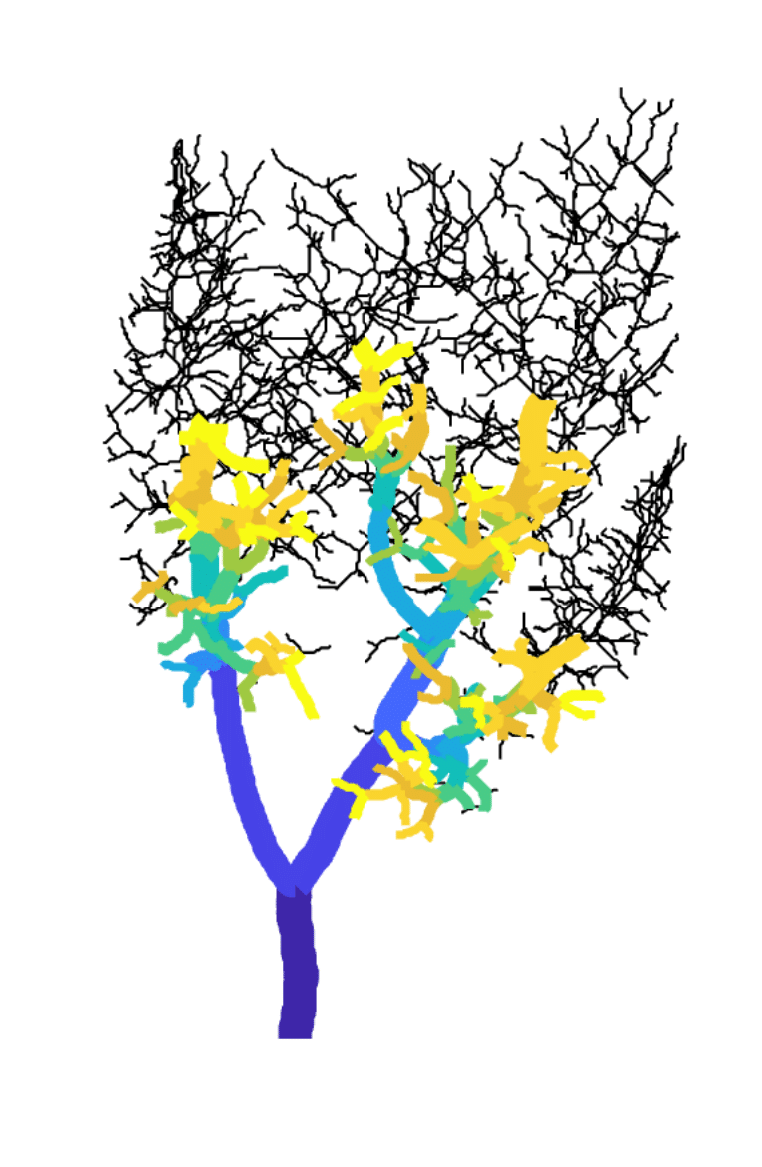}}
    \subfloat[]{\includegraphics[width =  0.2\textwidth]{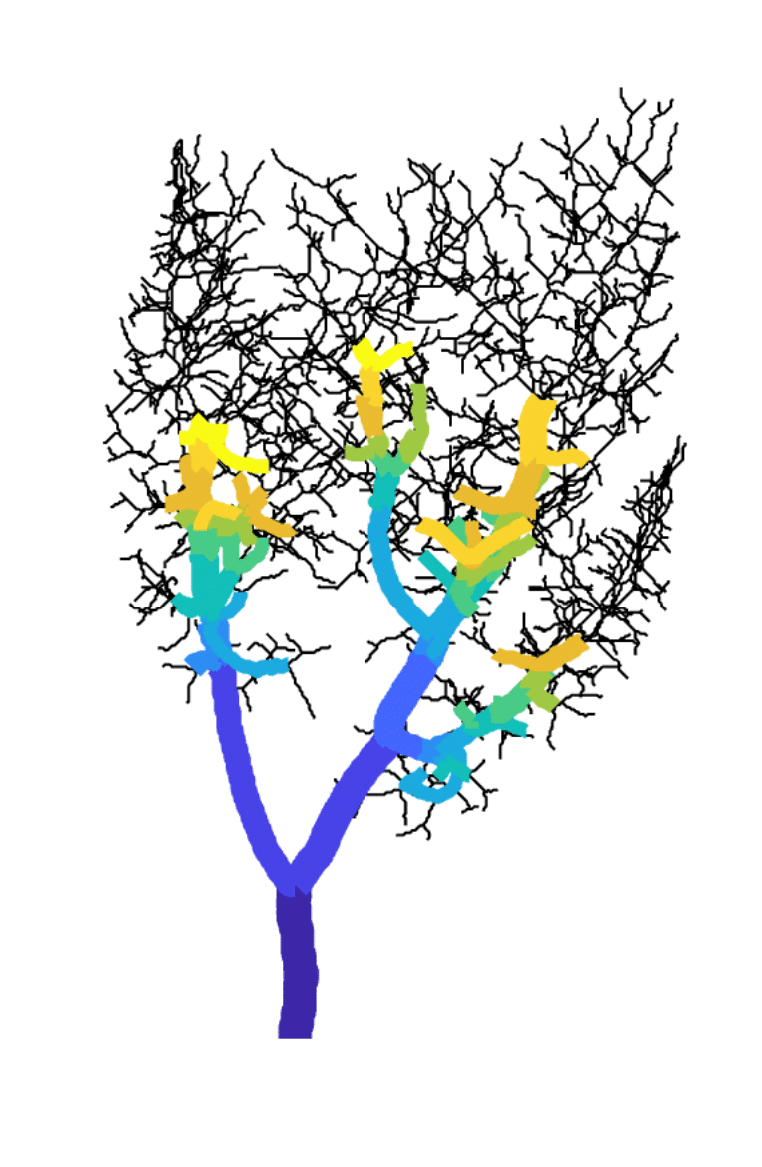}}
    \caption{(a) Segments of the first 10 generations from the parent are shown in thick, coloured lines. Black segments have been removed from the final tree. (b) Segments within the first 10 generations from the parent vessel that have a mean intra-segment radius of at least 3\;mm are shown with thick, coloured lines. The thin, black segments do not meet the inclusion criteria.
    } \label{fig:A:filter}
\end{figure}

The murine pulmonary networks are now in an appropriate format to use with the techniques described above. We have already seen the application of the Strahler order and generation count to the trees. The first tree (shown in panels (a)) contains 1976 segments in 29 generations. The first 10 generations after the parent vessel contain 362 of these segments. A tree containing no more than 10 branchings from the inlet and no vessel with mean radius larger than 0.3\;mm is made up of 113 segments. These trees are shown in Fig.~\ref{fig:A:filter}, where the left tree is filtered only by generation count, and the right tree is filtered by generation and radius. The black segments in both images are those that were discarded. The filtered trees seen below were created with a process identical to that employed to filter the coronary data and make, for example, Figs.~\ref{fig:rad} and \ref{fig:firstSix}.

We do not have surface the pulmonary meshes, so are unable to repeat the spatial analysis of the 3D data to find subdomains of the murine lungs perfused from a given vessel.

\bibliographystyle{plain}

\end{document}